\documentclass{jfm}
 \pdfoutput=1
\usepackage{graphicx}
\usepackage{newtxtext}
\usepackage{newtxmath}
\usepackage{natbib}
\usepackage{hyperref}
\hypersetup{
	colorlinks = true,
	urlcolor   = blue,
	citecolor  = black,
}
\usepackage{color}
\definecolor{magenta}{rgb}{1,0.5,0}
\definecolor{amethyst}{rgb}{0.6,0.4,0.8}
\definecolor{aureolin}{rgb}{0.99,0.93,0.0}
\definecolor{awesome}{rgb}{1.0,0.13,0.32}
\definecolor{ao-green}{rgb}{0.0, 0.5, 0.0}
\definecolor{applegreen}{rgb}{0.55, 0.71, 0.0}
\definecolor{armygreen}{rgb}{0.29, 0.33, 0.13}
\definecolor{cadmiumgreen}{rgb}{0.0, 0.42, 0.24}

\newcommand{\RomanNumeralCaps}[1]
\nolinenumbers

\title{A resolvent-based prediction framework for incompressible turbulent channel flow with limited measurements}

\author{Anjia Ying\aff{1},
	Tian Liang\aff{1},
	Zhigang Li\aff{1},
	Lin Fu\aff{1,2,3,4,}
	\corresp{\email{linfu@ust.hk}}
}

\affiliation{
	\aff{1} Department of Mechanical and Aerospace Engineering, The Hong Kong University of Science and Technology, Clear Water Bay, Kowloon, Hong Kong
	\aff{2} Department of Mathematics, The Hong Kong University of Science and Technology, Clear Water Bay, Kowloon, Hong Kong
	\aff{3} HKUST Shenzhen-Hong Kong Collaborative Innovation Research Institute, Futian, Shenzhen, PR China
	\aff{4} Center for Ocean Research in Hong Kong and Macau (CORE), The Hong Kong University of Science and Technology, Clear Water Bay, Kowloon, Hong Kong
}

\begin{document}
	\maketitle
	
	\begin{abstract}
		A new resolvent-based method is developed to predict the space-time properties of the flow field. To overcome the deterioration of the prediction accuracy with the increasing distance between the measurements and predictions in the Resolvent-Based Estimation (RBE), the newly proposed method utilizes the RBE to estimate the relative energy distribution near the wall rather than the absolute energy directly estimated from the measurements. Using this extra information from RBE, the new method modifies the energy distribution of the spatially uniform and uncorrelated forcing that drives the flow system by minimizing the norm of the cross-spectral density (CSD) tensor of the error matrix in the near-wall region in comparison with the RBE-estimated one, and therefore it is named as the Resolvent-informed White-noise-based Estimation (RWE) method. For validation, three time-resolved direct numerical simulation (DNS) datasets with the friction Reynolds numbers $Re_\tau = 180$, 550, and 950 are generated, with various locations of measurements ranging from the near-wall region ($y^+ = 40$) to the upper bound of the logarithmic region ($y/h \approx 0.2$) for the predictions. Besides the RWE, three existing methods, i.e., the RBE, the $\lambda$-model, and the White-noise-Based Estimation (WBE), are also included for the validation. The performance of the RBE and $\lambda$-model in predicting the energy spectra shows a strong dependence on the measurement locations. The newly proposed RWE shows a low sensitivity on $Re_{\tau}$ and the measurement locations, which may range from the near-wall region to the upper bound of the logarithmic region, and has a high accuracy in predicting the energy spectra. The RWE also performs well in predicting the space-time properties in terms of the correlation magnitude and the convection velocity. We further utilize the new method to reconstruct the instantaneous large-scale structures with measurements from the logarithmic region. Both the RWE and RBE perform well in estimating the instantaneous large-scale structure, and the RWE has smaller errors in the estimations near the wall. The structural inclination angles (SIAs) around $15^\circ$ are predicted by the RWE and WBE, which generally recover the DNS results.
	\end{abstract}
	
	\begin{keywords}
		
	\end{keywords}
	
	%%%%%%%%%%%%%%%%%%%%%%%%%%%%%%%%%%%%%%%%%%%%%%%%%%%%%%%%%%%%%%%%%%%%%%%%%%%%%%%%%%%%%%%%%%%%%%%%%%%%%%%%%%
	\section{Introduction}
	\label{sec:introduction}
	Credible predictions of turbulent flows have long been an essential concern for numerical and experimental studies. However, in many cases, only part of the flow information could be obtained due to the inaccuracy of the numerical models or the limitations of the experimental measurement techniques. For instance, in the numerical studies, the popular wall-modelled large-eddy simulation (WMLES) resolves only the large-scale flow motions beyond the local grid-scale and models the effect of the near-wall small-scale motions with RANS-like methods to balance the accuracy and cost for turbulence simulation, see, e.g., \citet{larsson2016large} and \citet{fu2021shock,fu2022prediction}. However, the velocity fluctuation in the near-wall region, which is an important turbulence property, is missing or inaccurate in WMLES  \citep{Bae2018}. As for the experimental studies, the measuring points are usually sparsely placed in space due to technical limitations, and important flow information might be lost. To derive the missing flow information in the unresolved or unmeasured regions, many researchers attempt to estimate the turbulence statistics from limited sets of available flow data, the methodologies of which could be generally categorized into the data-driven approaches \citep{Townsend1976,marusic2010predictive,baars2016spectral,guastoni2021convolutional} and the physics-based approaches \citep{mckeon2010critical,hwang2010linear}.
	
	The data-driven approaches predict the flow statistics based on the fundamental research on the statistical properties of wall-bounded turbulence, where the widely recognized attached eddy model (AEM) \citep{Townsend1976} and inner-outer interaction model (IOIM) \citep{marusic2010predictive,baars2016spectral} are proposed and extensively validated. The AEM considers the turbulence properties in the logarithmic region to be characterized by a collection of self-similar energy-containing eddies whose roots are attached to the wall. From the basic concept of attached eddies, many kinds of statistics of turbulence could be derived, such as the logarithmic profile of the variance of the streamwise velocity fluctuations. From AEM, the logarithmic laws could be further extended to describe the wall-normal distributions of higher-order even moments \citep{meneveau_marusic_2013,de2015scaling,yang2016hierarchical}. On the other hand, the IOIM states that the near-wall turbulence is influenced by the large-scale motions (LSM) and very-large-scale motions (VLSM) via the superposition and modulation effects, which also paves the way to predict the near-wall turbulence using the velocity data in the logarithmic region. Recently, the consistency between the AEM and the IOIM has been demonstrated by \citet{cheng2022consistency}, which enables isolating the attached eddies at a given single scale and further predicting their separated superposition effect in the near-wall region. With the isolated attached eddies at a given length scale, the turbulence properties can be further clarified \citep{cheng2022streamwise, cheng2023scale}. However, in practice, the IOIM needs the given data at the near-wall region to calculate the transfer kernel for predicting the footprint of the attached eddies in the near-wall region, which could not be directly applied for the predictions when the second-order flow statistics are not available. Meanwhile, the AEM only considers the attached eddies that are dominant in the logarithmic region, which is not the case in the inner and buffer layers where the smaller detached eddies are significant. Besides the above models describing the turbulence, the convolutional neural network (CNN) is also used for the data-driven prediction of the turbulence properties \citep{guemes2021coarse,guastoni2021convolutional}. For instance, using the model parameters trained from the existing datasets, \citet{guastoni2021convolutional} reconstruct the flow field with the shear stress measurements at the wall.
	
	In addition to the data-driven approaches, the physics-based ones predict the turbulence field based on the Navier-Stokes equations, which govern the flow dynamics. In general, the physics-based approaches rearrange the Navier-Stokes equations to the form of the linearized relationship between the nonlinear forcing (input) and the response of velocity, pressure, and temperature (output) \citep{mckeon2010critical,hwang2010linear}. Specifically, when the linearized relationship is defined in the frequency domain, the operator that builds the linear relationship is named the resolvent operator \citep{mckeon2010critical}. Taking Fourier transformation to the linearized Navier-Stokes equations in all the uniform spatial and temporal directions, the linearized relationship between the nonlinear forcing and the response at each scale is extracted, where the nonlinear forcing involves the convolutions from all the other scales \citep{mckeon2017engine}. So far, the relationship between the forcing and the response is equivalent to the original form of the Navier-Stokes equations without any assumptions. The response could be fully recovered as long as the nonlinear forcing is completely known \citep{morra2021colour}. However, the completed knowledge of the nonlinear forcing is unavailable as long as the flow data are not totally known. Despite this point, important turbulent properties can be extracted from the resolvent operator itself when the mean velocity profile is known. For instance, the forcing and response modes ordered by their gains could be obtained after taking singular value decomposition to the resolvent operator. When the gain of the leading mode dominates those of the sequential modes, which is referred to as the low-rank behaviour \citep{pickering2021optimal}, it can effectively construct a low-dimensional description of the turbulence \citep{sharma2013coherent, moarref2013model, mckeon2017engine}. Assuming the low-rank behaviour, \citet{beneddine2016conditions} predict the streamwise velocity spectra by fitting the amplitude of the leading response mode, which minimizes the square of the error between the predicted profile and the measurements at different positions. However, the energy of the leading resolvent mode is not always predominant over the following modes \citep{morra2021colour}, which means that only taking the leading mode inevitably sacrifices much information.
	
	Rather than considering only the leading mode, another group of physics-based approaches implicitly take advantage of the low-rank behaviour by treating the unknown nonlinear forcing as white in space \citep{hwang2010linear,morra2019relevance,madhusudanan2019coherent}. Since the white-noise assumptions imply that the energies of nonlinear forcing modes are equal to each other among all the modes \citep{towne2018spectral}, the energies of the response modes are proportional to the gains accordingly. By modeling a part of the forcing with the eddy viscosity terms \citep{cess1958survey,reynolds1972mechanics} in the linearized Navier-Stokes equations and assuming that the remaining forcing is white noise, the accuracy of prediction on the fluctuation statistics is much improved \citep{hwang2010linear,morra2019relevance}. Later, \citet{madhusudanan2019coherent} estimate the superposition effect of the large-scale structures on the near wall region using the eddy-viscosity model with the remaining forcing assumed as white noise. \citet{gupta2021linear} further improve the work of \citet{madhusudanan2019coherent} by modifying the forcing profile according to the wall-normal distributions of the eddy viscosity and flow scales.
	
	Besides the above approaches that assume a predefined forcing profile, the Kalman filter-based approaches \citep{hoepffner2005state,chevalier2006state} and resolvent-based approaches \citep{towne2020resolvent,martini2020resolvent} are also proposed. \citet{illingworth2018estimating} estimate the large-scale structures in wall turbulence with the H2-optimal approaches. \citet{illingworth2018estimating} also demonstrate significant improvement in the predictions when the eddy-viscosity term is introduced in the linearized Navier-Stokes equations. \citet{towne2020resolvent} predict the two-point space-time statistics in the near-wall region of turbulent channel flow by estimating the minimum forcing that could fully reproduce the measurements, namely the RBE. Since the approach only estimates the minimum forcing that recovers the observations, it underestimates the forcing that generates large responses at other positions but has minor effects on the measurements \citep{karban2022self}. In terms of practical applications, the measurements are not guaranteed to be located in the vicinity of the prediction region, which means that the direct applications of RBE are limited.
	Later, \citet{martini2020resolvent} generalizes the RBE approach by taking the forcing color and sensor noise into account when constructing the transfer function for estimation. An optimal linear estimator of the flow states can be obtained by taking the real forcing CSD tensors as input. In practice, the forcing statistics can be estimated from additional sensors or approximated from reasonable models. The improved RBE \citep{martini2020resolvent} recovers the original one \citep{towne2020resolvent} when the forcing is assumed to be white in space and the sensor noise is neglected.
	\citet{amaral2021resolvent} predict the turbulent channel flow with wall shear stress and pressure using the improved RBE, showing that the flow information can be better predicted when the actual forcing spectra are already known.	
	In practice, a method that can avoid rather predefined or already-known forcing statistics is indeed needed to fill the gap between theory and real applications.
	
	With the current physics-based methods, the predictions based on the white-noise assumptions \citep{morra2019relevance,madhusudanan2019coherent,gupta2021linear} could represent the general properties of the turbulence, but the accuracy is limited. On the other hand, without known forcing statistics, the RBE \citep{towne2020resolvent} performs well when the prediction layer is located near the measurement layer but becomes invalid when the prediction location moves far away from the measurement location. 
	In cases where the measurements are not close to the prediction region, both the methods based on white-noise assumptions and the RBE are not expected to provide accurate predictions on the flow information. However, these two kinds of methods could compensate for each other in a sense.
	The new method proposed in this study builds the skeleton of forcing based on the white noise to maintain the predicted response energy even when the measurement is far away from the prediction, while refines the forcing profile with near-wall relative energy profile from the RBE results. Through the above procedure, the advantages of these two kinds of methods are combined in the new approach.
	
	The remainder of this article is organized as follows. In section \ref{sec:review}, existing prediction methods are reviewed and discussed. In section \ref{sec:methodology}, the prediction method is derived and illustrated. In section \ref{sec:resluts}, the newly proposed method is validated by comparing the prediction with the results from the DNS data and existing prediction methods. Discussions and concluding remarks are presented in section \ref{sec:Conclusions}.
	
	%%%%%%%%%%%%%%%%%%%%%%%%%%%%%%%%%%%%%%%%%%%%%%%%%%%%%%%%%%%%%%%%%%%%%%%%%%%%%%%%%%%%%%%%%%%%%%%%%%%%%%%%%%
	\section{The existing methods}\label{sec:review}
	In this section, the mathematical description of the resolvent analysis is introduced, followed by a brief review of the existing methods to predict the flow field, including the WBE \citep{morra2019relevance}, the Wall-distance-dependent Model (W-model) and the Scale-dependent Model ($\lambda$-model) \citep{gupta2021linear}, as well as the RBE \citep{towne2020resolvent,martini2020resolvent}. Basic ideas of the to-be-reviewed methods will form the cornerstone of our newly proposed method derived in section \ref{sec:methodology}.
	
	%%%%%%%%%%%%%%%%%%%%%%%%%%%%%%%%%%%%%%%%%%%%%%%%%%%%%%%%%%%%%%%%%%%%%%%%%%%%%%%%%%%%%%%%%%%%%%%%%%%%%%%%%%
	\subsection{Mathematical description of the resolvent analysis}\label{subsec:math_resolvent}
	The incompressible Navier-Stokes equations are given by
	\begin{subeqnarray}
		\label{eq:nsequations}
		&\displaystyle\frac{\p \boldsymbol{u}}{\p t}+\boldsymbol{u}\cdot\bnabla\boldsymbol{u}
		= -\bnabla p+\frac{1}{Re_\tau}\bnabla\cdot(\bnabla\boldsymbol{u}+\bnabla\boldsymbol{u}^{\rm T}), \\ [4pt]
		&\bnabla\cdot\boldsymbol{u} = 0,
	\end{subeqnarray}
	where $Re_\tau = \frac{u_{\tau} h}{\nu}$ is the friction Reynolds number, $u_{\tau}$ is the friction velocity, $h$ is the half-channel height, $\nu$ is the kinematic viscosity, and the superscript ${}^{\rm T}$ denotes transpose. Following the previous studies \citep{illingworth2018estimating,morra2019relevance,towne2020resolvent}, the forcing $\boldsymbol{f}$ that contains the nonlinear interactions of velocity fluctuations while excluding the eddy-viscosity term is defined as
	\begin {equation}
	\label{eq:forcingdefinition}
	\boldsymbol{f}=
	-\boldsymbol{u}^\prime\cdot\bnabla\boldsymbol{u}^\prime
	-\frac{1}{Re_\tau}\bnabla\cdot[\frac{\nu_t}{\nu}(\bnabla\boldsymbol{u}^\prime+{\bnabla\boldsymbol{u}^\prime}^{\rm T})],
\end{equation}
where the superscript ${}^\prime$ denotes the fluctuation variable. Here, the eddy viscosity $\nu_t$ is calculated from the semi-empirical expression by \citet{cess1958survey} and reported by \citet{reynolds1972mechanics} as
\begin{equation}
	\label{eq:eddyviscosity}
	\displaystyle\nu_t=\frac{\nu}{2} \left\{ 1+\frac{\kappa^2 Re_\tau ^2}{9} 
	(2y-y^2)^2 (3-4y+2y^2)^2 [1-{\rm exp} (\frac{-Re_\tau y}{A})]^2 \right\} ^\frac{1}{2} -\frac{\nu}{2},
\end{equation}
where the constants $\kappa=0.426$ and $A=25.4$. By rearranging equation (\ref{eq:nsequations}), the linearized Navier-Stokes equations hold,
\begin{subeqnarray}
	\label{eq:linearnsequations}
	&\displaystyle\frac{\p \boldsymbol{u}^\prime}{\p t}+\overline{\boldsymbol{u}}\cdot\bnabla\boldsymbol{u}^\prime
	+\boldsymbol{u}^\prime\cdot\bnabla\overline{\boldsymbol{u}}+\bnabla p^\prime
	-\frac{1}{Re_\tau}\bnabla\cdot[\frac{\nu_T}{\nu}(\bnabla\boldsymbol{u}^\prime+{\bnabla\boldsymbol{u}^\prime}^{\rm T})]
	= \boldsymbol{f}, \\ [4pt]
	&\bnabla\cdot\boldsymbol{u}^\prime = 0,
\end{subeqnarray}
where $\overline{\boldsymbol{u}}$ is the mean velocity, and $\nu_T=\nu_t+\nu$ is the total viscosity. Note that equations (\ref{eq:linearnsequations}) are equivalent to the incompressble Navier-Stokes equations (\ref{eq:nsequations}).

From equations (\ref{eq:forcingdefinition}) and (\ref{eq:linearnsequations}), the inclusion of the eddy viscosity model into the linearized Navier-Stokes equations changes both the definition of forcing and the linearized relationship between the forcing and response. In \citet{mckeon2010critical}, the total forcing is defined as $\boldsymbol{f}= -\boldsymbol{u}^\prime\cdot\bnabla\boldsymbol{u}^\prime$. The forcing re-defined in this study is equivalent to the remaining portion of the forcing in \citet{mckeon2010critical} after excluding the eddy-viscosity term $\frac{1}{Re_\tau}\bnabla\cdot[\frac{\nu_T}{\nu}(\bnabla\boldsymbol{u}^\prime+{\bnabla\boldsymbol{u}^\prime}^{\rm T})]$. In the following sections, the tested prediction methods actually provide different approaches to model the re-defined forcing in (\ref{eq:forcingdefinition}). The total forcing, on the other hand, is equal to the summation of the portion that is modeled by the eddy viscosity terms and the remaining portion that is modeled by the prediction methods.

The linearized Navier-Stokes equations in each spatial scale $\boldsymbol{k}_{\rm s}$ are obtained by taking the Fourier transformation to equations (\ref{eq:linearnsequations}) in the uniform spatial directions. For instance, in the fully developed turbulent channel flow, the Fourier transformation is taken in the streamwise $(x)$ and spanwise $(z)$ directions.
The linearized equations (\ref{eq:linearnsequations}) at $\boldsymbol{k}_{\rm s} =  \left( k_x , k_z \right)$ can be therefore written in a discretized state-space form with $N$ points in the wall-normal direction as
\begin{subeqnarray}
	\label{eq:ss_linearnsequations}
	&\mathcal{M} \frac{\p \boldsymbol{q}_{\boldsymbol{k}_{\rm s}} (t)}{\p t}
	= \mathcal{A}_{\boldsymbol{k}_{\rm s}} \boldsymbol{q}_{\boldsymbol{k}_{\rm s}} (t) + \mathcal{B}\boldsymbol{f}_{\boldsymbol{k}_{\rm s}} (t), \\ [4pt]
	&\boldsymbol{m}_{\boldsymbol{k}_{\rm s}} (t) = \mathcal{C}\boldsymbol{q}_{\boldsymbol{k}_{\rm s}} (t) + \boldsymbol{n}_{\boldsymbol{k}_{\rm s}} (t),
\end{subeqnarray}
where
$\boldsymbol{q}_{\boldsymbol{k}_{\rm s}} (t) = \left[ \boldsymbol{u}^{\rm T}_{\boldsymbol{k}_{\rm s}} (t),p_{\boldsymbol{k}_{\rm s}} (t) \right]^{\rm T} $,
$\boldsymbol{m}_{\boldsymbol{k}_{\rm s}} (t) \in \mathbb{C}^{N_{\rm m}}$ is the system observation, 
$\boldsymbol{n}_{\boldsymbol{k}_{\rm s}} (t) \in \mathbb{C}^{N_{\rm m}}$ is the measurement noise, and $N_{\rm m}$ is the number of observations.
The expressions of the operators $\mathcal{M}$, $\mathcal{A}_{\boldsymbol{k}_{\rm s}}$, $\mathcal{B}$, and $\mathcal{C}$ are provided in Appendix \ref{appA}.
Further, taking the Fourier transformation to equation (\ref{eq:ss_linearnsequations}$a$) in the temporal direction, the following linear relationship at each spatio-temporal scale $\boldsymbol{k} = \left( \boldsymbol{k}_{\rm s},\omega \right)$ holds,
\begin{equation}
	\label{eq:resolventeqn1}
	\hat{\boldsymbol{q}}_{\boldsymbol{k}}
	=
	\mathcal{H}_{\boldsymbol{k}}
	\cdot
	\hat{\boldsymbol{f}}_{\boldsymbol{k}} \\,
\end{equation}
where $\hat{\boldsymbol{u}}_{\boldsymbol{k}}$, $\hat{p}_{\boldsymbol{k}}$ and $\hat{\boldsymbol{f}}_{\boldsymbol{k}}$ are the Fourier coefficients of velocity, pressure and forcing at scale $\boldsymbol{k}$, respectively, and the resolvent operator $\mathcal{H}_{\boldsymbol{k}}$ is expressed as,
\begin{equation}
	\label{eq:resolventoperator}
	\setlength{\arraycolsep}{0pt}
	\renewcommand{\arraystretch}{1.3}
	\setlength{\arraycolsep}{2.0pt}
	\mathcal{H}_{\boldsymbol{k}} = \left(-{\rm i}\omega
	\mathcal{M}
	- \mathcal{A}_{\boldsymbol{k}_{\rm s}} \right)^{-1},
\end{equation}
where ${\rm i}=\sqrt{-1}$. From equation (\ref{eq:resolventeqn1}), the velocity $\hat{\boldsymbol{u}}_{\boldsymbol{k}}$ and pressure $\hat{p}_{\boldsymbol{k}}$ are regarded as the response of the input forcing $\hat{\boldsymbol{f}}_{\boldsymbol{k}}$ through the linear operator $\mathcal{H}_{\boldsymbol{k}}$. Since we will focus on predicting the velocity field in this study, the linearized equation (\ref{eq:resolventoperator}) can be reduced to the following form,
\begin{equation}
	\label{eq:resolventeqn2}
	\hat{\boldsymbol{u}}_{\boldsymbol{k}}
	=
	\mathcal{R}_{\boldsymbol{k}}
	\cdot
	\hat{\boldsymbol{f}}_{\boldsymbol{k}},
\end{equation}
where $\mathcal{R}_{\boldsymbol{k}}$ is obtained from $\mathcal{H}_{\boldsymbol{k}}$ by deleting the rows corresponding to $\hat{p}_{\boldsymbol{k}}$ at the left-hand side of equation (\ref{eq:resolventoperator}) and columns corresponding to the constant 0 at the right-hand side. A typical application of equation (\ref{eq:resolventeqn2}) is to approximate the coherent structures of turbulence using the resolvent modes, which are obtained by taking the singular value decomposition to $\mathcal{R}_{\boldsymbol{k}}$, i.e.,
\begin{equation}
	\label{eq:resolventeqn_decomp}
	\hat{\boldsymbol{u}}_{\boldsymbol{k}} \\
	=
	\sum_{j=1}^\infty
	\left(
	\boldsymbol{\hat{\psi}}_{\boldsymbol{k},j}
	\sigma_{\boldsymbol{k},j}
	\boldsymbol{\hat{\varphi}}_{\boldsymbol{k},j}^\ast
	\right)
	\cdot
	\hat{\boldsymbol{f}}_{\boldsymbol{k}}
	=
	\sum_{j=1}^\infty\sigma_{\boldsymbol{k},j}
	\boldsymbol{\hat{\psi}}_{\boldsymbol{k},j}
	\left( \boldsymbol{\hat{\varphi}}_{\boldsymbol{k},j}^\ast
	\cdot
	\hat{\boldsymbol{f}}_{\boldsymbol{k}}\right)
	=
	\sum_{j=1}^\infty\sigma_{\boldsymbol{k},j}
	\boldsymbol{\hat{\psi}}_{\boldsymbol{k},j}
	\boldsymbol{\beta} _{\boldsymbol{k},j},
\end{equation}
where $	\sum_{j=1}^\infty
\left(
\boldsymbol{\hat{\psi}}_{\boldsymbol{k},j}
\sigma_{\boldsymbol{k},j}
\boldsymbol{\hat{\varphi}}_{\boldsymbol{k},j}^\ast
\right) = \mathcal{R}_{\boldsymbol{k}}$ is the result of singular value decomposition of the resolvent operator $\mathcal{R}_{\boldsymbol{k}}$, the response mode $\boldsymbol{\hat{\psi}}_{\boldsymbol{k},j}$ and forcing mode $\boldsymbol{\hat{\varphi}}_{\boldsymbol{k},j}$ are ordered by their singular value $\sigma_{\boldsymbol{k},j}$,
the expansion coefficient $\boldsymbol{\beta} _{\boldsymbol{k},j}=\left( \boldsymbol{\hat{\varphi}}_{\boldsymbol{k},j}^\ast \cdot \hat{\boldsymbol{f}}_{\boldsymbol{k}}\right)$ is the projection of forcing on the $j{\rm -th}$ resolvent forcing mode, and the superscript ${}^{\ast}$ denotes the Hermitian transpose.

Using equation (\ref{eq:resolventeqn2}), the CSD tensors can be calculated as,
\begin{equation}
	\label{eq:resolvent_CSD}
	S_{\boldsymbol{uu},\boldsymbol{k}}
	=
	\left\langle \hat{\boldsymbol{u}}_{\boldsymbol{k}} \hat{\boldsymbol{u}}_{\boldsymbol{k}}^{\ast} \right\rangle 
	=
	\mathcal{R}_{\boldsymbol{k}}
	\cdot
	\left\langle \hat{\boldsymbol{f}}_{\boldsymbol{k}} \hat{\boldsymbol{f}}_{\boldsymbol{k}}^{\ast} \right\rangle 
	\cdot
	\mathcal{R}_{\boldsymbol{k}}^{\ast}
	=
	\mathcal{R}_{\boldsymbol{k}}
	S_{\boldsymbol{ff},\boldsymbol{k}}
	\mathcal{R}_{\boldsymbol{k}}^{\ast},
\end{equation}
where $\left\langle \cdot \right \rangle $ denotes the ensemble average. As in equation (\ref{eq:resolvent_CSD}), the CSD tensor of the response is fully determined when the forcing CSD $S_{\boldsymbol{ff},\boldsymbol{k}}$ is known. Thus, in many existing methods (e.g., \citet{morra2019relevance},  \citet{towne2020resolvent}, \citet{gupta2021linear}), the estimating or modeling object is actually $S_{\boldsymbol{ff},\boldsymbol{k}}$, after which the response $S_{\boldsymbol{uu},\boldsymbol{k}}$ can be directly derived from $S_{\boldsymbol{ff},\boldsymbol{k}}$ via equation (\ref{eq:resolvent_CSD}). In the following, the typical methods that make use of the resolvent analysis to predict the turbulence field will be introduced.

%%%%%%%%%%%%%%%%%%%%%%%%%%%%%%%%%%%%%%%%%%%%%%%%%%%%%%%%%%%%%%%%%%%%%%%%%%%%%%%%%%%%%%%%%%%%%%%%%%%%%%%%%%
\subsection{The white-noise-based estimation}\label{subsec:WBE}
With the simplest assumption, the portion of forcing that excludes the eddy viscosity terms, as defined in equation (\ref{eq:forcingdefinition}), can be modeled to be white as in some researches \citep{hwang2010linear,madhusudanan2019coherent,morra2019relevance}, which means it is spatially uniform and uncorrelated.
Note that, in this approach, the total forcing as defined by \citet{mckeon2010critical} is the summation of the eddy-viscosity portion and the white-noise-assumed portion.
With the white-noise assumption of the WBE approach, the CSD tensor of forcing at each scale $\boldsymbol{k}$ can be expressed as,
\begin{equation}
	\label{eq:white_forcing}
	S_{\boldsymbol{ff},\boldsymbol{k},{\rm {WBE}}}=
	E_{\boldsymbol{k}} \cdot \mathcal{I},
\end{equation}
where $E_{\boldsymbol{k}}$, as the energy of forcing, keeps constant at each node and in each direction. Substituting equation (\ref{eq:white_forcing}) into (\ref{eq:resolvent_CSD}) and considering the resolvent modes in equation (\ref{eq:resolventeqn_decomp}), it can be deduced that,
\begin{equation}
	\label{eq:resolvent_WBE_decompose}
	S_{\boldsymbol{uu},\boldsymbol{k},{\rm {WBE}}}
	=
	E_{\boldsymbol{k}}
	(\mathcal{R}_{\boldsymbol{k}}
	\mathcal{R}_{\boldsymbol{k}}^{\ast})
	=
	E_{\boldsymbol{k}}
	\sum_{j=1}^\infty\sigma_{\boldsymbol{k},j}^2
	(\boldsymbol{\hat{\psi}}_{\boldsymbol{k},j}
	\boldsymbol{\hat{\psi}}_{\boldsymbol{k},j}^{\ast}).
\end{equation}
If we define $(\boldsymbol{\hat{\psi}}_{\boldsymbol{k},j}\boldsymbol{\hat{\psi}}_{\boldsymbol{k},j}^{\ast})$ as the CSD of the response at the $j$-th mode, the resultant response CSD can be interpreted as the linear summation of the CSDs of all the resolvent modes weighted by the gains $\sigma_{\boldsymbol{k},j}^2$. 
The white forcing, as the simplest form, can be utilized to estimate the coherent structures \citep{hwang2010linear,madhusudanan2019coherent,gupta2021linear} and the spectra of turbulence \citep{morra2019relevance}. However, the resultant accuracy of prediction with the initial white forcing is far from engineering usage \citep{towne2020resolvent} even if the forcing is partially modeled by the eddy-viscosity term.

%%%%%%%%%%%%%%%%%%%%%%%%%%%%%%%%%%%%%%%%%%%%%%%%%%%%%%%%%%%%%%%%%%%%%%%%%%%%%%%%%%%%%%%%%%%%%%%%%%%%%%%%%%
\subsection{The wall-distance-dependent and scale-dependent models}\label{subsec:L-model}
To improve the prediction accuracy of the white forcing, \citet{gupta2021linear} propose the Wall-distance-dependent Model (W-model) and Scale-dependent Model ($\lambda$-model) by modifying the profile of the forcing according to the profile of the eddy viscosity and the flow scales. The W-model and $\lambda$-model maintain the diagonal property of $S_{\boldsymbol{ff},\boldsymbol{k}}$ as in equation (\ref{eq:white_forcing}), while modifying the vertical energy distribution. Extended from the work of \citet{jovanovic2005componentwise} in laminar flow, \citet{gupta2021linear} propose the W-model by assuming that the vertical profile of forcing energy is proportional to that of the eddy viscosity, i.e.,
\begin{equation}
	\label{eq:W-model}
	W_{\boldsymbol{k}}
	=
	E_{\boldsymbol{k}} \nu_t,
\end{equation}
where $W_{\boldsymbol{k}}$ denotes the diagonal of $S_{\boldsymbol{ff},\boldsymbol{k}}$, i.e., the energy profile of forcing at $\boldsymbol{k}$. Based on the fact that the nonlinear interaction of turbulence is scale-dependent \citep{cho2018scale}, \citet{gupta2021linear} further propose the $\lambda$-model with the modified eddy viscosity, i.e.,
\begin{equation}
	\label{eq:lambda-model}
	\nu_{t,{\boldsymbol{k}}}
	=
	\frac{\lambda}{\lambda + \lambda_m} \nu_t,
\end{equation}
where $\lambda = 2\pi / (k_x^2 + k_z^2)^{0.5}$, and $\lambda_m (y) = 50/{Re_\tau}+(2-50/{Re_\tau}) {\rm{tanh}} (6y)$. The forcing energy profile is thereby calculated by $W_{\boldsymbol{k}}=E_{\boldsymbol{k}} \nu_{t,{\boldsymbol{k}}}$.
The strategies of modifying the forcing profiles by \citet{gupta2021linear} efficiently improve the prediction of the vertical spatial correlation of streamwise velocity fluctuations in turbulent channel flow. However, the W-model and $\lambda$-model cannot accurately predict the energy distribution of velocity fluctuation in the near-wall region, as will be further discussed in the following sections.
%%%%%%%%%%%%%%%%%%%%%%%%%%%%%%%%%%%%%%%%%%%%%%%%%%%%%%%%%%%%%%%%%%%%%%%%%%%%%%%%%%%%%%%%%%%%%%%%%%%%%%%%%%
\subsection{The resolvent-based estimation}\label{subsec:RBE}
The RBE method \citep{towne2020resolvent} estimates the minimum L2-norm forcing that can fully recover the measured signal. Denoting $\boldsymbol{m}$ as the measured set of variables and $\boldsymbol{u}$ as the complete set of variables from all across the computational domain, where $\boldsymbol{m}=\mathscr{C} \boldsymbol{u}$, the RBE could be briefly expressed as,
\begin{equation}
	\label{eq:RBE_forcing}
	S_{\boldsymbol{ff},\boldsymbol{k},{\rm{RBE}}}= T_{ \boldsymbol{ff}, \boldsymbol{k}} \cdot S_{\boldsymbol{mm},\boldsymbol{k}} \cdot T_{ \boldsymbol{ff}, \boldsymbol{k}}^{\ast},
\end{equation}
where the estimation operator $T_{ \boldsymbol{ff}, \boldsymbol{k}} = (\mathscr{C} \mathcal{R}_{\boldsymbol{k}})^{\dagger}$, $S_{\boldsymbol{mm},\boldsymbol{k}} = \mathscr{C}S_{\boldsymbol{uu},\boldsymbol{k}}\mathscr{C}^{\rm T}$ is the CSD tensor of the measured variables at scale $\boldsymbol{k}$, $S_{\boldsymbol{ff},\boldsymbol{k},{\rm{RBE}}}$ is the estimated CSD tensor of forcing, the expression of the observation matrix $\mathscr{C}$ can be found in equation (\ref{eq:C_2}) in Appendix \ref{appA}, and the superscript ${}^{\dagger}$ denotes the pseudo-inverse. With the estimated forcing, the CSD tensor of the complete set of variable $\boldsymbol{u}$ is calculated by
\begin{equation}
	\label{eq:RBE_response}
	S_{ \boldsymbol{uu}, \boldsymbol{k},{\rm RBE}}= \mathcal{R}_{\boldsymbol{k}}S_{\boldsymbol{ff},\boldsymbol{k},{\rm{RBE}}}{\mathcal{R}_{\boldsymbol{k}}}^{\ast}.
\end{equation}
The RBE has been validated to be efficient for predicting the field where the turbulence is highly correlated with the measured reference points \citep{towne2020resolvent,yang2020numerical}. On the other hand, since the RBE method only estimates the ``observed" forcing, its prediction deteriorates with the decrease of correlation between the signals of measurements and the predicted locations.

Later, \citet{martini2020resolvent} demonstrate that the estimated forcing in the original version of RBE in equation (\ref{eq:RBE_forcing}) is actually the stationary point of the error matrix $\overline{ \boldsymbol{\hat{e}}_{f,\boldsymbol{k}} \boldsymbol{\hat{e}}_{f,\boldsymbol{k}}^{\dagger} }$, where the error $\boldsymbol{\hat{e}}_{f,\boldsymbol{k}}$ is defined as the difference between the estimated forcing and the white-noise-assumed forcing here. Since the real forcing is not white, the estimation of the original RBE is not optimal to minimize the relative error between the estimated forcing and the real one. \citet{martini2020resolvent} then propose the improved RBE that incorporates the effect of the forcing color and the measurement noise, i.e.,
\begin{equation}
	\label{eq:Opt_RBE_forcing}
	T_{ \boldsymbol{ff}, \boldsymbol{k},{\rm {opt}}}
	=S_{\boldsymbol{ff},\boldsymbol{k}}(\mathscr{C} \mathcal{R}_{\boldsymbol{k}})^{\dagger}\left[(\mathscr{C} \mathcal{R}_{\boldsymbol{k}}) S_{\boldsymbol{ff},\boldsymbol{k}} (\mathscr{C} \mathcal{R}_{\boldsymbol{k}})^{\dagger}
	+S_{nn,\boldsymbol{k}}
	\right]^{-1},
\end{equation}
by which the stationary point of the error matrix $\overline{ \boldsymbol{\hat{e}}_{f,\boldsymbol{k}} \boldsymbol{\hat{e}}_{f,\boldsymbol{k}}^{\dagger} }$ between the estimated forcing and the real one is obtained.
$S_{nn,\boldsymbol{k}}$ is the CSD tensor of the measurement noise at scale $\boldsymbol{k}$, which is set as zero since the DNS data at a given reference layer can be directly provided as measurements without introducing additional errors in this study.
When the real forcing color from additional sensors is used to inform the transfer function, the optimized RBE performs better than the original RBE in estimating the turbulent channel flow given the same amount of measurements. However, the optimized RBE needs the knowledge of forcing statistics obtained from additional sensors or forcing models, which is not considered in this study. Thus, the RBE method mentioned in the following refers to its original version by \citet{towne2020resolvent}, which will be used to develop our newly proposed methods and provide comparison results.

The above existing methods can be categorized into two groups. First, the WBE, W-model, and $\lambda$-model assume that the forcing is uncorrelated in space, then explicitly model the forcing energy with predefined profiles. These approaches aim to describe the forcing statistics throughout the flow field. On the other hand, the information of the measurements is only utilized to determine the overall forcing energy $E_{\boldsymbol{k}}$, while the estimated relative energy distribution of forcing is independent of the measurements. The advantage of these methods with predefined profiles is that the predicted response can generally reflect the energy distribution of the response to some extent, no matter how far the measurement layers are located from the prediction region. However, the performance of the prediction is highly dependent on the specific form of the predefined profiles, which implies that the accuracy of this group of methods could not be as good as expected, even if the measurements and predictions are close to each other, as can be seen in section \ref{sec:resluts}. Second, the RBE infers the forcing statistics from the measurements without imposing assumptions on the form of the forcing profile. The RBE has been validated to be efficient for predicting the field where the turbulence is highly correlated with the measured reference points \citep{towne2020resolvent,yang2020numerical}. However, since it only estimates the ``observed" forcing, its prediction accuracy deteriorates with the increasing distance between the measurements and the predicted locations. 

%%%%%%%%%%%%%%%%%%%%%%%%%%%%%%%%%%%%%%%%%%%%%%%%%%%%%%%%%%%%%%%%%%%%%%%%%%%%%%%%%%%%%%%%%%%%%%%%%%%%%%%%%%
\section{Derivations of the Resolvent-informed White-noise-based Estimation method}\label{sec:methodology}
%%%%%%%%%%%%%%%%%%%%%%%%%%%%%%%%%%%%%%%%%%%%%%%%%%%%%%%%%%%%%%%%%%%%%%%%%%%%%%%%%%%%%%%%%%%%%%%%%%%%%%%%%%
From the above discussions of the existing methods, the group of methods assuming predefined profiles and the RBE compensate with each other in a sense. Specifically, the group of methods assuming predefined profiles performs better in estimating the general energy distribution of the forcing and response when there is a long distance between the measurements and the prediction region. On the other hand, the RBE is efficient in providing reasonable estimations when the measurements are located near the prediction. According to \citet{holford_lee_hwang_2023}, the energy spectra of turbulence can be well recovered using the spatially uncorrelated forcing with optimal profiles. In this study, following the strategy of modifying the spatially uncorrelated forcing profile, we aim to propose an adaptive method to adjust the spatially uniform and uncorrelated forcing profile based on reliable inference informed by the RBE.

As already discussed, when the measured reference layer is not close to the prediction region, the RBE cannot be directly applied for prediction due to the deterioration of accuracy. However, the RBE can, instead, be utilized to estimate the relative energy profile of the response with respect to an assumed reference layer that is closer to the wall than the actual reference layer. Based on the basic RBE formula (\ref{eq:RBE_forcing}-\ref{eq:RBE_response}), the predicted relative CSD tensor of velocity $u_i$ near the wall can be estimated as,
\begin{equation}
	\label{eq:reducedRBE_cospectra_relative}
	\hat{S}_{ u_i u_i , \boldsymbol{k},{\rm RBE}}=
	\frac{S_{ u_i u_i , \boldsymbol{k},{\rm RBE}}}{S_{mm,\boldsymbol{k}}}= \mathcal{R}_{\boldsymbol{k}}(\mathscr{C} \mathcal{R}_{\boldsymbol{k}})^{\dagger} \cdot (\mathscr{C} \mathcal{R}_{\boldsymbol{k}})^{\dagger\ast}{\mathcal{R}_{\boldsymbol{k}}}^{\ast},
\end{equation}
where $S_{mm,\boldsymbol{k}}$ should be the scalar energy of $u_i$ at a single reference height so that it can be eliminated from the denominator by the numerator. Note that there is no requirement on the specific value of the reference height in equation (\ref{eq:reducedRBE_cospectra_relative}), a pretty high accuracy can be obtained for predicting the relative CSD tensor. This desirable property of the RBE in estimating the relative response statistics provides a standard for the modification of the initially assumed white forcing profile. To be distinguished from the actual reference layer, which is denoted as $y_{\rm R}$, the assumed reference layer used for estimating the relative CSDs in equation (\ref{eq:reducedRBE_cospectra_relative}) is denoted as the quasi-reference layer $y_{\rm Q}$ in this article.
To obtain the relative CSD tensor used for the optimization, the value of $y_{\rm Q}$ should be determined first.
According to \citet{towne2020resolvent}, the accuracy of RBE decreases as the wavenumbers and frequency increase. As the wavenumbers and frequency are closely related to the flow scale, we choose to determine the quasi-reference layer according to the flow scale in the wall-normal direction. The purpose of this step is to let the RBE provide a reasonable estimation of the widest possible wall-normal extent.
To quantify the wall-normal scale, the linear coherence spectrum (LCS) $\gamma^2$ \citep{baars2016spectral} is introduced here, i.e.,
\begin{equation}
	\label{eq:LCS}
	\gamma^2 (\boldsymbol{k}) = \frac{\left| \left\langle  \hat{u} (y_{\rm Q}^+) \overline{\hat{u} (y_{\rm P}^+)} \right\rangle  \right|^2 }
	{\left\langle \left|  \hat{u} (y_{\rm Q}^+) \right|^2 \right\rangle \left\langle \left|  \hat{u} (y_{\rm P}^+) \right|^2 \right\rangle},
\end{equation}
where $\hat{u}$ is the Fourier coefficient of $u$ at scale $\boldsymbol{k}$, the overline denotes the complex conjugate, and $y_{\rm P}^+ = 15$ corresponds to the height of the near-wall inner peak \citep{smits2011high}. Since the LCS at a given distance increases as the flow scale enlarges, the LCS could be an effective index to quantify the flow scale. Based on the value of LCS preliminarily estimated from the WBE, the quasi-reference layer $y_{\rm Q}$ is set as the height beyond the inner peak where $\gamma^2 (\boldsymbol{k}) = 0.3$. When the height of $y_{\rm Q}$ exceeds $y_{\rm R}$, it will be set as $y_{\rm R}$ instead. With the above procedure, the value of $y_{\rm Q}$ is adaptively determined. Details of the implementation of the LCS calculation and discussions about the impact of the threshold LCS on the prediction accuracy are provided in Appendix \ref{appC}.

With the information of the turbulence statistics below $y_{\rm Q}$ from the RBE, our goal is to minimize the relative error between the estimated CSD of the velocity $u$ from the modified forcing and the RBE-estimated one.
As will be revealed in section \ref{subsec:RMS_profile}, the RBE just provides the reliable prediction of the relative energy profile below the quasi-reference layer $y_{\rm Q}$, which does not work well when $y \ge y_{\rm Q}$.
Thus, the modification range of the forcing is restricted below $y_{\rm Q}$.
For the region beyond the quasi-reference layer, there is no reliable information to further improve the forcing profile there. Thus, a conservative strategy is adopted by setting the forcing to be unity at each node for $y \ge y_{\rm Q}$, which corresponds to the spatially uniform forcing, as also assumed by the WBE reviewed in section \ref{subsec:WBE}. Derivations of the explicit relationship between the relative energy profiles of forcing and response are provided in Appendix \ref{appD}.
For each scale $\boldsymbol{k}=(k_x , k_z , \omega)$, the norm minimization problem is considered as follows, i.e.,
\begin{equation}
	\begin{aligned}
		\label{eq:minmimize_norm}
		& \mathop{{\rm{minimize}}}\limits_{W_{\boldsymbol{k}}}& &\left\|  \frac{S_{uu,{\boldsymbol{k}}} (W_{\boldsymbol{k}}) |_{y<y_{\rm Q}} }{S_{uu, \boldsymbol{k} , y_{\rm Q}} (W_{\boldsymbol{k}})} - \hat{S}_{ uu, \boldsymbol{k},{\rm RBE}} |_{y<y_{\rm Q}}\right\|,& \\
		& ~{\rm{subject~to~}}& &~0 \le W_{\boldsymbol{k}}(y) \le 1, ~~ \frac{{\rm d} W_{\boldsymbol{k}}(y)}{{\rm d}y} \ge 0,~~&\forall y \in [0 , y_{\rm Q}),& \\
		&&&W_{\boldsymbol{k}}(y) = 1,~~&\forall y \in [y_{\rm Q},2h],
	\end{aligned}
\end{equation}
where the norm $\| \bcdot \|$ is defined as $\left\|  \bcdot \right\| = \int_{0}^{y_{\rm Q}} (\bcdot)^2 {\rm d} y$.
This constrained optimization problem is solved with the interior point method \citep{momoh1999review}. By minimizing the norm of the error matrix, the energy spectrum of the modified response is optimized with respect to the RBE-estimated one.
After the relative profile of $W_{\boldsymbol{k}}$ is obtained from the optimization, the forcing profile will be multiplied by a unified coefficient $E_{\boldsymbol{k}}$ to match the response energy at the reference layer for all the velocity components of $u$, $v$, and $w$, respectively.

\begin{figure}\centering% Requires \usepackage{graphicx}
	{\includegraphics[width=5.0in, angle=0]{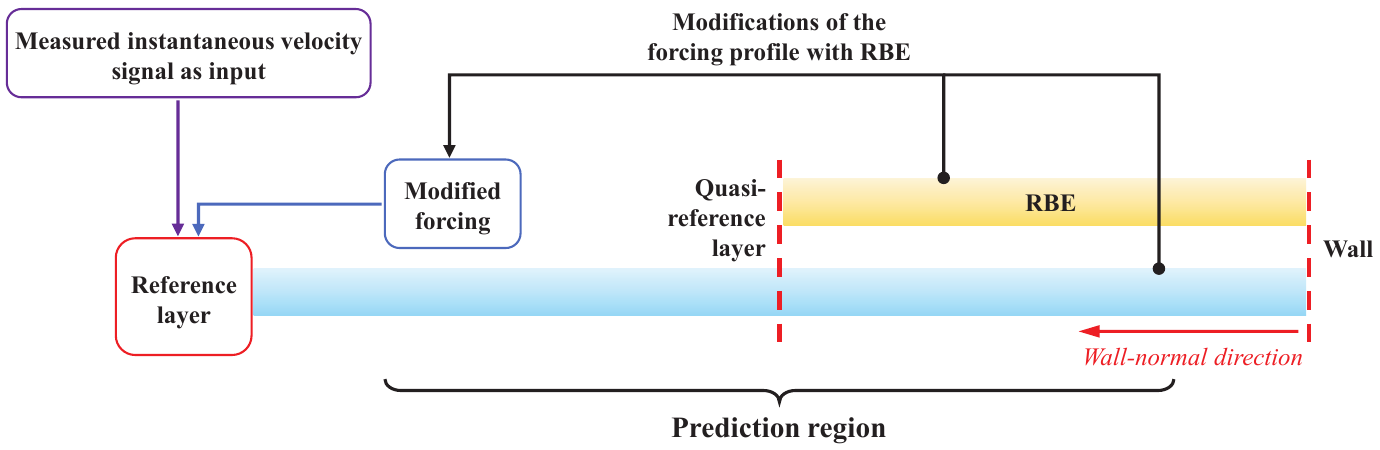}}
	\caption{Schematic sketch of the RWE.}
	\label{fig:Sketch}
\end{figure}

\begin{figure}\centering% Requires \usepackage{graphicx}
	{\includegraphics[width=4.0in, angle=0]{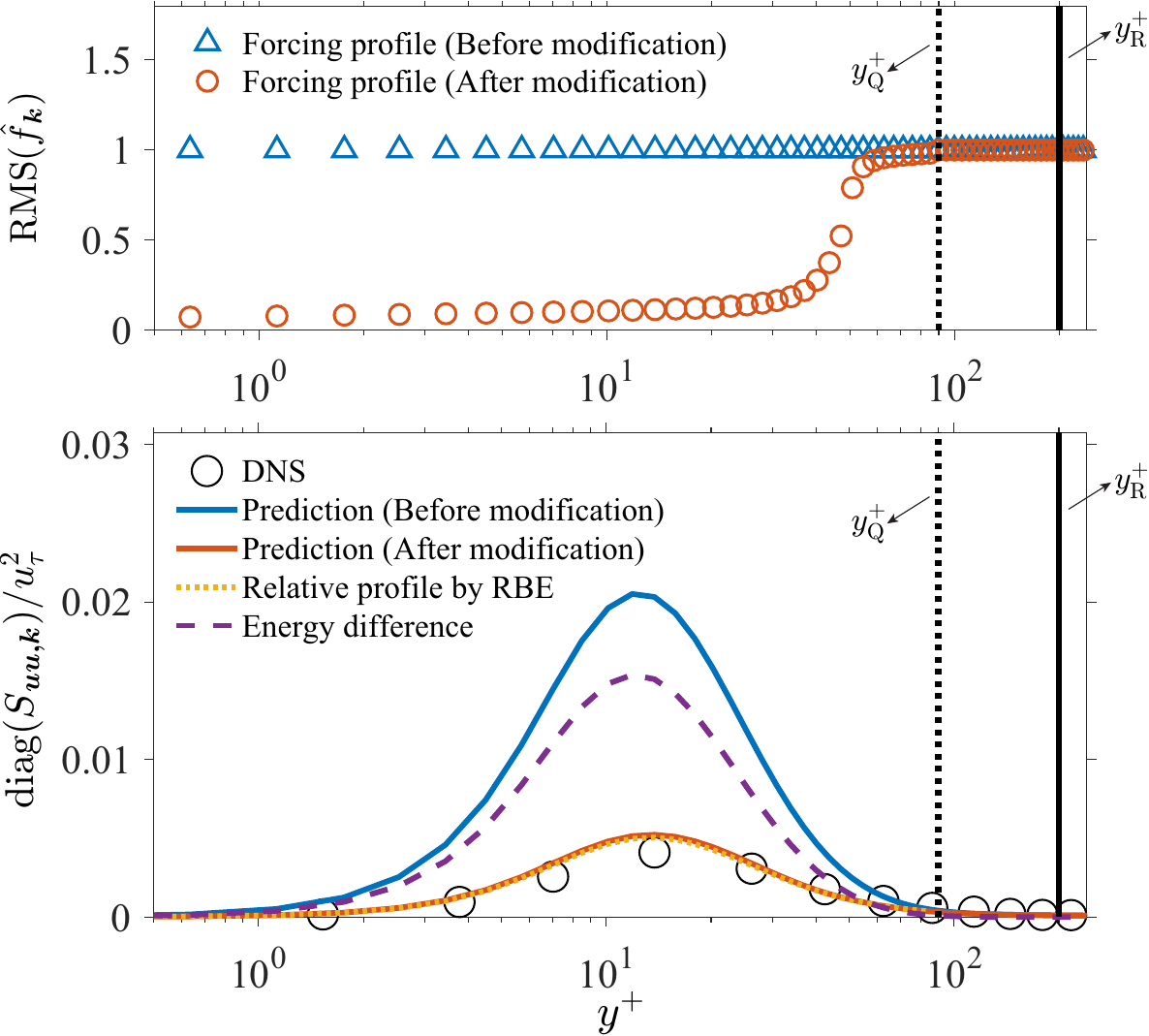}}
	\caption{Sketch of the modification process at $(k_x, k_z, \omega)=(3.0/h, 30/h, 1.3u_c / h)$. ($a$) Energy profile of forcing; ($b$) Energy profile of the streamwise velocity.}
	\label{fig:Sketch_Modification}
\end{figure}

The above construction process of the newly proposed RWE method is sketched in figure \ref{fig:Sketch}. To illustrate the actual modification process of the forcing and its effects on the response energy profile, the scale with $(k_x, k_z, \omega)=(3.0/h, 30/h, 1.3u_c / h)$ corresponding to the large-scale motions \citep{smits2011high} is selected, as in figure \ref{fig:Sketch_Modification}, where $u_c$ is the mean velocity at the half-channel height $h$. The forcing energy is uniformly distributed in the vertical direction before the modifications, with the corresponding response energy profile much larger than the DNS results in the near-wall region, as figure \ref{fig:Sketch_Modification}$(b)$. The height of the quasi-reference layer $y_{\rm Q}^+$ is determined according to the LCS value defined in equation (\ref{eq:LCS}). During the modification procedure, the forcing energy is reduced in the near-wall region to let the predicted relative energy profile approach that by the RBE till the norm of the error matrix in equation (\ref{eq:minmimize_norm}) reaches the minimum under the constraints. The forcing and response profiles after modification are denoted in the orange color in figure \ref{fig:Sketch_Modification}$(a)$ and $(b)$, respectively. The response energy profile after modification matches well with the DNS and RBE results.

In the next section, validations of the RWE will be conducted in terms of the prediction capability of the near-wall statistics as well as the instantaneous flow field with the DNS data and several representative existing prediction methods. The WBE \citep{morra2019relevance} and RBE \citep{towne2020resolvent} will be included in the following validations, since they provide the initial forcing profile and the reference for optimization for the newly proposed RWE method, respectively. Besides, the $\lambda$-model \citep{gupta2021linear} considers the effects of flow scale on the estimated forcing profile, which is found to perform better than the W-model and WBE (named as B-model by \citet{gupta2021linear}) for estimating the large-scale motions in the near-wall region. Thus, the $\lambda$-model will also be included in the following validations. Note that the $\lambda$-model is originally used in cases where temporal information of velocity is unknown. The forcing is thus assumed to be white in time in those cases \citep{gupta2021linear}, the response CSD tensor of which can be obtained via the algebraic Lyapunov equation. Meanwhile, in this study, the $\lambda$-model will be applied in the time-resolved cases instead, where the flow is estimated at each spatio-temporal scale quantified by $\boldsymbol{k}=(k_x , k_z , \omega)$. The modified eddy viscosity and forcing profiles in the $\lambda$-model will be calculated by equation (\ref{eq:lambda-model}) at each scale $(k_x , k_z)$, keeping consistent with the original version.

%%%%%%%%%%%%%%%%%%%%%%%%%%%%%%%%%%%%%%%%%%%%%%%%%%%%%%%%%%%%%%%%%%%%%%%%%%%%%%%%%%%%%%%%%%%%%%%%%%%%%%%%%%%%%%%%%%%%%%%%%
\section{Results}\label{sec:resluts}
In this section, the DNS data with three friction Reynolds numbers equal to 180, 550, and 950 are used to provide reference measurements at corresponding locations and validate the tested methods in predicting the flow properties of turbulent channel flows.

%%%%%%%%%%%%%%%%%%%%%%%%%%%%%%%%%%%%%%%%%%%%%%%%%%%%%%%%%%%%%%%%%%%%%%%%%%%%%%%%%%%%%%%%%%%%%%%%%%%%%%%%%%%%%%%%%%%%%%%%%
\subsection{Descriptions of the DNS database}\label{subsec:DNS_Database}
The code used to compute the extensively validated DNS database for channel flows \citep{del2003spectra,hoyas2008reynolds} is utilized to generate the time-resolved channel flow data with $Re_\tau$ = 180, 550, and 950. Details of the DNS setups are listed in table \ref{tab:Channel_DNS}. To provide time-resolved results, the sampling time intervals $\Delta t^+=\Delta t \cdot \frac{u_{\tau}^2}{\nu}$ are set as 2.13, 4.81, and 5.09 for cases with $Re_{\tau} = 180$, 550 and 950, respectively. The normalized total simulation time $(u_\tau \cdot T)/h$ is larger than 5.0 in each case to obtain statistically convergent results.
To assess the DNS dataset generated in this study, comparisons of the mean and root-mean-squared velocity profiles with the open source DNS database \citep{del2003spectra,hoyas2008reynolds} are provided in Appendix \ref{appB}.

To process the DNS data, the flow field is divided into blocks with a spatial domain of sizes ${L_x}/h = 4\pi$, ${L_z}/h = \pi$, and ${L_y}/h = 1$ in each case. Given that the turbulent channel flow is statistically symmetric about the centerline $y=h$, the flow data at $y=y_0$ will be utilized together with that at $y=2h-y_0$ when investigating the flow at $y=y_0$. The time periods of the blocks are set as $80\Delta t$, $80\Delta t$, and $120\Delta t$ for cases with $Re_{\tau} = 180$, 550 and 950, respectively, with $75\%$ overlap in the temporal direction. The rectangular window function is used when conducting spectral analyses. With the above setups for data processing, the numbers of blocks are 1128, 752, and 360 for cases with $Re_{\tau} = 180$, 550, and 950, respectively.
\begin{table}
	\begin{center}
		\renewcommand\arraystretch{1.5}
		
		\begin{tabular}{lccccccc}
			$Re_{\tau}$  & $Nx$   &   $Nz$ &   $Ny$ & ${L_x}/h$ & ${L_z}/h$ & ${L_y}/h$ & $(u_\tau \cdot T)/h$\\[3pt]
			180   &768  &512  &97  & $12{\pi}$& $4{\pi}$ & 2 & 11.45\\
			550   &1536 &1536 &257 & $8{\pi}$ & $4{\pi}$ & 2 & 8.80 \\
			950   &3072 &2304 &385 & $8{\pi}$ & $3{\pi}$ & 2 & 5.39\\
		\end{tabular}
		
		\caption{Parameters of the incompressible channel DNS setups.}
		\label{tab:Channel_DNS}
	\end{center}
\end{table}
%%
%%%%%%%%%%%%%%%%%%%%%%%%%%%%%%%%%%%%%%%%%%%%%%%%%%%%%%%%%%%%%%%%%%%%%%%%%%%%%%%%%%%%%%%%%%%%%%%%%%%%%%%%%%%%%%%%%%%%%%%%%
\subsection{Case settings}\label{subsec:Case_settings}
Six cases are set to validate the prediction methods, as summarized in table \ref{tab:Case_list}. Besides the Reynolds number $Re_{\tau}$, the height of the reference layer $y_{\rm R}$ where the measurements are obtained is also treated as an independent variable to test the sensitivities of the methods to the location of measurements, which ranges from the near-wall region at $y^+ = 40$ to the upper bound of the logarithmic region at $y/h \approx 0.2$. The wall-normal direction $y$ is discretized with 129, 201, and 257 Chebyshev polynomials in cases with $Re_\tau = 180$, 550, and 950, respectively, with no-slip boundary conditions are applied at the walls.
\begin{table}
	\begin{center}
		\renewcommand\arraystretch{1.5}
		
		\begin{tabular}{lcccccc}
			{Case} &  180-40  &  550-40   &   550-100 &  950-40 &  950-100 &  950-200\\[3pt]
			$Re_{\tau}$   & 180  & 550  & 550 & 950 & 950  & 950\\
			$y_{\rm R}^+$   & 40 & 40 & 100 & 40 & 100 & 200\\
			$y_{\rm R}/h$   & 0.22 & 0.073 & 0.18 & 0.042 & 0.11 & 0.21\\
		\end{tabular}
		
		\caption{Case settings for validation.}
		\label{tab:Case_list}
	\end{center}
\end{table}
%%
%%%%%%%%%%%%%%%%%%%%%%%%%%%%%%%%%%%%%%%%%%%%%%%%%%%%%%%%%%%%%%%%%%%%%%%%%%%%%%%%%%%%%%%%%%%%%%%%%%%%%%%%%%%%%%%%%%%%%%%%%
\subsection{Reynolds stress profiles}\label{subsec:RMS_profile}
In this section, the root-mean-squared (RMS) velocities and the Reynolds shear stress (RSS) are investigated to study the ensemble effect of fluctuations with all the spatio-temporal scales along the height.
Specifically, the RMS velocities quantify the fluctuation energies of $u^{\prime}$, $v^{\prime}$, and $w^{\prime}$, which are important indexes to validate the DNS (e.g., \citet{cheng2019identity}) or prediction methods (e.g., \citet{towne2020resolvent}).
On the other hand, the RSS $\left\langle u^{\prime} v^{\prime}  \right\rangle $ quantifies the correlation of $u^{\prime}$ and $v^{\prime}$, which is closely related to the skin friction of wall-bounded turbulence \citep{fukagata2002contribution} and has been investigated via the resolvent analysis for flow control \citep{luhar2015framework,nakashima2017assessment}.

\begin{figure}\centering% Requires \usepackage{graphicx}
	{\includegraphics[width=5in, angle=0]{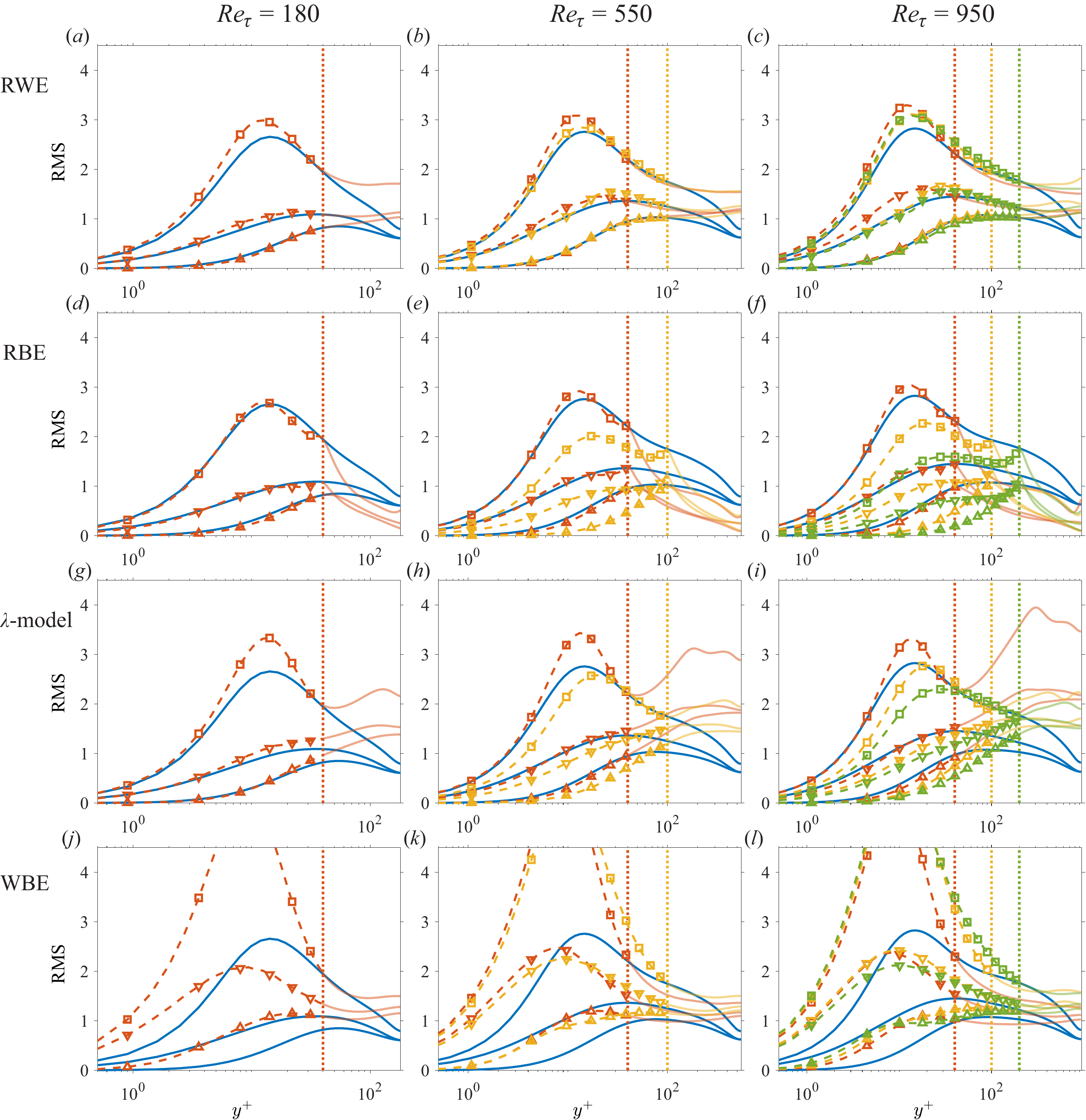}}
	\caption{Comparisons of the RMS profiles for cases with $Re_\tau = 180$ ($a,d,g,j$), 550 ($b,e,h,k$) and 950 ($c,f,i,l$). The solid lines denote the DNS results. The dashed lines denote the predictions from the RWE ($a,b,c$), $\lambda$-model ($d,e,f$), RBE ($g,h,i$), and WBE ($j,k,l$), where the lines with square, lower triangles and upper triangles denote the RMSs of the streamwise velocity, spanwise velocity, and vertical velocity, respectively. The vertical dotted lines denote the reference layers corresponding to the dashed lines with the same colors.}
	\label{fig:RMS_profile}
\end{figure}

When $Re_\tau = 180$, the only reference layer at $y^+ = 40$ is used for the predictions. According to the evaluations of RBE by \citet{yang2020numerical}, the best performance of RBE in terms of estimating the RMS profile of the streamwise velocity is achieved when $y^+ = 39$, which is very close to $y^+ = 40$ used in the current case. As in figure \ref{fig:RMS_profile}$(a)$, the RMS profiles predicted by RBE are pretty consistent with the DNS results in all three directions, with only $2.0 \%$ relative error at the inner peak of the streamwise RMS velocity profile.
The RWE also performs well when $Re_\tau = 180$ with a relative error of $13.2 \%$ at the inner peak. On the other hand, the predictions from the $\lambda$-model and WBE both deviate quite a lot from the DNS results, with maximum errors of $25.7 \%$ and $109.7 \%$ at the inner peaks of the streamwise RMS velocity. Note that this case is considered to be the least challenging one with the smallest Reynolds number and the closest reference layer to the wall, the large prediction errors in the $\lambda$-model and WBE results indicate that the predefined forcing profiles cannot properly model the forcing effects if not adjusted by additional flow information, especially in the near-wall region.

\begin{table}
	\begin{center}
		\renewcommand\arraystretch{1.5}
		\begin{tabular}{lcccccc}
			Case               &  180-40   &  550-40   &   550-100 & 950-40  &  950-100 &  950-200 \\[3pt]
			RWE                & 13.2      & 11.82     &  3.3      & 16.6    & 10.1     & 8.9 \\
			RBE                & 2.0       & 5.9       & -28.0     & 7.7     & -20.6    & -47.4 \\
			$\lambda$-model    & 25.7      & 24.5      & -8.5      & 17.5    & -8.0     & -24.0 \\
			WBE                & 109.7     & 122.1     & 113.4     & 103.6   & 115.2    & 115.7 \\
		\end{tabular}
		\caption{Summary of the relative percentage errors of the predicted streamwise RMS velocity at the inner peak.}
		\label{tab:RE_RMS}
	\end{center}
\end{table}

With the increase of $Re_\tau$, the RBE continues performing well when the reference layers are located at $y^+ = 40$, with the maximum error of $7.7 \%$ at the inner peak in Case 950-40. However, as the reference layer moves away from the wall, the RBE-predicted energy decreases rapidly, which is considered to be attributed to the decrease of coherence between the signals at the reference layer and the near-wall region. 
Especially, when the reference layer is located at $y^+ = 200$ with $Re_\tau = 950$, the energy peak can be barely observed in the RBE-predicted streamwise RMS profile. 
Like the RBE, the energies predicted by the $\lambda$-model also tend to decrease when the reference layer lifts upward, which overestimates and underestimates the energy with $y_{\rm R}^+ = 40$ and $y_{\rm R}^+ = 200$, respectively. When $y_{\rm R}^+ = 100$, the $\lambda$-model-predicted results match fairly well with the DNS results. The WBE overestimates the energies of the fluctuations of all the velocity components in all the tested cases, which indicates that the WBE cannot be directly used to estimate the energy magnitude in the near-wall region. Despite this, the capacity of WBE to estimate the distributions of the relative energy spectra should be further investigated in the following sections. Compared with the above methods, the RWE performs steadily well in all the cases with various Reynolds numbers and choices of the reference layers, with the maximum error equal to $16.6 \%$ at the inner peak of streamwise RMS velocity in Case 950-40. 

Besides the RMS velocity profiles below the reference layer, those in the higher region in each case are also depicted in figure \ref{fig:RMS_profile} with translucent curves. From the RWE results, the predicted profiles match well with the DNS results in the wall-normal range of $y \in (y_{\rm R},0.2h) $ for all the RMS velocities with different Reynolds numbers and reference layers.
Moreover, for the same Reynolds number, the profiles predicted by the RWE nearly overlap with each other for each Reynolds number, which again demonstrates the insensitivity of the RWE results to the choice of reference layer.
For the region that is higher than the upper bound of the logarithmic region, i.e., $y > 0.2h$, the RWE results become larger than those from the DNS, which should be attributed to the mismatch of the RWE-modeled forcing profile in the outer layer with the real forcing in the DNS results.
On the other hand, the other three considered methods cannot provide satisfying results at $y > y_{\rm R}$. For the RBE, the magnitudes of the RMS velocities shrink rapidly when the prediction location lifts up from the reference layer. On the contrary, the $\lambda$-model results become obviously larger than the DNS results at $y > y_{\rm R}$, which implies that the $\lambda$-model overestimates the forcing in the logarithmic region. The WBE underestimates the fluctuation energies at $y \in (y_{\rm R},0.2h) $, which in turn overestimates the magnitudes in the outer layer.

\begin{figure}\centering% Requires \usepackage{graphicx}
	{\includegraphics[width=5in, angle=0]{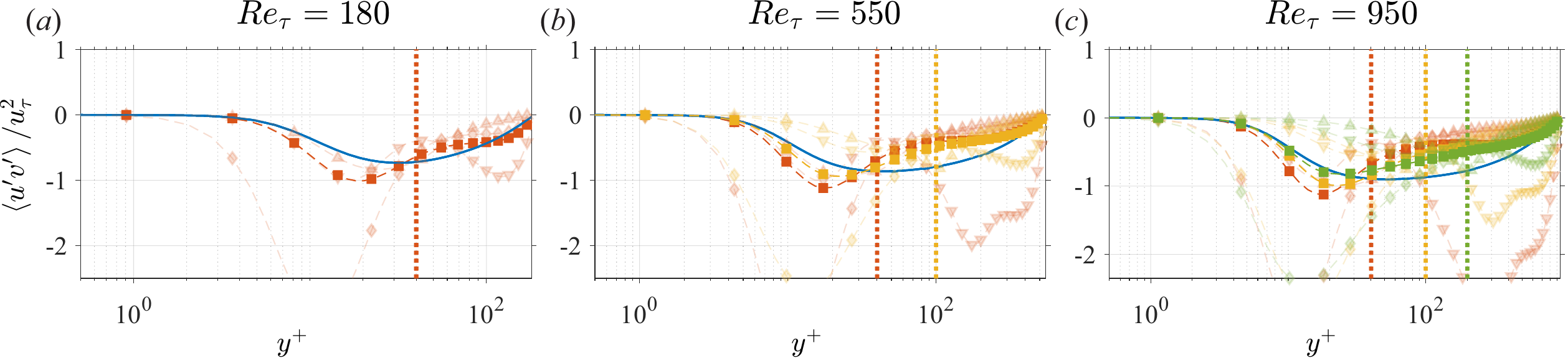}}
	\caption{Comparisons of the Reynolds shear stress profiles for cases with $Re_\tau = 180$ ($a$), 550 ($b$) and 950 ($c$). The solid lines and dashed lines denote the DNS results and predictions, respectively. The lines with square, upper triangles, lower triangles, and diamonds denote RWE, RBE, $\lambda$-model, and WBE results, respectively, where only the RWE results are depicted with opaque colors. The vertical dotted lines denote the reference layers corresponding to the dashed lines with the same colors.}
	\label{fig:UV_profile}
\end{figure}

The RSS profiles are depicted in figure \ref{fig:UV_profile}. The RWE-predicted RSS profiles, which are obtained from the measurements of the streamwise velocity at the reference layers, are fairly consistent with the DNS results in all the cases.
The RBE results that are depicted with upper triangles match well with the DNS results when $y_{\rm R}^+ = 40$. However, when the reference layer increases to 100 and 200, the magnitudes of the RBE-predicted RSS profiles become obviously underestimated.
In the meantime, the $\lambda$-model and WBE neither provide satisfying results with different Reynolds numbers and reference heights.

From the above discussions, the impacts of the friction Reynolds numbers and measuring locations on the prediction accuracy of the tested methods can be concluded as follows. When the $Re_\tau$ increases, the accuracy of prediction with the same $y_{\rm R}^+$ by the RBE and RWE does not show an obvious difference.  Meanwhile, the variation of the measuring locations significantly influences the prediction accuracy of all the methods except for the newly proposed RWE. Specifically, the increasing height of the reference layer reduces the predicted fluctuation energy at a given height for the RBE and $\lambda$-model, the errors of which become minimum when $y_{\rm R}^+ = 40$ and 100, respectively. Since the measurements are usually not guaranteed to be fixed at an ideal location in the WMLES or experiments, the insensitivity of the choices of reference layers implies a large potential of RWE to practical engineering applications.

%%%%%%%%%%%%%%%%%%%%%%%%%%%%%%%%%%%%%%%%%%%%%%%%%%%%%%%%%%%%%%%%%%%%%%%%%%%%%%%%%%%%%%%%%%%%%%%%%%%%%%%%%%%%%%%%%%
\subsection{One-dimensional energy spectra of the fluctuation velocities}\label{subsec:1D-Auto-spectra}
Besides the RMS profiles, the energy spectra, which describe the energy distributions among different scales, provide a more comprehensive picture to display the predicted flow properties.
In this section, the distributions of one-dimensional energy spectra at different heights as functions of $k_x$, $k_z$, or $\omega$ will be studied. For brevity, only the results of streamwise velocity $u$, as the most interesting physical quantity, in the cases where the reference layers are the farthest away from the wall for the corresponding Reynolds numbers, i.e., Cases 180-40, 550-100, and 950-200, are chosen for the validations in this section.

\begin{figure}\centering% Requires \usepackage{graphicx}
	{\includegraphics[width=5.0in, angle=0]{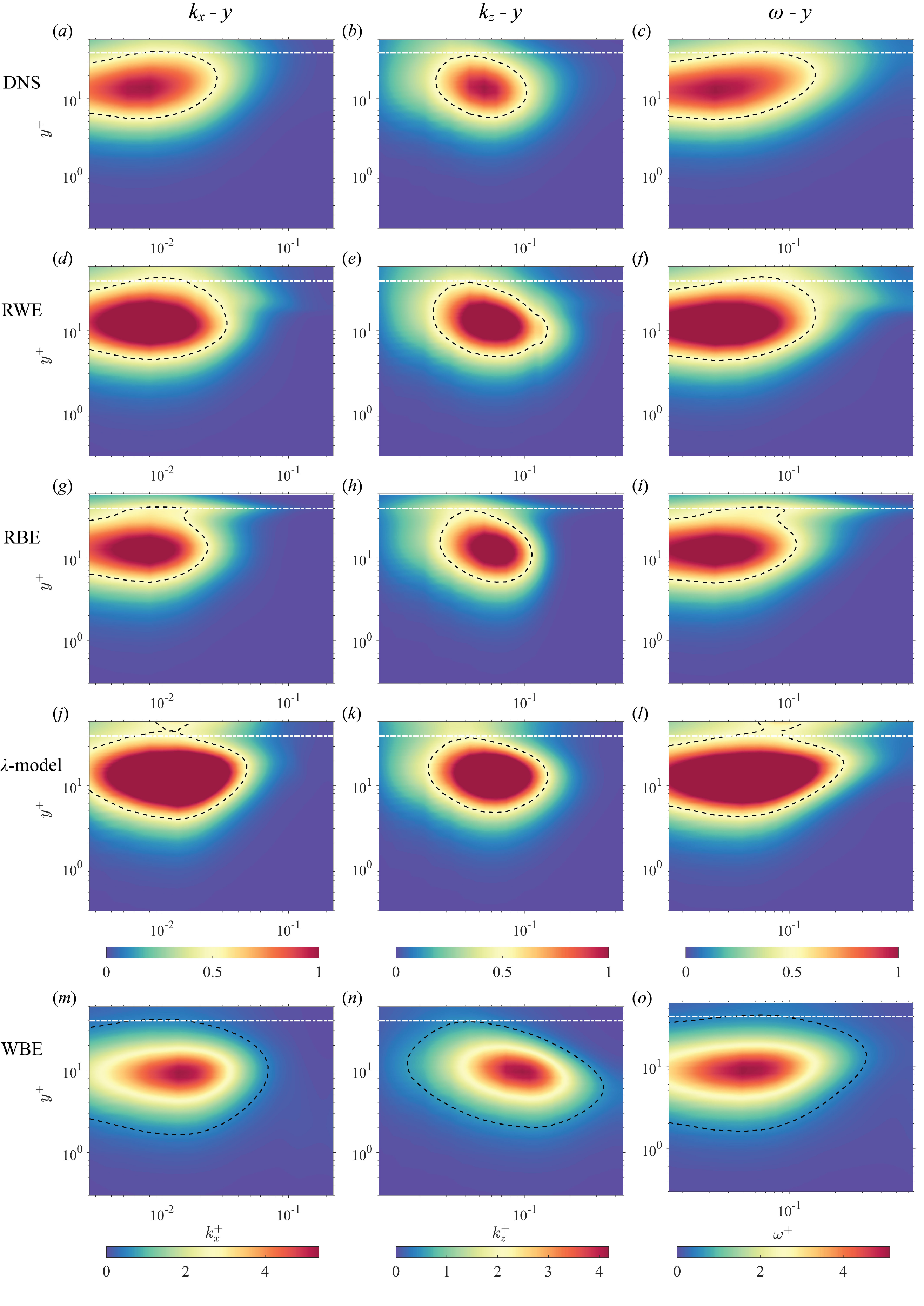}}
	\caption{Premultiplied energy spectra of streamwise velocity as a function of the wall-normal distance $y$ and the streamwise wavenumber $k_x$ ($a,d,g,j,m$), spanwise wavenumber $k_z$ ($b,e,h,k,n$), and frequency $\omega$ ($c,f,i,l,o$) in Case 180-40, from the results of DNS ($a,b,c$), RWE ($d,e,f$), RBE ($g,h,i$), $\lambda$-model ($j,k,l$), and WBE ($m,n,o$). The black dashed lines denote the contour of $0.5 S_{{\rm DNS,max}}$, and the white dash-doted lines denote the height of the reference layer. The values shown in the figures are normalized by $S_{{\rm DNS,max}}$.}
	\label{fig:autospectra_kx-y_Case 180-40_uu}
\end{figure}

\begin{figure}\centering% Requires \usepackage{graphicx}
	{\includegraphics[width=5.0in, angle=0]{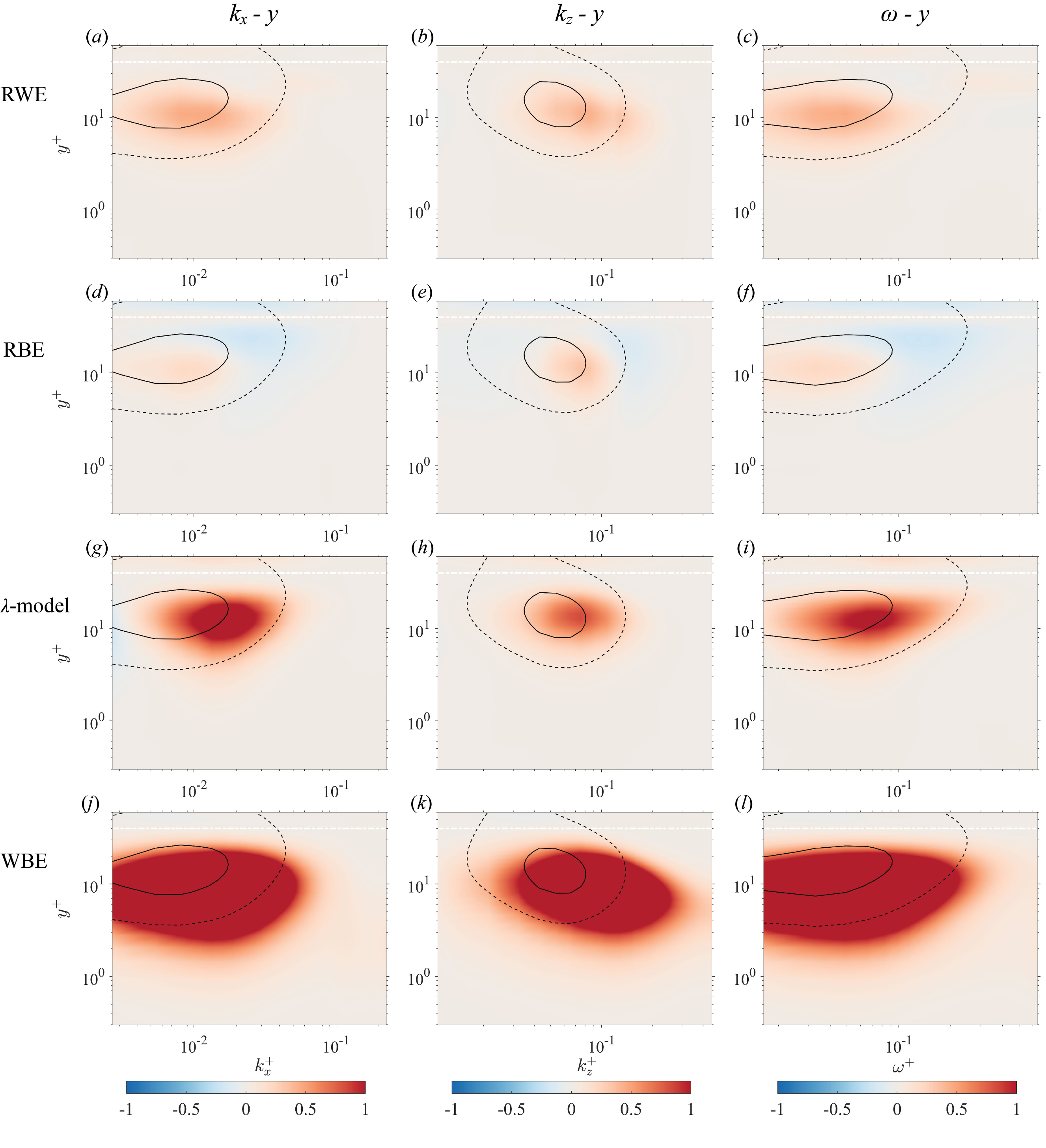}}
	\caption{Relative error of the premultiplied energy spectra of streamwise velocity as a function of the wall-normal distance $y$ and the streamwise wavenumber $k_x$ ($a,d,g,j,m$), spanwise wavenumber $k_z$ ($b,e,h,k,n$), and frequency $\omega$ ($c,f,i,l,o$) in Case 180-40, from the results of DNS ($a,b,c$), RWE ($d,e,f$), RBE ($g,h,i$), $\lambda$-model ($j,k,l$), and WBE ($m,n,o$). The values shown in the figures are normalized by $S_{{\rm DNS,max}}$. The black solid and dashed lines denote the contours of $0.75 S_{{\rm DNS,max}}$ and $0.25 S_{{\rm DNS,max}}$, respectively.}
	\label{fig:RE_autospectra_kx-y_Case 180-40_uu}
\end{figure}

%%%%%%%%% k-y for uu
Figure \ref{fig:autospectra_kx-y_Case 180-40_uu} depicts the premultiplied one-dimensional energy spectra of the streamwise velocity $u$ with the reference layer located at $y^+ = 40$, where the location of the reference layer is denoted with the white dashed line in each case.
The black dashed line denoting the contour of $0.5 S_{{\rm DNS,max}}$ is marked in each case to highlight the energy-concentrating region, where $S_{{\rm DNS,max}}$ is the maximum premultiplied spectral energy of the DNS result in the same case. The values of the premultiplied energy are normalized by $S_{{\rm DNS,max}}$.
When the reference layer is located at $y^+ = 40$ with $Re_{\tau}=180$, the RBE works well in predicting the energy distributions at different heights and flow scales, with the maximum error lower than $0.25 S_{{\rm DNS,max}}$ according to figure \ref{fig:RE_autospectra_kx-y_Case 180-40_uu} that depicts the relative error. This result is consistent with the previous work that also uses the RBE to predict the energy spectra with the reference layer at $y^+ = 37$ \citep{towne2020resolvent}. 
The energies predicted by the RWE also match fairly well with the DNS results, whose maximum error is about $0.46S_{{\rm DNS,max}}$.
On the other hand, the relative errors that are larger than $1.0S_{{\rm DNS,max}}$ are observed in the results of the $\lambda$-model and WBE. For the $\lambda$-model, the maximum error occurs at the heights around the inner peak $y^+ = 15$ and the scales that are slightly smaller than the energy-concentrating scales predicted by the DNS, as denoted by the black dashed lines. This indicates that the deviations between the $\lambda$-model and DNS results mainly concentrate at small scales near the inner peak. On the other hand, the WBE overestimates the energy within a wide wall-normal range from $y^+ = 3$ to 20, which indicates that the WBE results cannot be directly applied to the predictions of the magnitude of the energy spectra.

Although the WBE overestimates the magnitude of the energy, it is still worth investigating the relative spectral energy predicted by the WBE. To depict the relative spectral energy distributions in the WBE results, we set the upper bound of the color bar as the maximum spectral energy predicted by the WBE rather than $S_{\rm DNS,max}$. The relative distributions of the energies predicted by WBE are roughly consistent with those from the DNS, while deviations could also be found, e.g., the inner peak for the streamwise velocity from WBE is located at about $y^+ \approx 10$, which should be at $y^+ \approx 15$ according to the DNS results.

Among the above-discussed methods, the RBE provides the most accurate results, followed by the RWE. Although the RWE result does not totally recover the RBE results, it provides a better result over the WBE and $\lambda$-model.


\begin{figure}\centering% Requires \usepackage{graphicx}
	{\includegraphics[width=5.0in, angle=0]{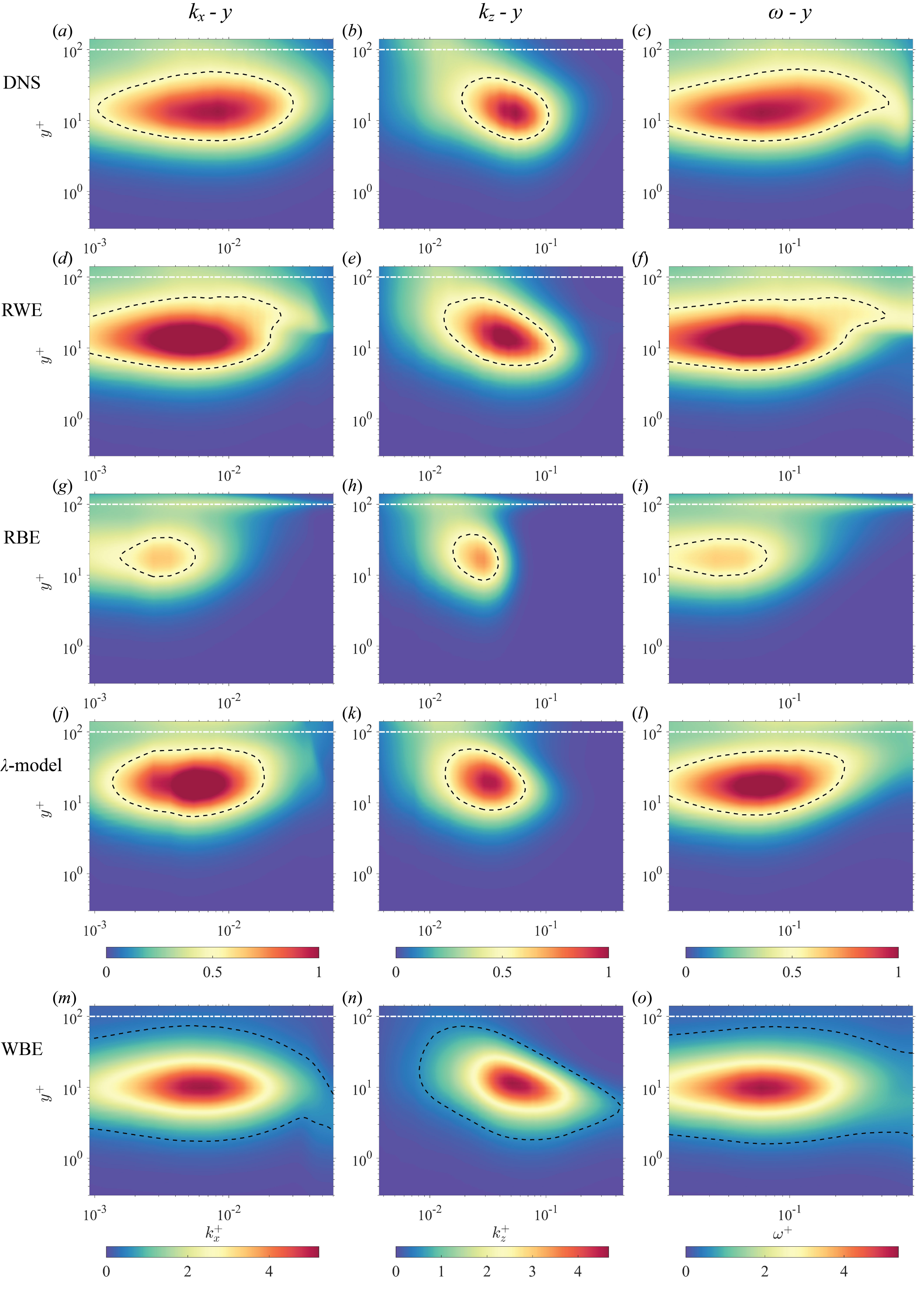}}
	\caption{Same as figure \ref{fig:autospectra_kx-y_Case 180-40_uu}, but in Case 550-100.}
	\label{fig:autospectra_kx-y_Case 550-100_uu}
\end{figure}

\begin{figure}\centering% Requires \usepackage{graphicx}
	{\includegraphics[width=5.0in, angle=0]{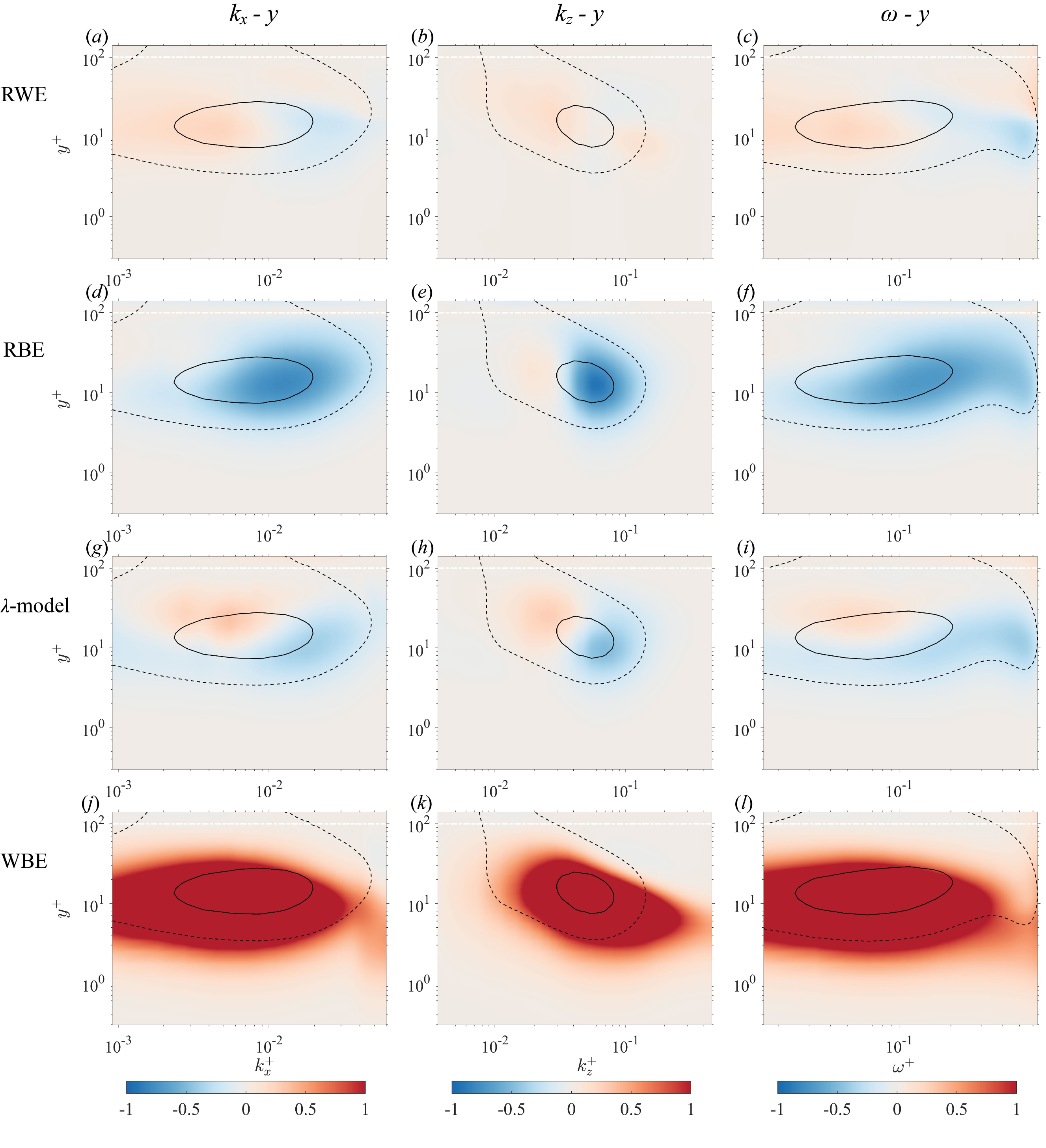}}
	\caption{Same as figure \ref{fig:RE_autospectra_kx-y_Case 180-40_uu}, but in Case 550-100.}
	\label{fig:RE_autospectra_kx-y_Case 550-100_uu}
\end{figure}

\begin{figure}\centering% Requires \usepackage{graphicx}
	{\includegraphics[width=5.0in, angle=0]{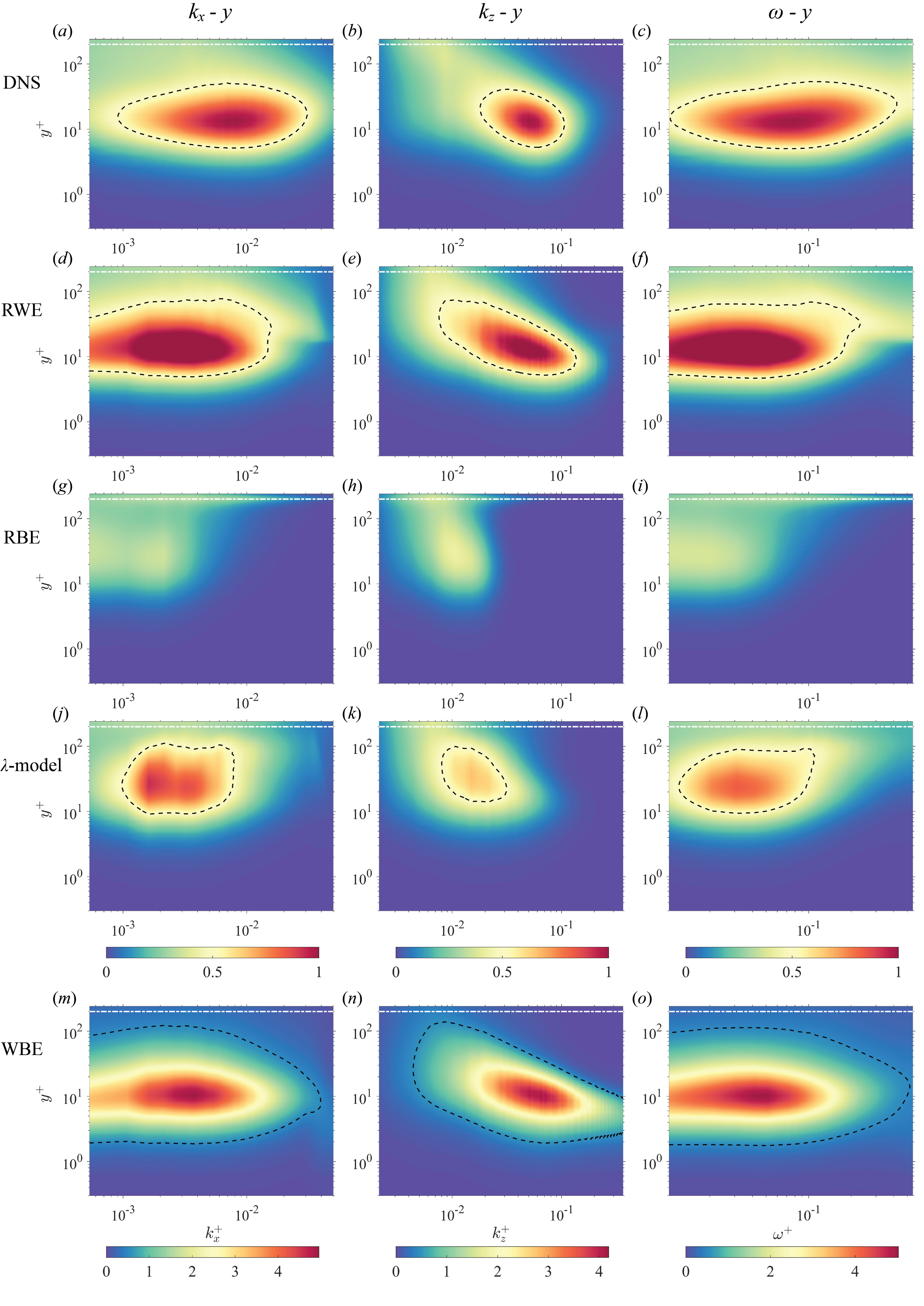}}
	\caption{Same as figure \ref{fig:autospectra_kx-y_Case 180-40_uu}, but in Case 950-200.}
	\label{fig:autospectra_kx-y_Case 950-200_uu}
\end{figure}

\begin{figure}\centering% Requires \usepackage{graphicx}
	{\includegraphics[width=5.0in, angle=0]{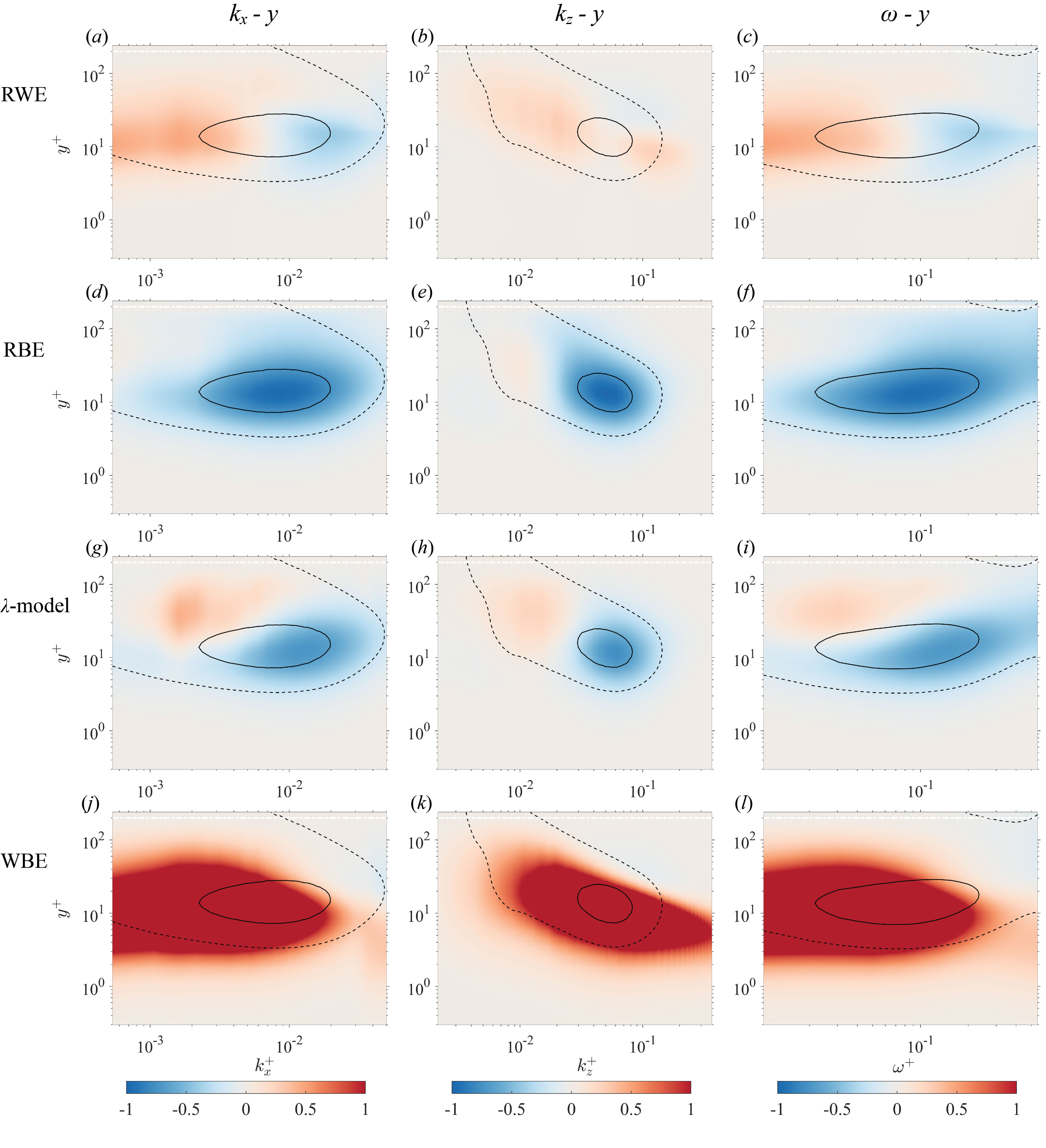}}
	\caption{Same as figure \ref{fig:RE_autospectra_kx-y_Case 180-40_uu}, but in Case 950-200.}
	\label{fig:RE_autospectra_kx-y_Case 950-200_uu}
\end{figure}

With the increase of the reference height $y_{\rm R}$, as in figures \ref{fig:autospectra_kx-y_Case 550-100_uu} and \ref{fig:autospectra_kx-y_Case 950-200_uu}, the prediction accuracy of RBE deteriorates, obviously. The RBE-predicted energy-concentrating regions denoted by the black dashed lines shrink towards the vicinity around $y^+ = 15$ in Case 550-100, which finally disappear in Case 950-200, indicating that the near-wall energy is seriously underestimated by RBE when $y_{\rm R}^+ = 200$.
On the other hand, as the reference layer lifts up, the magnitude of the $\lambda$-model-predicted energy decreases gradually. In Case 550-100, the $\lambda$-model results match well with the DNS results, with the relative error lower than $0.35 S_{{\rm DNS,max}}$, as in figure \ref{fig:RE_autospectra_kx-y_Case 550-100_uu}. When $y_{\rm R}^+$ increases to 200, the energy magnitude predicted by the $\lambda$-model continues decreasing and consequently becomes lower than the DNS results by more than $0.75 S_{{\rm DNS,max}}$ in the small scales at around $y^+ = 15$, as in figure \ref{fig:RE_autospectra_kx-y_Case 950-200_uu}. The WBE significantly overestimates the magnitude of energy, while the relative energy distribution roughly reflects the patterns of the DNS results. Compared to the WBE, the $\lambda$-model has a higher accuracy, which highlights the importance of including the scale effects when determining the forcing profile.
Meanwhile, the RWE keeps the highest accuracy in the streamwise energy distributions with $k_x^+$, $k_z^+$, and $\omega^+$, with the maximum error of $0.33 S_{{\rm DNS,max}}$. The good performance of the RWE supports the capability of the adaptive modification process as derived in section \ref{sec:methodology}.


\begin{figure}\centering% Requires \usepackage{graphicx}
	{\includegraphics[width=5.3in, angle=0]{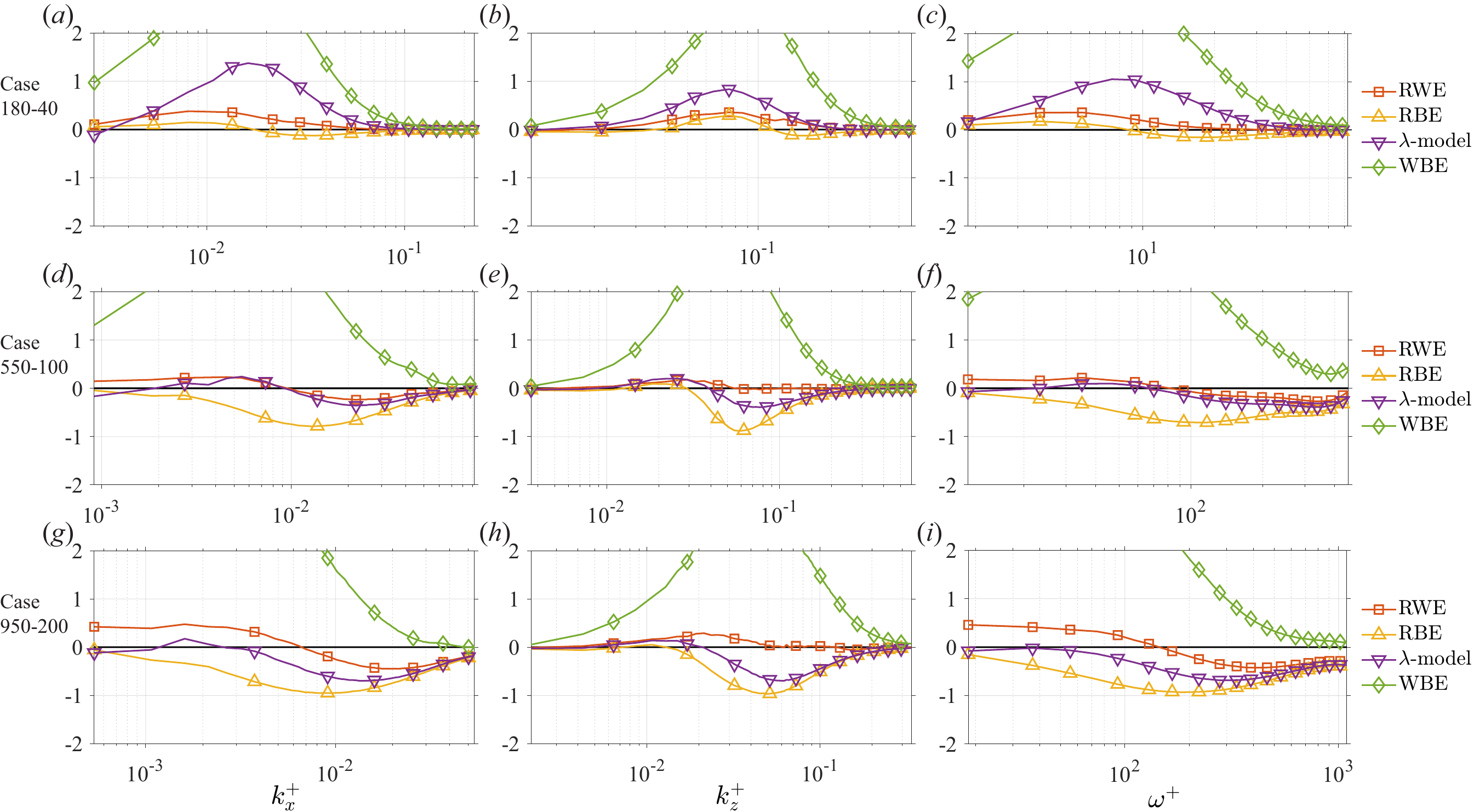}}
	\caption{Relative error of the predicted energy at $y^+ = 15$ along $k_x^+$ ($a, d, g$), $k_z^+$ ($b, e, h$), and $\omega^+$ ($c, f, i$) in Cases 180-40 ($a, b, c$), 550-100 ($d, e, f$), and 950-200 ($g, h, i$). The values of the curves are normalized by $S_{{\rm DNS,max}}$.}
	\label{fig:RE_Line_15}
\end{figure}

In order to quantitatively analyze the patterns of the near-wall predictions from different methods as the reference layer moves away from the wall, the relative errors at the inner peak ($y^+ = 15$) and along different scales quantified by $k_x^+$, $k_z^+$, and $\omega^+$ are shown in figure \ref{fig:RE_Line_15}. The values are normalized by $S_{{\rm DNS,max}}$, as consistent with those in figures \ref{fig:RE_autospectra_kx-y_Case 180-40_uu}, \ref{fig:RE_autospectra_kx-y_Case 550-100_uu}, and \ref{fig:RE_autospectra_kx-y_Case 950-200_uu}. The green curves denoting the relative errors of the WBE results deviate too much from the reference zero value, and thus they cannot be fully depicted in the figures. The predicted energies from the RBE and $\lambda$-model show a clear trend to decrease compared to the DNS results as the reference layer lifts up. For instance, in Case 180-40, the spectral energy predicted by the $\lambda$-model is larger than the DNS results at almost all the scales, whose maximum error is more than $1.37 S_{{\rm DNS,max}}$ with a positive value. When the reference layer lifts up to $y^+ = 200$, the maximum relative error of the $\lambda$-model predictions becomes $0.7 S_{{\rm DNS,max}}$ with a negative value instead. With the increasing height of the reference layer, the RBE and $\lambda$-model achieve their best performance with $y_{\rm R}^+ = 40$ and 100, respectively. On the other hand, the relative error of the RWE results is pretty small in the cases tested. Meanwhile, with the increase of flow scales, the spectral energy predicted by the RWE tends to increase compared to the DNS results, which can also be observed in the predictions from the other tested methods.

%%

%%%%%%%%%%%%%%%%%%%%%%%%%%%%%%%%%%%%%%%%%%%%%%%%%%%%%%%%%%%%%%%%%%%%%%%%%%%%%%%%%%%%%%%%%%%%%%%%%%%%%%%%%%%%%%%%%%
\subsection{Two-dimensional spectra of the fluctuation velocities}\label{subsec:2D-Auto-spectra}
In this section, the two-dimensional auto- and cross-spectra of the fluctuation velocities in the near-wall region ($y^+ = 10$) are analyzed to investigate the energy distributions and correlations at different spatial scales.
For brevity, the results for the auto-spectra of the streamwise velocity and the cross-spectra of the streamwise and wall-normal velocities are presented for the analyses.

\begin{figure}\centering% Requires \usepackage{graphicx}
	{\includegraphics[width=5.0in, angle=0]{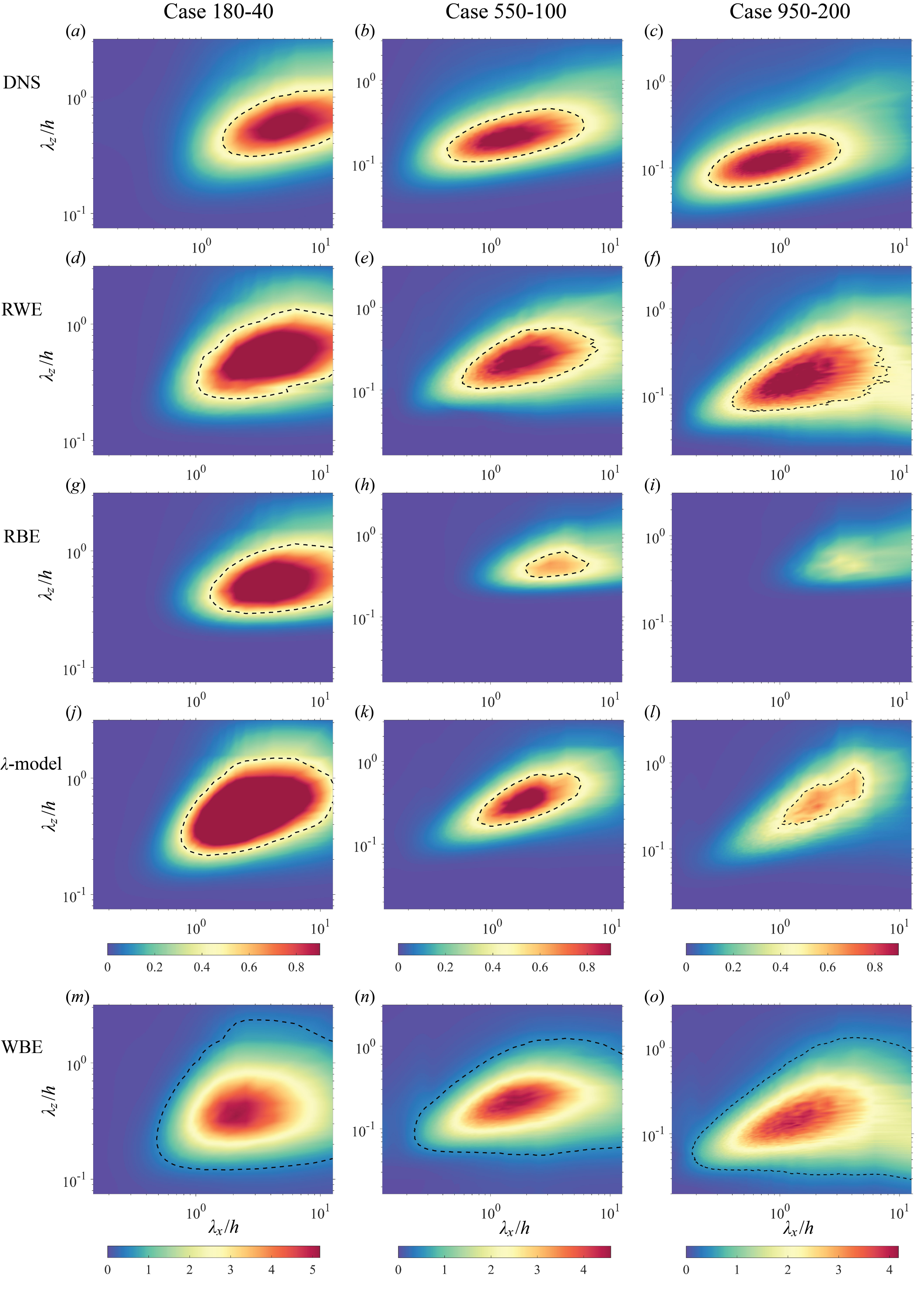}}
	\caption{Premultiplied energy spectra as a function of the streamwise wavelength $\lambda_x$ and spanwise wavelength $\lambda_z$ from the results of DNS ($a,b,c$), RWE ($d,e,f$), W-model ($g,h,i$), RBE ($j,k,l$), and WBE ($m,n,o$), at $y^+ = 10$ in Case 180-40 ($a,d,g,j,m$), Case 550-100 ($b,e,h,k,n$), and Case 950-200 ($c,f,i,l,o$). The values shown in the figures are normalized by $S_{{\rm DNS,max}}$.}
	\label{fig:autospectra_kxkz_Case_950-200_streamwise}
\end{figure}

\begin{figure}\centering% Requires \usepackage{graphicx}
	{\includegraphics[width=5.0in, angle=0]{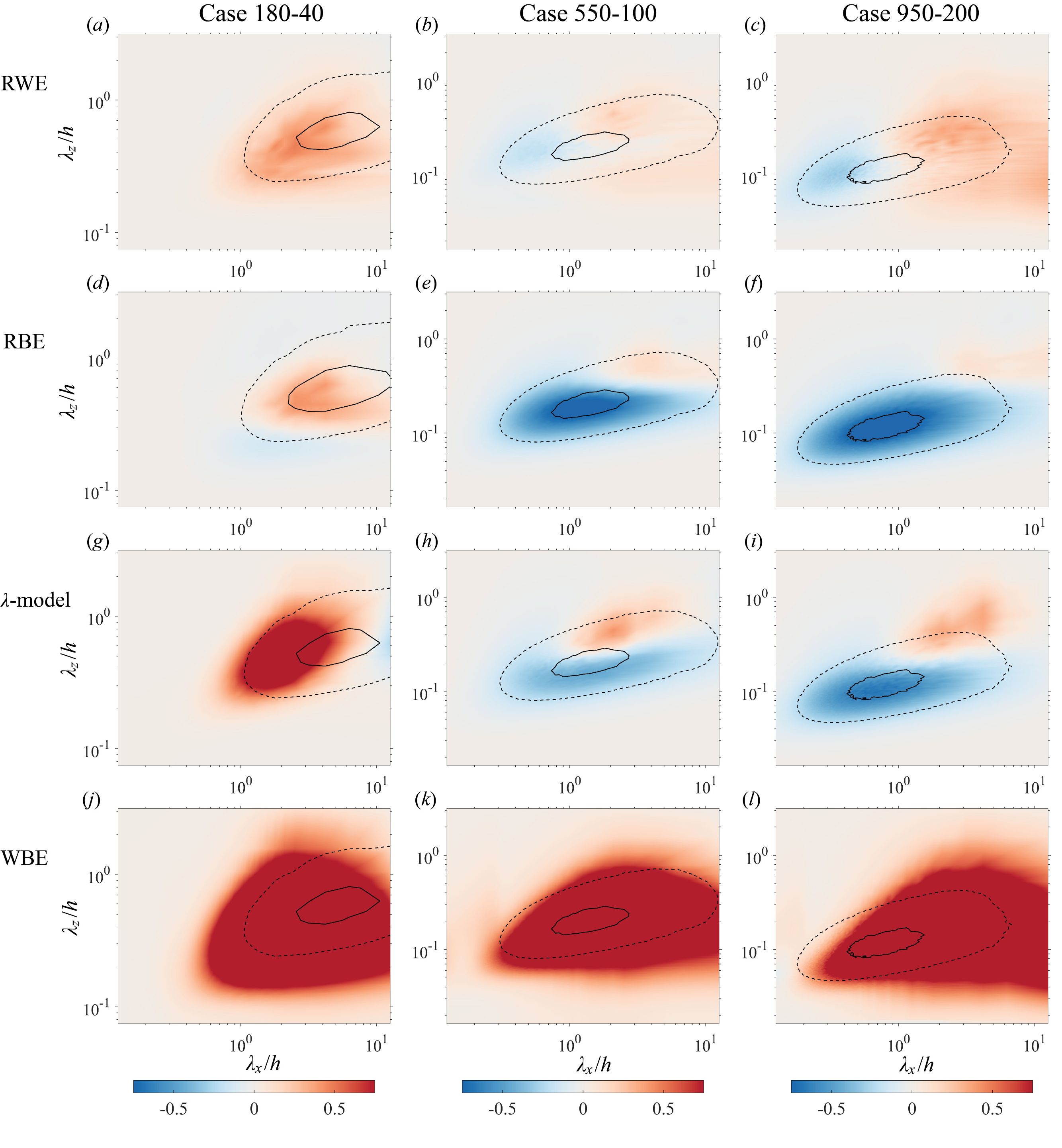}}
	\caption{Relative errors of the predicted premultiplied energy spectra normalized by the maximum energy spectra density from DNS as a function of the streamwise wavelength $\lambda_x$ and spanwise wavelength $\lambda_z$ from the results of RWE ($a,b,c$), W-model ($d,e,f$), RBE ($g,h,i$), and WBE ($j,k,l$), at $y^+ = 10$ in Case 180-40 ($a,d,g,j$), Case 550-100 ($b,e,h,k$), and Case 950-200 ($c,f,i,l$). The values shown in the figures are normalized by $S_{{\rm DNS,max}}$.}
	\label{fig:relative_deviation_autospectra_kxkz_Case_950-200_streamwise}
\end{figure}

First, the results of streamwise velocity in Cases 180-40, 550-100, and 950-200 at the near-wall plane at $y^+ = 10$ are presented in figure \ref{fig:autospectra_kxkz_Case_950-200_streamwise}. In order to show the relative energy spectrum predicted by the WBE, the separate color bar with its upper bound set as the maximum spectral energy predicted by the WBE is used. Meanwhile, the relative errors of the premultiplied spectra are depicted in figure \ref{fig:relative_deviation_autospectra_kxkz_Case_950-200_streamwise}, with the black solid and dashed lines denoting the contours of $0.75 S_{{\rm DNS,max}}$ and $0.25 S_{{\rm DNS,max}}$ of the DNS results, respectively. When $Re_\tau = 180$ with $y_{\rm R}^+ = 40$, it is observed that the predictions of RWE and RBE are relatively consistent with the DNS results. On the other hand, the $\lambda$-model and WBE obviously overestimate the spectral energy. This observation is consistent with that in the one-dimensional spectra as discussed in section \ref{subsec:1D-Auto-spectra}.
With the increase of the reference height $y_{\rm R}^+$, the predicted energies from the RBE and $\lambda$-model decrease correspondingly. When $y_{\rm R}^+ = 200$, only a small amount of the RBE-predicted energy is observed in the large-scale region with $\lambda_x /h \ge 2$, which indicates that the RBE becomes invalid when the reference layer in this case. In section \ref{subsec:1D-Auto-spectra}, we have observed that the $\lambda$-model-predicted one-dimensional energy distributions match well with the DNS results with a wide range of $k_x^+$ and $k_z^+$ values. However, when considering the energy distributions as binary functions of $\lambda_x$ and $\lambda_z$, the $\lambda$-model results are not consistent with the DNS. From figure \ref{fig:relative_deviation_autospectra_kxkz_Case_950-200_streamwise}, the energy predicted by the $\lambda$-model is obviously lower than the DNS result, in the energetic region denoted by the black dashed lines. In the RWE results, the energy-concentrating regions that are surrounded by the black dashed lines in figure \ref{fig:autospectra_kxkz_Case_950-200_streamwise} match well with the DNS results. From figure \ref{fig:relative_deviation_autospectra_kxkz_Case_950-200_streamwise}, the largest relative error in the RWE results occurs in Case 180-40, which is equal to $0.48 S_{\rm DNS,max}$. In Cases 550-100 and 950-200, the relative error is always lower than $0.4 S_{\rm DNS,max}$.
As consistent with the one-dimensional results, the WBE results are overestimated by more than 4 times in magnitude for all the cases. Whereas, the WBE results fairly reflect the energy distributions from the DNS.


\begin{figure}\centering% Requires \usepackage{graphicx}
	{\includegraphics[width=5.0in, angle=0]{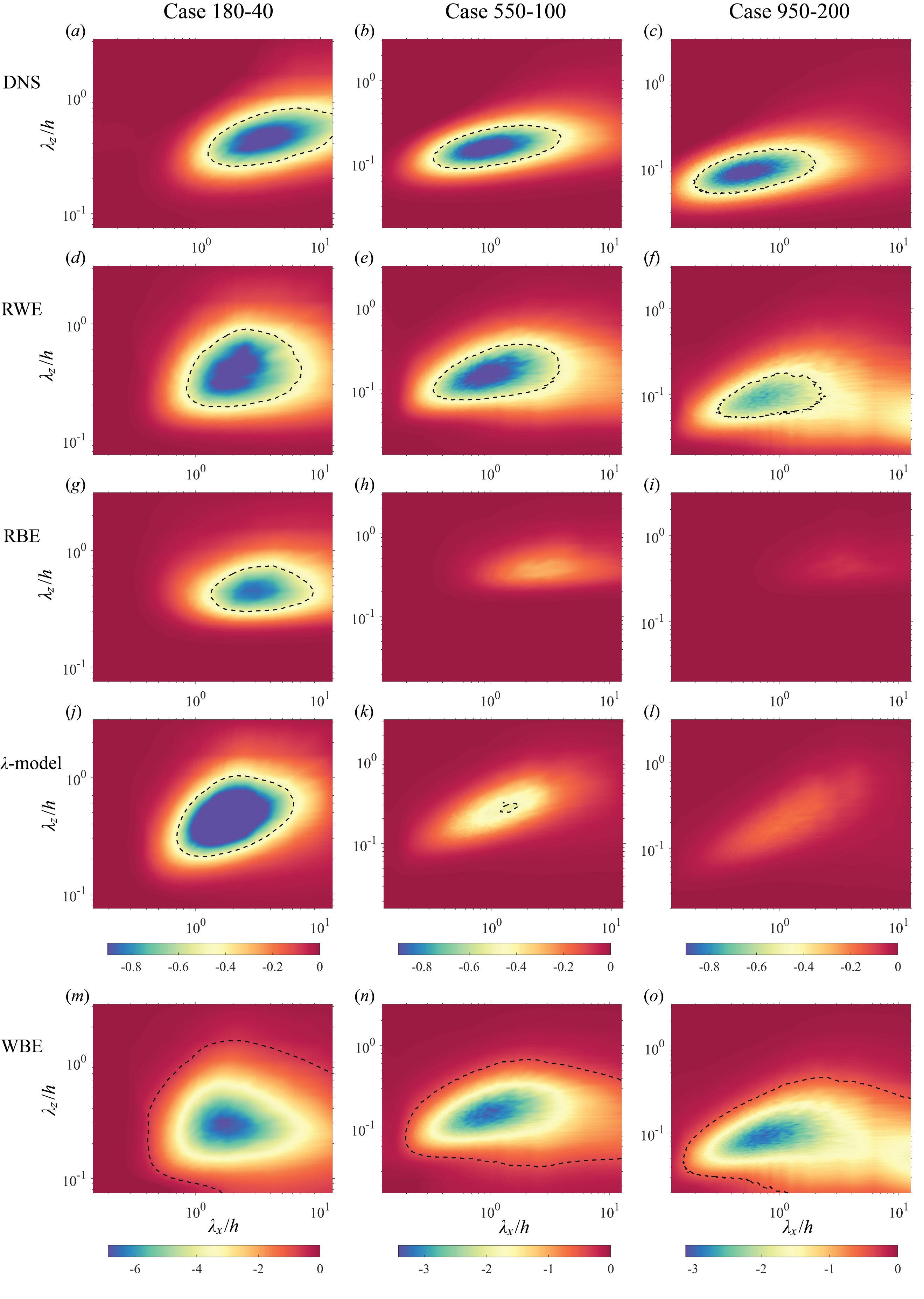}}
	\caption{The same as figure \ref{fig:autospectra_kxkz_Case_950-200_streamwise}, but for the premultiplied cross-spectra between $u^{\prime}$ and $v^{\prime}$ at $y^+ = 10$. The values shown in the figures are normalized by $\left|  S_{{\rm DNS}} \right|_{\rm max}$.}
	\label{fig:cross-spectra_kxkz_Case_950-200_streamwise}
\end{figure}

\begin{figure}\centering% Requires \usepackage{graphicx}
	{\includegraphics[width=5.0in, angle=0]{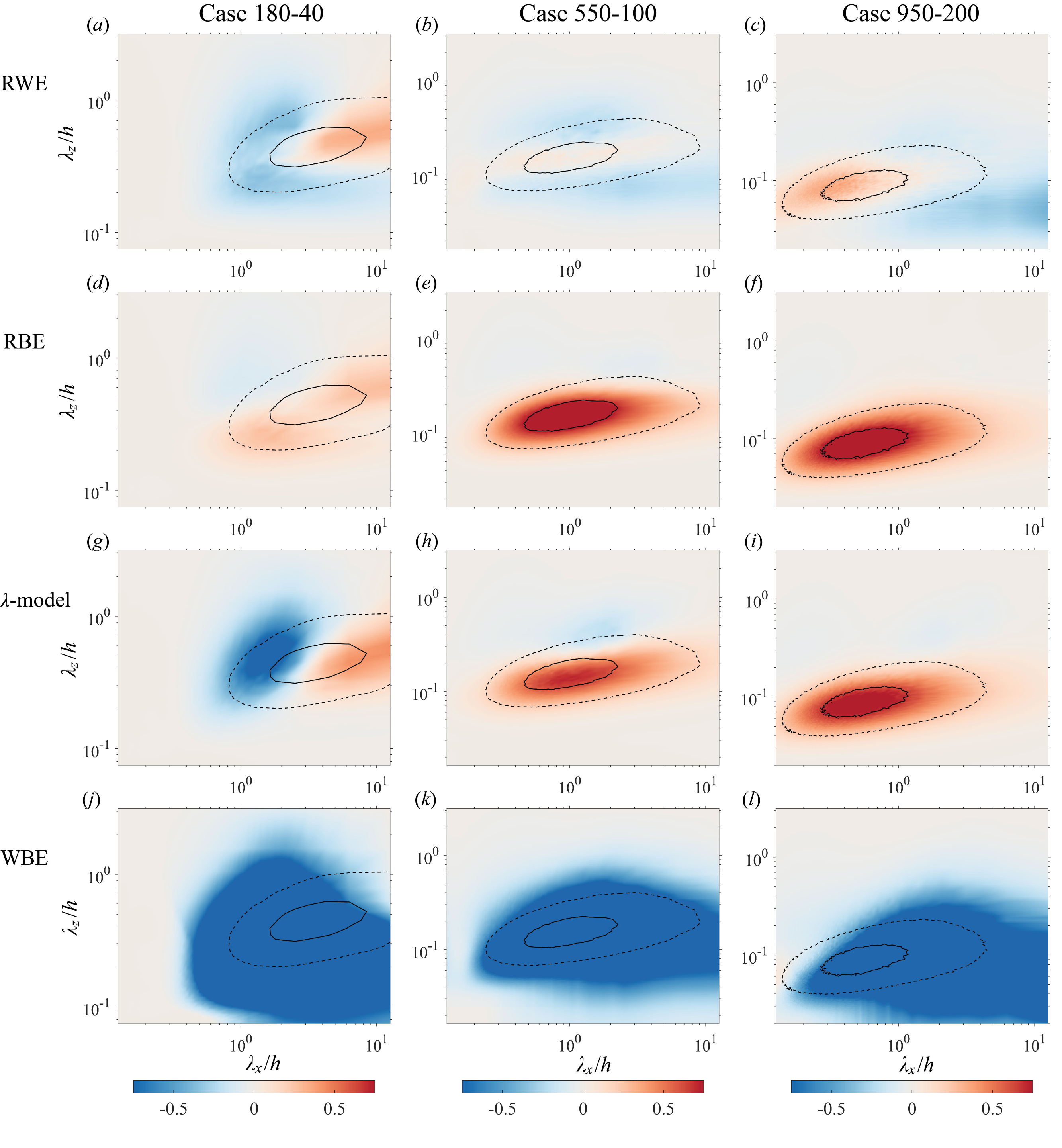}}
	\caption{The same as figure \ref{fig:relative_deviation_autospectra_kxkz_Case_950-200_streamwise}, but for the relative errors of the premultiplied cross-spectra between $u^{\prime}$ and $v^{\prime}$ at $y^+ = 10$. The values shown in the figures are normalized by $\left|  S_{{\rm DNS}} \right|_{\rm max}$.}
	\label{fig:relative_deviation_cross-spectra_kxkz_Case_950-200_streamwise}
\end{figure}

Then, the cross-spectra between $u^{\prime}$ and $v^{\prime}$ at $y^+ = 10$ are investigated, as shown in figure \ref{fig:cross-spectra_kxkz_Case_950-200_streamwise} with the relative errors presented in figure \ref{fig:relative_deviation_cross-spectra_kxkz_Case_950-200_streamwise}.
The values in figures \ref{fig:cross-spectra_kxkz_Case_950-200_streamwise} and \ref{fig:relative_deviation_cross-spectra_kxkz_Case_950-200_streamwise} are normalized by the peak absolute value of the DNS results, as denoted by $\left|  S_{{\rm DNS}} \right|_{\rm max}$ here.
The region with large values of negative correlations between the streamwise and wall-normal velocities are observed in each case from the DNS results, which is slightly offset towards smaller flow scales than the auto-spectra as in figure \ref{fig:autospectra_kxkz_Case_950-200_streamwise}.
In Case 180-40, the RBE provides the most accurate result of the cross-spectra. The RWE result is larger in magnitude than the DNS results for about $0.3 \left|  S_{{\rm DNS}} \right|_{\rm max} $ in the energy-concentrating region in Case 180-40. Meanwhile, the relative errors of the RWE results are much smaller than those in the $\lambda$-model and WBE results. As the reference layer lifts up to $y^+ = 100$ and 200, both the RBE and $\lambda$-model obviously underestimate the correlations between $u^{\prime}$ and $v^{\prime}$. Meanwhile, the RWE keeps a relatively low prediction error in the estimation of the cross-spectra in Cases 550-100 and 950-200.

%%
%%%%%%%%%%%%%%%%%%%%%%%%%%%%%%%%%%%%%%%%%%%%%%%%%%%%%%%%%%%%%%%%%%%%%%%%%%%%%%%%%%%%%%%%%%%%%%%%%%%%%%%%%%%%%%%%%%%%%%
\subsection{Space-time properties of the fluctuation velocities near the wall}\label{subsec:Space-time}
The space-time correlation is an important property of turbulence \citep{he2017space}. According to Taylor's frozen hypothesis \citep{taylor1938spectrum}, the spatial distributions of the turbulent motions are carried past a fixed point by the convection velocity without significant changes. However, Taylor's frozen hypothesis cannot accurately describe the complex space-time properties of turbulence \citep{he2006elliptic,he2017space}. In this section, we will test the capability of the newly proposed RWE method in estimating the space-time properties of turbulence in the near-wall region.

\begin{figure}\centering% Requires \usepackage{graphicx}
	{\includegraphics[width=5.0in, angle=0]{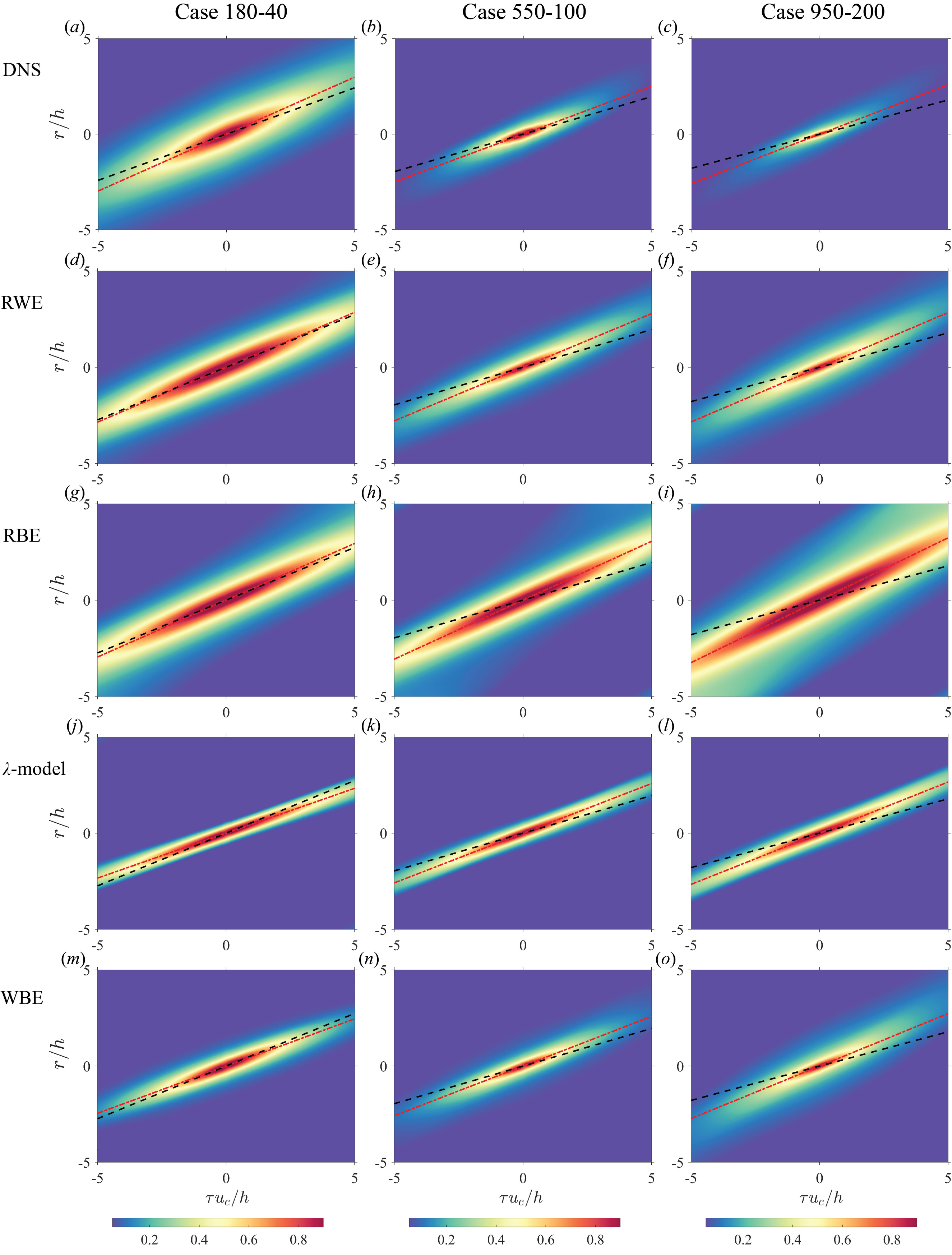}}
	\caption{Space-time correlation of the streamwise velocity from the results of DNS ($a,b,c$), RWE ($d,e,f$), $\lambda$-model($g,h,i$), RBE ($j,k,l$), and WBE ($m,n,o$) at $y^+ = 10$ in Cases 180-40 ($a,d,g,j,m$), 550-100 ($b,e,h,k,n$), and 950-200 ($c,f,i,l,o$). The slope of the black dashed lines denotes the mean streamwise velocity at the prediction layers, while the red dot-dashed lines denote the convection velocity calculated by equation (\ref{eq:Convection_U}).}
	\label{fig:Space-time_streamwise}
\end{figure}

The space-time correlation $R\left(r,\tau \right)$ is defined as the correlation of the velocity fluctuation at $(x,t)$ and $(x+r,t+\tau)$, which can be calculated by,
\begin{equation}
	\label{eq:Correlation_function}
	R\left(r,\tau \right)
	= \frac{\left\langle u'(x,t) u'(x+r,t+\tau)\right\rangle }{\sqrt{\left\langle u'(x,t)^2 \right\rangle \left\langle u'(x+r,t+\tau)^2 \right\rangle}}
	=\frac{\sum_{\boldsymbol{k}} S_{uu,\boldsymbol{k}} e^{{\rm i}(k_x r - \omega \tau)}}{\sum_{\boldsymbol{k}} S_{uu,\boldsymbol{k}}},
\end{equation}
for stationary stochastic processes in the $x$ and temporal direction such as the fully developed turbulent channel flow. With equation (\ref{eq:Correlation_function}), the space-time correlation can be estimated using the predicted spectra $S_{uu,\boldsymbol{k}}$. 
The space-time correlations of the streamwise velocity obtained from different methods are depicted in figure \ref{fig:Space-time_streamwise}, where the slopes of the black and red dashed lines respectively denote the local mean streamwise velocity $\left\langle u \right\rangle$ and the convection velocity $U_c$, as defined by
\begin{equation}
	\label{eq:Convection_U}
	U_c = \mathop{\rm{arg~max}}\limits_{U} \int_{-\infty}^{\infty}R\left(Ut,t\right)dt,
\end{equation}
following \citet{choi1990space}.
From the DNS results in figure \ref{fig:Space-time_streamwise}($a$-$c$), the magnitude of the space-time correlation at a given spatio-temporal point increases with the increasing $Re_\tau$, which is due to inner-flow scaling \citep{kunkel2006study} for the flows in the near-wall region. In all three cases, the red dashed line that denotes the convection velocity $U_c$ does not overlap with the black one that denotes the local mean streamwise velocity $\left\langle u \right\rangle$, where $U_c > \left\langle u \right\rangle$. The deviations of $U_c$ and $\left\langle u \right\rangle$ in the near-wall region are consistent with previous observations \citep{kim1993propagation, del2003spectra}. Since the convection velocity $U_c$ quantifies the propagation of the coherent structures \citep{kim1993propagation,he2017space}, the near-wall flow structures are propagated with a larger velocity than the local streamwise velocity.

In the predicted results from the four considered methods, quite diverse patterns are observed. In Case 180-40, the RWE and RBE results match well with the DNS results in terms of the correlation magnitude. However, the convection velocities predicted by the RWE and RBE are slightly lower than the DNS results, as the slopes of the predicted red dashed lines are smaller than those of the DNS. On the other hand, the space-time correlations predicted by the $\lambda$-model and WBE deviate even larger from the DNS results. For the $\lambda$-model, the regions where the space-time correlations are larger than 0.2 concentrate at the vicinity of the red dashed line corresponding to the convection velocity, which conflicts with the turbulence nature \citep{he2017space}. The space-time correlation predicted by the WBE is smaller than that from the DNS at a given spatio-temporal point. Moreover, the convection velocities predicted by the $\lambda$-model and WBE appear to be smaller than the local mean streamwise velocity, with relative errors larger than those of the RWE and RBE results in magnitude.

When the reference height $y_{\rm R}$ increases from 40 to 200, the space-time correlations predicted by RWE tend to become larger compared to the DNS results, which indicates that the large-scale motions are overestimated as the reference layer lifts up. The RBE-predicted results show a much larger overestimation of the space-time correlations, with the dark red region where the correlation magnitude is larger than 0.6 occupying the depicted area when $y_{\rm R}^+=200$. Meanwhile, the convection velocity predicted by the RBE is obviously larger than the DNS results in Case 950-200. As discussed in figure \ref{fig:autospectra_kxkz_Case_950-200_streamwise}, only the turbulent motions with large scales with $\lambda_x /h \ge 2$ are estimated by the RBE at $y^+ = 10$ in Case 950-200, which means that the space-time correlations predicted by the RBE are dominated by the large-scale flow motions with high spatio-temporal coherence. The $\lambda$-model-predicted correlations are large in magnitude in the region close to the red dashed line that denotes the convection velocity, which totally deviates from the patterns of the DNS results. The space-time correlation distributions predicted by the WBE also appear to increase in magnitude as the reference layer lifts up.

%%

\begin{figure}\centering% Requires \usepackage{graphicx}
	{\includegraphics[width=5.2in, angle=0]{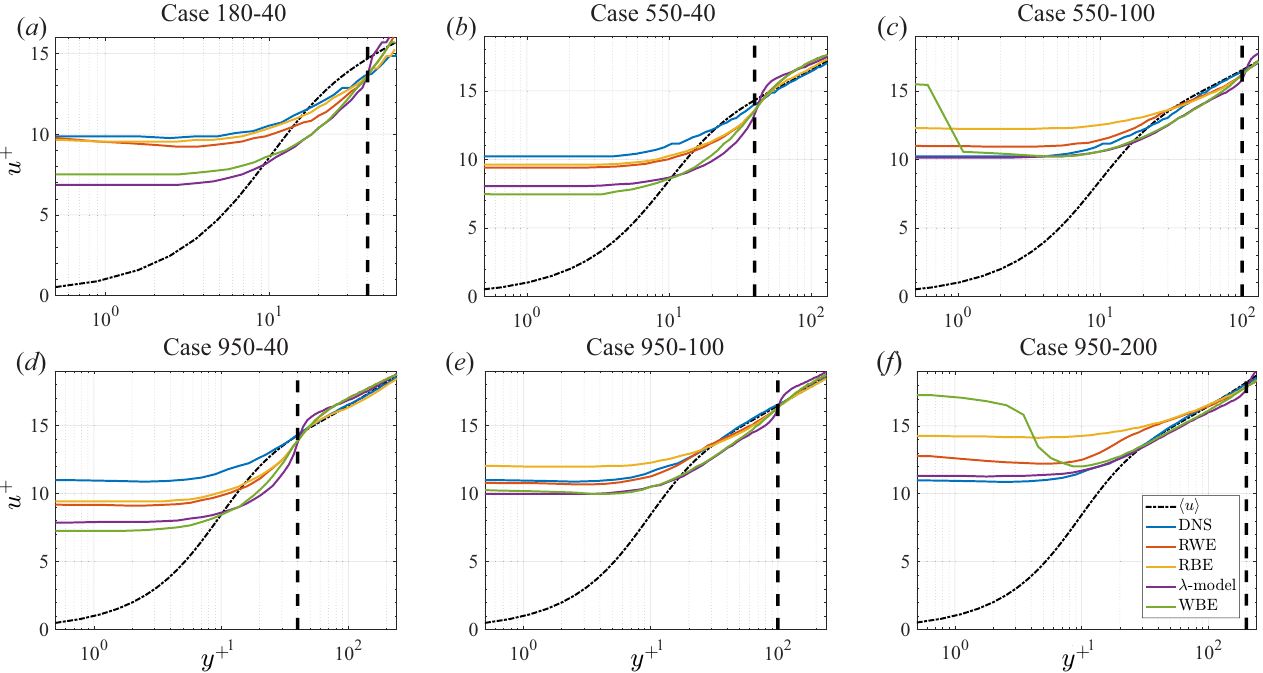}}
	\caption{Convection velocity as a function of $y^+$ for Cases 180-40 ($a$), 550-40 ($b$), 550-100 ($c$), 950-40 ($d$), 950-100 ($e$) and 950-200 ($f$).}
	\label{fig:Convection_U}
\end{figure}

\begin{figure}\centering% Requires \usepackage{graphicx}
	{\includegraphics[width=5.2in, angle=0]{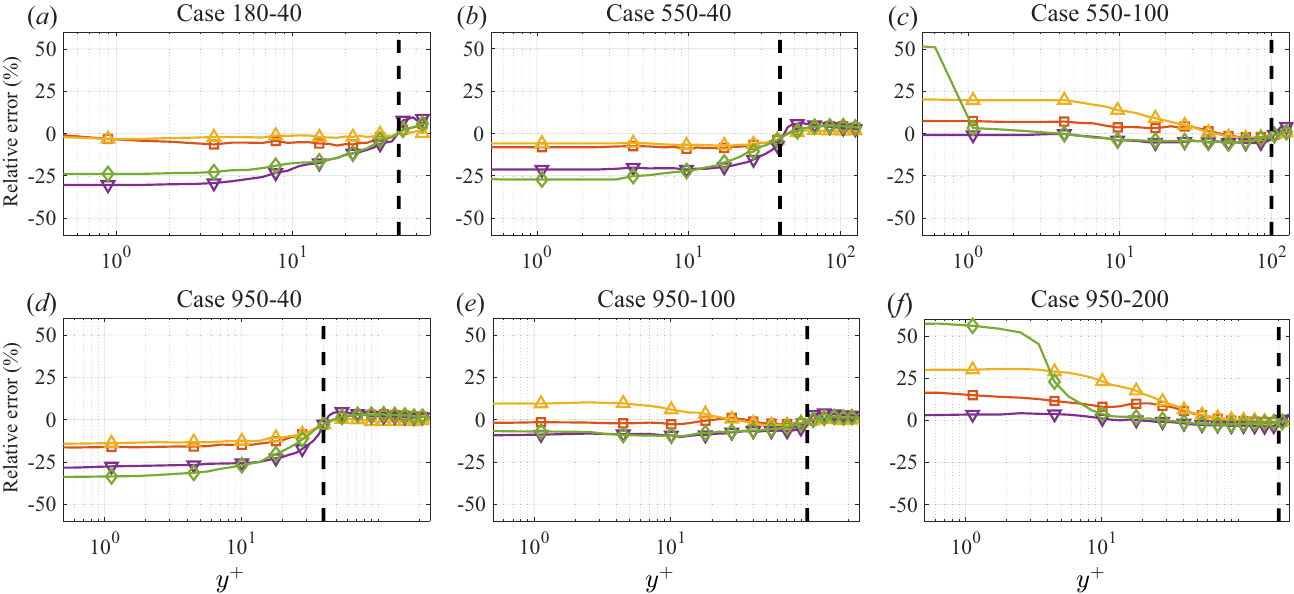}}
	\caption{Relative error of the convection velocity as a function of $y^+$ for Cases 180-40 ($a$), 550-40 ($b$), 550-100 ($c$), 950-40 ($d$), 950-100 ($e$) and 950-200 ($f$).}
	\label{fig:RE_Convection_U}
\end{figure}

%%%%%%%%% Convection velocity
From the above discussions on the space-time properties, it is noticed that the predicted near-wall convection velocities are quite different among the results from different methods. To further investigate the variations of the predicted convection velocity with different Reynolds numbers and reference heights, figure \ref{fig:Convection_U} shows convection velocity $U_c$ as a function of $y^+$ in all the 6 cases with $Re_{\tau}=180-950$ and the reference layer, ranging from $y^+ = 40$ to $y/h = 0.2$. 
The relative errors of the predicted convection velocities for different methods along the height are shown in figure \ref{fig:RE_Convection_U}.
Distinct characteristics are observed in the convection velocities predicted by different methods. 
For RBE, the convection velocity matches well with the DNS results when $y_{\rm R}^+ = 40$ with all the three considered friction Reynolds numbers that are equal to 180, 550, and 950. In Case 950-40, the RBE-predicted convection velocity is lower than the DNS results near the wall by about $14 \%$, which is still the most accurate compared to the results from other prediction methods.
With the increase of the reference height, the near-wall convection velocity by RBE increases and thus deviates from the DNS results, which is attributed to the underestimated near-wall small-scale flows with smaller convection velocities compared to those of the large-scale motions.
On the other hand, the WBE and $\lambda$-model underestimate the convection velocity when $y_{\rm R}^+ = 40$, which should be induced by the overestimated small-scale flows near the wall. With the increase of reference height, the predicted convection velocities by $\lambda$-model and WBE increase correspondingly, just like the phenomenon observed in the RBE results. A nonphysical increase of $U_c$ when approaching the wall in the WBE results is observed at the near-wall region in Cases 550-100 and 950-200, making the WBE-predicted near-wall convection velocity the largest among all the methods.
The $\lambda$-model provides the most accurate predictions of the convection velocity in Cases 550-100 and 950-200.
In all the tested cases, the RWE steadily provides fairly well results in predicting the convection velocity. When $y_{\rm R}^+ = 40$ and 100, the relative errors between the RWE and the DNS results are less than $15\%$. When $y_{\rm R}^+ = 200$, the RWE-predicted convection velocity shows an enlarged deviation with respect to the DNS results with the maximum error equal to $16.4 \%$ near the wall, which is still smaller than that in the WBE and RBE results.

%%

%%%%%%%%%%%%%%%%%%%%%%%%%%%%%%%%%%%%%%%%%%%%%%%%%%%%%%%%%%%%%%%%%%%%%%%%%%%%%%%%%%%%%%%%%%%%%%%%%%%%%%%%%%%%%%%%%%%%%%
\subsection{Instantaneous large-scale flow structure of the streamwise velocity near the wall}\label{subsec:Instataneous_Footprint}
According to the inner-outer interaction model \citep{marusic2010predictive,cheng2022consistency}, the footprint of the large-scale motions on the lower region can be calculated as
\begin{equation}\label{eq:IOIM}
	u_{L}^{+}\left(x^{+}, y^{+}, z^{+}\right)=F_{x}^{-1}\left\{H_{L}\left(\lambda_{x}^{+}, y^{+}\right) F_{x}\left[u_{o}^{+}\left(x^{+}, y_{o}^{+}, z^{+}\right)\right]\right\},
\end{equation}
where $u_{o}^{+}$ is the streamwise velocity fluctuation at $y_o^+$ in the logarithmic region, and $F_x$ and $F_x^{-1}$ denote FFT and inverse FFT in the streamwise direction, respectively. \citet{gupta2021linear} extend the transfer kernel $H_L$ to quantify the correlation between $u^+(y^+)$ and $u_{o}^{+}(y_o^+)$ at a given two-dimensional length scale of $(\lambda_{x}^{+} , \lambda_{z}^{+})$ rather than just the streamwise scale $\lambda_{x}^{+}$. The re-defined transfer kernel $H_L$ can be calculated as
\begin{equation}
	\begin{aligned}
		\label{eq:HL}
		H_{L}\left(\lambda_{x}^{+}, \lambda_{z}^{+}, y^{+}\right)
		&=\frac{\left\langle\hat{u}\left(\lambda_{x}^{+}, \lambda_{z}^{+}, y^{+}, z^{+}\right) \overline{\hat{u}}\left(\lambda_{x}^{+},\lambda_{z}^{+}, y_{0}^{+}, z^{+}\right)\right\rangle}{\left\langle\hat{u}\left(\lambda_{x}^{+}, \lambda_{z}^{+}, y_{0}^{+}, z^{+}\right) \hat{u}\left(\lambda_{x}^{+}, \lambda_{z}^{+}, y_{0}^{+}, z^{+}\right)\right\rangle}
		=\frac{S_{uu,y y_o}(k_{x}, k_{z})}{S_{uu,y_o y_o}(k_{x}, k_{z})} \\
		&=\frac{\sum_{\omega} S_{uu,y y_o}(k_{x}, k_{z}, \omega)}{\sum_{\omega} S_{uu,y_o y_o}(k_{x}, k_{z}, \omega)}.
	\end{aligned}
\end{equation}
Since all the tested methods can directly predict the CSD tensor of the velocity from the measurements, the transfer kernel $H_{L}$ can be estimated from the auto-spectra coefficient $S_{uu,y_o y_o}(k_{x}, k_{z}, \omega)$ and the cross-spectra coefficient $S_{uu,y y_o}(k_{x}, k_{z}, \omega)$ from the estimated CSD tensor. In \citet{gupta2021linear}, the large-scale fluctuations are estimated with the linear model as in equation (\ref{eq:IOIM}). Following \citet{gupta2021linear}, we will also conduct the estimation of large-scale structures using the transfer function estimated from the tested methods.
Case 950-200 where the reference layer is the farthest away from the wall among all the tested cases will be used to investigate the estimated footprint at two prediction heights of $y^+ = 100$ and 10 that correspond to the logarithmic region and the near-wall region, respectively.

\begin{figure}\centering% Requires \usepackage{graphicx}
	{\includegraphics[width=5.2in, angle=0]{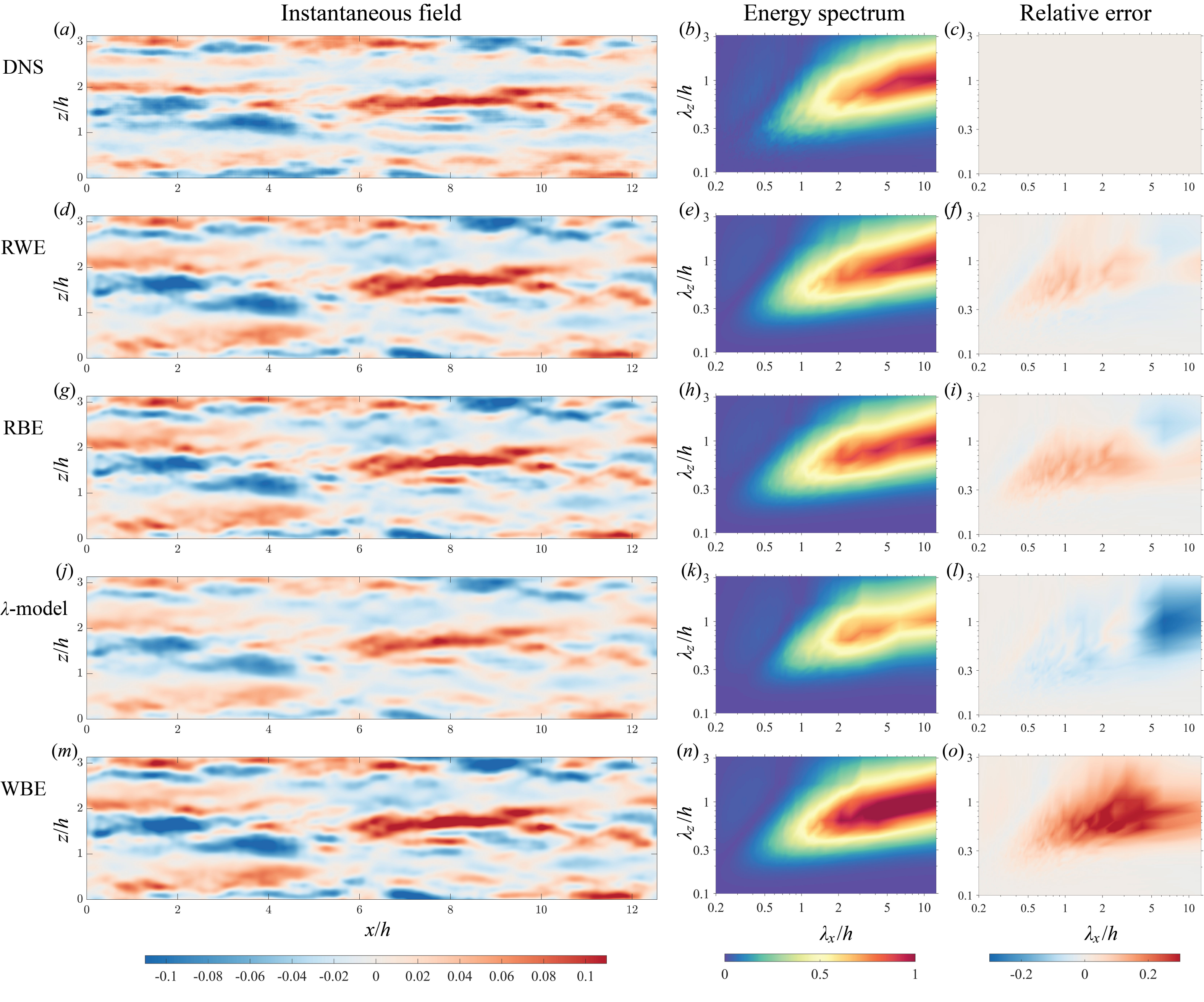}}
	\caption{Comparisons of the predicted footprint of the large scale motions of $y_{\rm R}^+ = 200$ to the prediction layer of $y^+ = 100$ in Case 950-200 from DNS ($a,b,c$), RWE ($d,e,f$), W-model ($g,h,i$), RBE ($j,k,l$), and WBE ($m,n,o$). Panels ($a,d,g,j,m$) are the instantaneous flow field, ($b,e,h,k,n$) are the energy spectra, and ($c,f,i,l,o$) are the relative error normalized by the maximum spectral energy in the DNS results. The values of fluctuation velocity shown in ($a,d,g,j,m$) are normalized by the mean velocity at $y = h$. The value of spectral energy and relative error shown in ($b,e,h,k,n$) and ($c,f,i,l,o$) are normalized by $S_{{\rm DNS,max}}$, respectively.}
	\label{fig:Footprint_y_950_100}
\end{figure}

Figure \ref{fig:Footprint_y_950_100} shows the estimated large-scale structures of the streamwise velocity fluctuation at $y^+ = 100$. From the instantaneous flow field shown in the left column, the RWE and RBE predictions well reflect the characteristics of the instantaneous flow structures. The good performance of RWE and RBE can also be seen in the spectral space, where the relative errors are smaller compared to those of the $\lambda$-model and WBE. Specifically, the relative errors in the RWE results are smaller than those in the RBE results, with a maximum error of $0.13 S_{{\rm DNS,max}}$ in the premultiplied energy spectra. On the other hand, the WBE and $\lambda$-model overestimate and underestimate the fluctuation energy at $y^+ = 100$, respectively. 

\begin{figure}\centering% Requires \usepackage{graphicx}
	{\includegraphics[width=5.2in, angle=0]{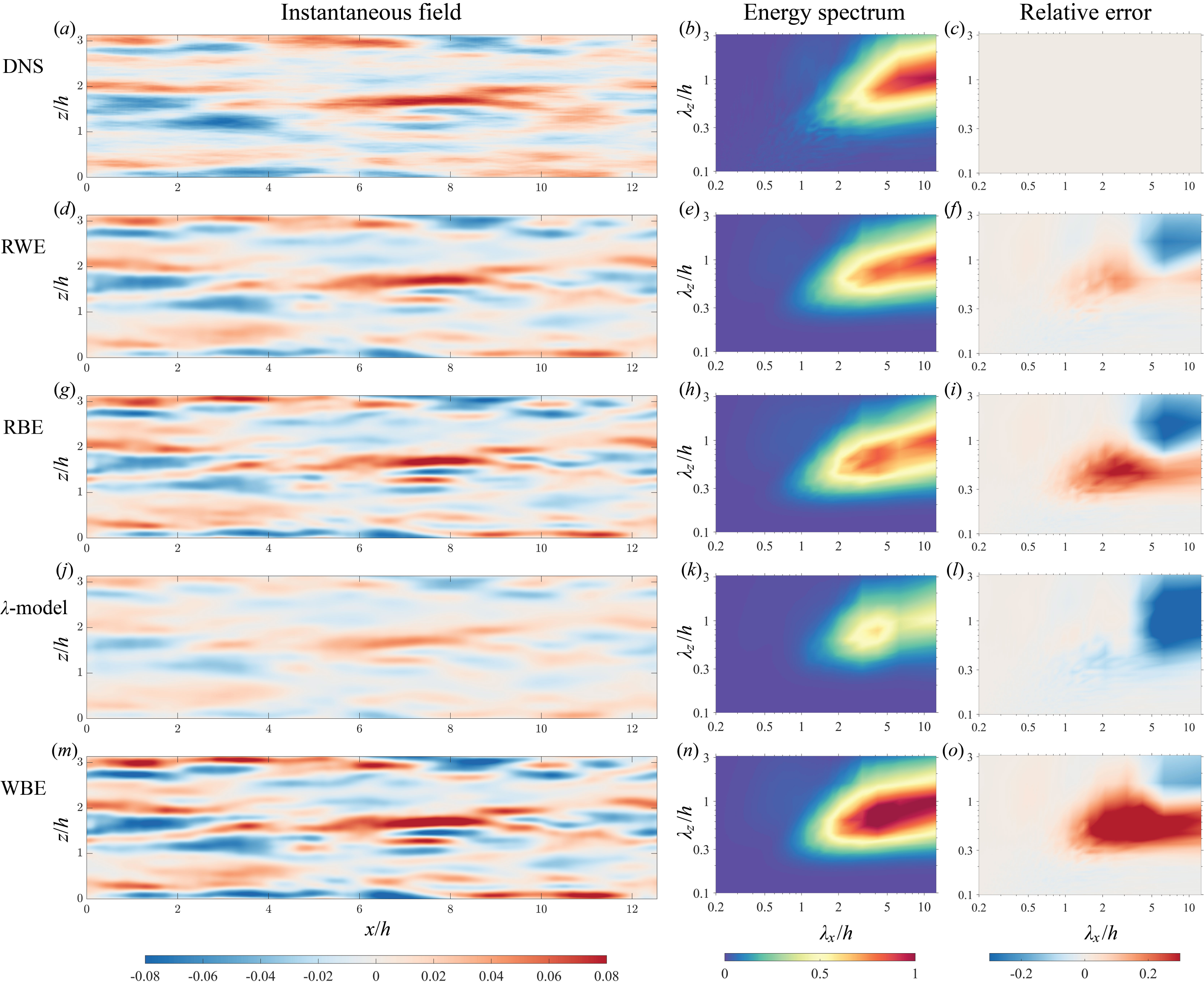}}
	\caption{Same as figure \ref{fig:Footprint_y_950_100}, but with the prediction layer at $y^+ = 10$.}
	\label{fig:Footprint_y_950_10}
\end{figure}

As the prediction layer approaches the wall, the magnitude of relative errors between the predictions from all the tested methods becomes larger, as in figure \ref{fig:Footprint_y_950_10}. For the RWE methods, the largest magnitude of error equal to $0.26 S_{{\rm DNS,max}}$ appears at $y^+ = 10$ for the large-scale motions with $(\lambda_x /h, \lambda_z /h) = (6.2, 1.6)$, where the predicted energy is lower than the DNS result. Meanwhile, another extreme point for error from RWE is found at $(\lambda_x /h, \lambda_z /h) = (2.5, 0.6)$ with the positive value equal to $0.17 S_{{\rm DNS,max}}$.
The relative errors in the RBE predictions reach the extreme points at $(\lambda_x /h, \lambda_z /h) = (6.2, 1.6)$ and $(2.5, 0.4)$. At the extreme points, the errors from RBE are equal to $-0.34 S_{{\rm DNS,max}}$ and $0.32 S_{{\rm DNS,max}}$, both of which are larger than those from RWE at the same points.
The $\lambda$-model-predicted energy is lower than the DNS for the scales with $\lambda_x /h \ge 6$, with the relative error larger than $0.5 S_{{\rm DNS,max}}$, which indicates that the fluctuation energy predicted by the $\lambda$-model is much overestimated in the near-wall region.
As for the WBE, the relative error is larger than $0.3 S_{{\rm DNS,max}}$ when $\lambda_x /h \ge 1.8$ and $0.4 \le \lambda_z /h \le 0.8$.
Considering the above comparisons, the RWE performs better than the other tested methods, with the smallest relative error in the estimated energy spectra at both the logarithmic region and the near-wall region.

In \citet{gupta2021linear}, the $\lambda$-model and WBE (namely B-model) are also used to predict the large-scale motions in the near-wall region, with the assumption that the input forcing is white in time, which leads to the same conclusion that the energy predicted by the $\lambda$-model is lower than that from the WBE. However, when applied in time-resolved cases as in this study, the energies predicted by the $\lambda$-model and WBE both tend to decrease compared to those using the white-in-time forcing in \citet{gupta2021linear}, which causes the obvious underestimation of energy by the $\lambda$-model. The deviations between the results generated by the forcing with different temporal properties indicate that the $\lambda$-model could be improved by considering the effect of temporal scales when applied in the time-resolved cases. On the other hand, the validity of the RWE-kind method for estimating large-scale motions in cases where only the spatial data are available should be further tested.

From the instantaneous flow field for large-scale motions as discussed above,  the structures are observed to delay backward as the prediction layer approaches the wall. The spatial delay of the flow structure is quantified by the SIA \citep{marusic2007reynolds}, which is defined as,
\begin{equation}
	\label{eq:Inclination_angle}
	\theta = {\rm{arctan}} \left[( y/({\Delta} x^{\ast}) \right],
\end{equation}
where ${\Delta} x^{\ast} =  \mathop{\rm{arg~max}}\limits_{{\Delta} x}R_{\tau u} \left( \Delta x \right) $, and
\begin{equation}
	\label{eq:Correlation_incline}
	R_{\tau u} \left( \Delta x \right) 
	= \frac{\left<\tau \left( t \right) u \left( t+\Delta x / U \right)\right>}{\sqrt{\left< \tau^2 \right>\left< u^2 \right>}}
	= \frac{\sum_{\boldsymbol{k}} S_{\tau u , \boldsymbol{k}} e^{{\rm i} k_x \Delta x}}{\sigma_\tau \sigma_u}
\end{equation}
is the correlation function between the wall shear stress and the streamwise velocity fluctuation with a spatial delay of $\Delta x$, and $S_{\tau u , \boldsymbol{k}}$ is the cross-spectra between the wall shear and the velocity fluctuations at ${\boldsymbol{k}}$.
To further investigate the ability of the tested methods in predicting the spatial distribution of the flow structures, the SIAs from the predictions are analyzed in the following. 
Since the logarithmic region does not exist clearly for the channel flow with $Re_\tau = 180$, only the Cases 550-100, 950-100, and 950-200 will be discussed.

\begin{figure}\centering% Requires \usepackage{graphicx}
	%\subfigure[]
	%\sidesubfloat[]
	{\includegraphics[width=5.0in, angle=0]{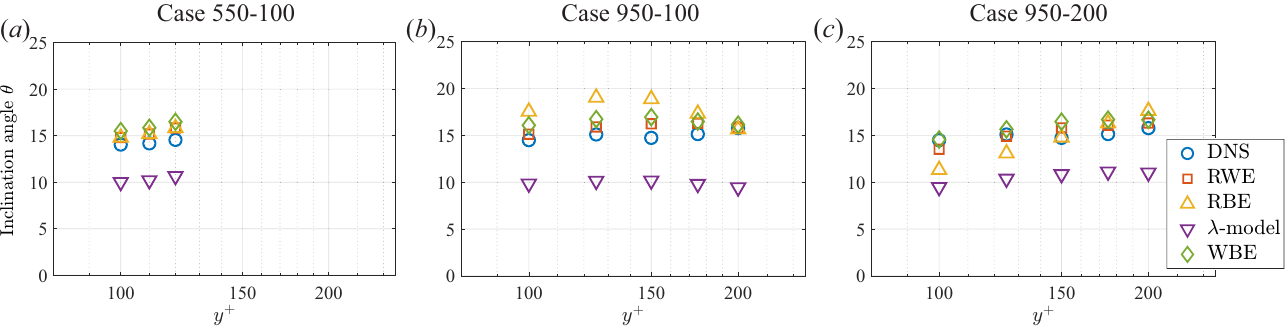}}
	\caption{Structure inclination angles of the predicted fluctuation velocity in ($a$) Case 550-100, ($b$) Case 950-100 and ($c$) Case 950-200.}
	\label{fig:Inclination_angles}
\end{figure}

Figure \ref{fig:Inclination_angles} depicts the SIAs in the tested cases based on the reference positions located inside the logarithmic region from $y^+ = 100$ to $y/h = 0.2$ \citep{jimenez2018coherent}.
According to the DNS results, the inclination angles keep almost constant around $15^{\circ}$ at reference positions with different heights, which is consistent with the results in \citet{marusic2007reynolds}.
In Case 550-100, all the tested methods fairly predict the inclination angle of around $15^{\circ}$ except for the $\lambda$-model. In fact, the $\lambda$-model-predicted SIAs are always around $10^{\circ}$ in all the three cases shown in figure \ref{fig:Inclination_angles}.
In Case 950-100, the RBE-predicted SIA is larger than the DNS result by about $3^{\circ}$ when $y^+ = 100$. As the height lifts up, the SIA predicted by RBE reaches the maximum value of about $19^{\circ}$ when $y_{\rm R}^+  = 125$, then gradually decreases and finally overlaps with the DNS results when $y_{\rm R}^+  = 200$. The RWE and WBE perform well in Case 950-100, with the root-mean-squared error compared to the DNS result equal to $0.96^{\circ}$ and $1.55^{\circ}$, respectively.
When the reference layer moves to $y^+ = 200$, as in figure \ref{fig:Inclination_angles}$(c)$, the RBE-predicted SIA is accurate when $y^+ \ge 175$. However, when the horizontal layers approach the wall, the SIA predicted by the RBE decreases, which becomes around $12^{\circ}$ when $y^+ = 100$. On the other hand, the RWE has the highest accuracy, with the root-mean-squared error equal to $0.81^{\circ}$. Meanwhile, the WBE-predicted results are also close to the DNS results, whose root-mean-squared error is $1.14^{\circ}$. 

Based on the above analyses of the flow structures predicted by the tested methods, the RWE has the best performance in estimating both spectral energy distributions and the structural inclination angles of the large-scale motions of the turbulent channel flow. Meanwhile, the WBE also performs well in predicting the SIAs, indicating that the relative distribution of the coherent structures can be fairly estimated by the WBE which just utilizes the white-noise-assumed forcing.

%%%%%%%%%%%%%%%%%%%%%%%%%%%%%%%%%%%%%%%%%%%%%%%%%%%%%%%%%%%%%%%%%%%%%%%%%%%%%%%%%%%%%%%%%%%%%%%%%%%%%%%%%%%%%%%%%%%%%%
\section{Conclusions}\label{sec:Conclusions}
In this study, a new method for predicting the turbulent channel flow is proposed, which is denoted as RWE. The RWE modifies the spatially uniform and uncorrelated forcing profile based on the relative value of response energy in the near-wall region estimated by the RBE \citep{towne2020resolvent}. Comprehensive validations are made to investigate the ability of RWE to predict the spectra of the incompressible turbulent channel flow. Time-resolved DNS datasets with $Re_\tau = 180$, 550, and 950 are used to provide limited layers of reference measurements for the predictions. The DNS results are further used as a standard to evaluate the accuracy of different prediction methods, including the RWE, $\lambda$-model \citep{gupta2021linear}, RBE, and the method using the WBE. To test the sensitivity of the prediction methods on the locations of measurements and the friction Reynolds numbers, several reference layers ranging from $y^+ = 40$ at the near-wall region to $y/h \approx 0.2$ at the upper bound of the logarithmic region are set in cases with $Re_\tau = 180$, 550 and 950, resulting in a total number of 6 cases for the validations.

The root-mean-squared (RMS) profiles of the velocity fluctuations are first tested for all 6 cases. The RMS profiles predicted by the RWE are pretty consistent with the DNS results. Specifically, the relative deviations of the streamwise RMS peaks between the RWE and DNS results are less than $11 \%$ in all the tested cases, indicating that this newly proposed method is robust in the friction Reynolds numbers and the measurement locations. On the other hand, the performances of the RBE and $\lambda$-model show a dependence on the specific measurement locations. The prediction accuracy of RBE is close to that of RWE when the reference layer is located at $y^+ = 40$ that is in the vicinity of the wall. With the increase of the reference height, the accuracy of RBE near the wall deteriorates rapidly. The near-wall energy predicted by the $\lambda$-model also decreases when the reference layer lifts. The $\lambda$-model reaches its best performance with the reference layer at $y^+ = 100$, and overestimates and underestimates the near-wall energy when $y^+ = 40$ and 200, respectively. In all the tested cases, the WBE largely overestimates the near-wall energy, indicating that the WBE is not suitable to be directly used for the estimations of near-wall energy magnitude.

The prediction capability of the considered methods is further investigated by examining the energy spectra. The RBE continuously shows a good prediction capacity at specific reference layers at $y^+ = 40$. Meanwhile, it is observed that the relative spectra predicted by the WBE can fairly reflect the energy distribution of the DNS results. Compared with the other tested methods, the RWE better predicts the energy distribution of velocity fluctuations at different scales along the heights ranging from the wall to the reference layer. On the other hand, compared to the DNS results, the predicted energy of all the considered methods tends to decrease as the flow scale decreases, while the RWE keeps the minimum relative error.

The space-time properties of the predicted flow field are also investigated, including the magnitude of the space-time correlation and the convection velocity. In Case 950-200, the magnitude of space-time correlations is overestimated by all the methods, while the RWE and WBE provide the closest results compared to the DNS. On the other hand, the RBE obviously overestimates the correlation magnitude, especially at the near-wall region $y^+ = 10$ due to the underestimation of the small-scale motions.
The convection velocity of the coherent structures is also studied. The RBE and RWE results match well with the DNS when the reference layer is located at $y^+ = 40$ that is close to the wall. However, the predicted convection velocity in the near-wall region by the RBE and RWE increases with the increasing height of the reference layer, which reaches the maximum error when $y_{\rm}^+ = 200$. In Case 950-200, the RWE-predicted convection velocity is larger than the DNS result by $16.4 \%$ at the near-wall region, which is smaller than the WBE and RBE.

The instantaneous flow field near the wall can be predicted based on the estimated CSD sensor and the reference velocity signal in the logarithmic region. The large-scale structures are well predicted by the RWE and RBE. On the other hand, the $\lambda$-model and WBE underestimate and overestimate the energies of the near-wall large-scale motions, respectively. The SIAs from the tested methods are researched to further investigate the characteristics of the predicted flow structures. The WBE and RWE perform well in predicting the SIAs in Cases 550-100 and 950-200 when compared with the DNS results, i.e., at around $15^{\circ}$.

From the above discussions, the strategy to modify the forcing profile informed by the RBE shows a strong potential in predicting the turbulent statistics of incompressible channel flows. Meanwhile, there are some discussions on the application scope of the current RWE method and its future extensions.
(1) In this study, we focus on estimating the near-wall statistics based on the measurements in the higher region, especially for the cases where the reference layer is located in the logarithmic region. A related problem could be the estimation of the flow patterns in the logarithmic region based on the wall measurements, which is explored by \citet{amaral2021resolvent} using the RBE and \citet{guastoni2021convolutional} using the CNN. Due to the restrictions of $y_{\rm Q} \le y_{\rm R}$ when determining the quasi-reference layer, the current RWE will degenerate to WBE when $y_{\rm R} \rightarrow 0$. To effectively make use of the wall measurements, the way to determine $y_{\rm Q}$ in the current RWE should be modified, which can be explored in future studies.
(2) The current RWE approach only uses one type of observation at one reference layer. It is clear that the prediction accuracy could be further improved with different measurements at multiple reference layers with suitable modeling approaches. Such a topic is to be explored in the future work.
(3) The method used in this study relies on the availability of time-resolved measurement data. The RWE could be more practical if it can predict the flow with the measurement data that is sparse in time, which might be realized using the assumptions of the stochastic forcing as used by \citep{hwang2010linear, gupta2021linear}.
(4) Possible extensions of the RWE to other types of turbulent flows, such as the pipe flows, boundary layers, and jet flows, could be explored in the future. For wall-bounded turbulence, the basic steps of the RWE can be directly applied. 
According to \citet{monty2009comparison} and \citet{lee2013comparison}, the differences between the channel/pipe flows and boundary layers are obvious in the very-large-scale motions, which are energetic in the outer region.
In the inner region where the RWE is applied, the differences in flow structures are not as obvious as those in the outer region. 
Thus, the RWE is believed to be valid in estimating the wall-bounded turbulence, including the turbulent pipe flows and boundary layers.
For jets, the approach of RWE developed in this study could not be directly applied, since the flow pattern is different from that of the wall-bounded turbulence. In \citet{schmidt2018spectral}, the inclusion of the eddy viscosity has been demonstrated to largely improve the agreement with the resolvent modes and the SPOD modes, which implies the potential of linear estimation of jets in the future. However, to our knowledge, the applicability of the resolvent-based estimation for jets from limited measurements is not fully investigated. Future studies are needed to explore the applications of the RWE-like methods in estimating the jets.
(5) The validity of the RWE in predicting the compressible or stratified turbulence has not been tested, which should be comprehensively investigated in the future.

%%%%%%%%%%%%%%%%%%%%%%%%%%%%%%%%%%%%%%%%%%%%%%%%%%%%%%%%%%%%%%%%%%%%%%%%%%%%%%%%%%%%%%%%
%---------------------------------------------------------------------------------------% 

%
\section*{Declaration of Competing Interest}

The authors declare that they have no known competing financial interests or personal relationships that could have appeared to influence the work reported in this paper. 

\section*{Data availability}
The data that support the findings of this study are available on request from the corresponding author, LF.

\section*{Acknowledgments}
%---------------------------------------------------------------------------------------% 
Lin Fu acknowledges the fund from the Research Grants Council (RGC) of the Government of Hong Kong Special Administrative Region (HKSAR) with RGC/ECS Project (No. 26200222) and RGC/GRF Project (No. 16201023), the fund from CORE as a joint research center for ocean research between QNLM and HKUST, and the fund from the Project of Hetao Shenzhen-Hong Kong Science and Technology Innovation Cooperation Zone (No. HZQB-KCZYB-2020083).

\appendix
\section{Linearized Navier-Stokes operator}\label{appA}
Recall the discretized state-space form of the linearized Navier-Stokes equations as
\begin{subeqnarray}
	\label{eq:ss_linearnsequations_appA}
	&\mathcal{M} \frac{\p \boldsymbol{q}_{\boldsymbol{k}_{\rm s}} (t)}{\p t}
	= \mathcal{A}_{\boldsymbol{k}_{\rm s}} \boldsymbol{q}_{\boldsymbol{k}_{\rm s}} (t) + \mathcal{B}\boldsymbol{f}_{\boldsymbol{k}_{\rm s}} (t), \\ [4pt]
	&\boldsymbol{m}_{\boldsymbol{k}_{\rm s}} (t) = \mathcal{C}\boldsymbol{q}_{\boldsymbol{k}_{\rm s}} (t) + \boldsymbol{n}_{\boldsymbol{k}_{\rm s}} (t).
\end{subeqnarray}
The operators $\mathcal{M} \in \mathbb{C}^{4N \times 4N}$, $\mathcal{A}_{\boldsymbol{k}_{\rm s}}  \in \mathbb{C}^{4N \times 4N}$, $\mathcal{B} \in \mathbb{C}^{4N \times 3N}$ in equation (\ref{eq:ss_linearnsequations_appA}$a$) are expressed as
\begin{equation}
	\label{eq:explan_operator_MABC}
	\begin{aligned}
		&\mathcal{M} = \left[
		\begin{array}{cccc}
			\mathcal{I}_N & 0 & 0 & 0 \\
			0 & \mathcal{I}_N & 0 & 0 \\
			0 & 0 & \mathcal{I}_N & 0 \\
			0 & 0 & 0 & 0 \\
		\end{array}  \right], 
		{~~}
		\mathcal{A}_{\boldsymbol{k}_{\rm s}} = 
		\left[
		\begin{array}{cccc}
			\mathcal{L}_{\boldsymbol{k}_{\rm s}}  & -\bnabla ^{\rm T} \\
			\bnabla & 0  \\
		\end{array}  \right],
		{~~}
		\mathcal{B} = 
		\left[
		\begin{array}{cccc}
			\mathcal{I}_N & 0 & 0 \\
			0 & \mathcal{I}_N & 0 \\
			0 & 0 & \mathcal{I}_N \\
			0 & 0 & 0 \\
		\end{array}  \right].
	\end{aligned}
\end{equation}
The spatial linearized Navier-Stokes operator $\mathcal{L}_{\boldsymbol{k}_{\rm s}} \in \mathbb{C}^{3N \times 3N}$ in equation (\ref{eq:explan_operator_MABC}) for the momentum equations is calculated as,
\begin{equation}
	\mathcal{L}_{\boldsymbol{k}_{\rm s}} = L_0 + {\rm i}{k_x}{L_x} + {L_y} \mathscr{D} + {\rm i}{k_z}{L_z} + {L_2}(- k_x^2 + \mathscr{D}^2 - k_z^2),
\end{equation}
\label{eq:app:resolventoperator}
with
\begin{equation}
	\label{eq:explan_operator_1}
	\begin{aligned}
		&L_x = \left[
		\begin{array}{ccc}
			\overline{\mathscr{u}}  & -\mathscr{v}_T^{\prime} & 0 \\
			0 & \overline{\mathscr{u}} & 0 \\
			0	& 0	&	\overline{\mathscr{u}} \\
		\end{array}  \right], 
		{~~}
		L_y = 
		\left[ \begin{array}{ccc}
			-\mathscr{v}_T^{\prime}  & 0 & 0 \\
			0 & -2\mathscr{v}_T^{\prime} & 0 \\
			0	& 0	&	-\mathscr{v}_T^{\prime} \\
		\end{array}  \right],
		{~~}
		L_z =
		\left[ \begin{array}{ccc}
			0  & 0 & 0 \\
			0 & 0 & 0 \\
			-\mathscr{v}_T^{\prime}	& 0	&	0 \\
		\end{array}  \right], \\
		&L_0 =
		\left[ \begin{array}{ccc}
			0  & \mathscr{u}^{\prime} & 0 \\
			0 & 0 & 0 \\
			0	& 0	&	0 \\
		\end{array}  \right],
		{~~}
		L_2 =
		\left[ \begin{array}{ccc}
			-\mathscr{v}_T  & 0 & 0 \\
			0 & -\mathscr{v}_T & 0 \\
			0	& 0	&	-\mathscr{v}_T \\
		\end{array}  \right],
	\end{aligned}
\end{equation}
where $\overline{\mathscr{u}} = {\rm diag}\left[ \overline{\boldsymbol{u}} \right] $,  $\overline{\boldsymbol{u}}$ is the mean streamwise velocity profile, $\mathscr{v}_T = \frac{1}{Re_{\tau}} {\rm diag}\left[ \frac{1}{\nu} \nu_T \right]$, $\mathscr{v}_T^{\prime} = \frac{1}{Re_{\tau}} {\rm diag}\left[ \frac{1}{\nu} \nu_T^{\prime} \right] $,
$\mathscr{D}$ and ${}^{\prime}$ denote ${\p} / {\p y}$.
The Hamiltonian operator $\bnabla \in \mathbb{C}^{N \times 3N}$ in equation (\ref{eq:resolventoperator}) is calculated as,
\begin{equation}
	\label{eq:explan_operator_2}
	\bnabla = \left[
	\begin{array}{ccc}
		{\rm i}{k_x}\mathcal{I}_N  & \mathscr{D} & {\rm i}{k_z}\mathcal{I}_N \\
	\end{array}  \right]. 
\end{equation}

The observation matrix $\mathcal{C}$ in equation (\ref{eq:ss_linearnsequations_appA}$b$) is determined by the location of the measured variable. In this study, the state matrix is assembled as
\begin{equation}
	\label{eq:state_m}
	\boldsymbol{q}_{\boldsymbol{k}_{\rm s}} (t)
	= \left[ \boldsymbol{u}^{\ast}_{\boldsymbol{k}_{\rm s}} (t),p^{\ast}_{\boldsymbol{k}_{\rm s}} (t) \right]^{\ast}
	= \left[ u^{\ast}_{\boldsymbol{k}_{\rm s}} (t),v^{\ast}_{\boldsymbol{k}_{\rm s}} (t),w^{\ast}_{\boldsymbol{k}_{\rm s}} (t),p^{\ast}_{\boldsymbol{k}_{\rm s}} (t) \right]^{\ast}.
\end{equation}
Denoting $\left( u_1,u_2,u_3 \right) = \left( u,v,w  \right) $, when the measured variable is the velocity $u_i~(i=1,2,3)$ at the $n$-th node, then the observation matrix $\mathcal{C} \in \mathbb{C}^{1 \times 4N}$ should be expressed as
\begin{equation}
	\label{eq:C_1}
	\mathcal{C} = \left[  \mathcal{O}_{1 \times \left[ {(i-1)N + n-1}\right]}  ~~ 1 ~~ \mathcal{O}_{ 1 \times \left[{4N-(i-1)N-n}\right]}  \right].
\end{equation}
Note that, in other cases, such as when the state matrix is assembled in another approach or the measurements are located at multiple wall-normal locations, the expression of the measurement operator should be redetermined in an analogous manner. For instance, in equation (\ref{eq:RBE_forcing}), the resolvent operator $\mathcal{R}_{\boldsymbol{k}}$ is utilized, which excludes the rows corresponding to the pressure state. In this case, when the measured variable is the velocity $u_i~(i=1,2,3)$ at the $n$-th node, the observation matrix as denoted by $\mathscr{C}$ should be calculated by
\begin{equation}
	\label{eq:C_2}
	\mathscr{C} = \left[  \mathcal{O}_{1 \times \left[ {(i-1)N + n-1}\right]}  ~~ 1 ~~ \mathcal{O}_{ 1 \times \left[{3N-(i-1)N-n}\right]}  \right].
\end{equation}

\section{Assessment of the DNS dataset generated in this study}\label{appB}
\begin{figure}\centering% Requires \usepackage{graphicx}
	{\includegraphics[width=5.3in, angle=0]{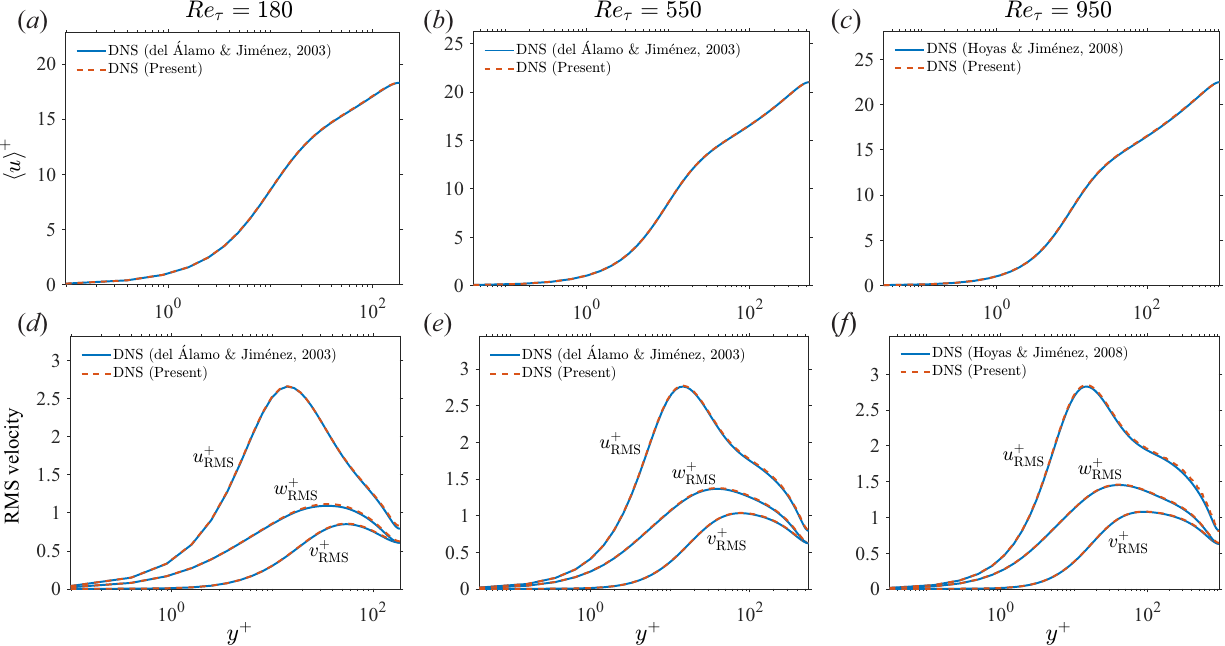}}
	\caption{Comparisons of the mean and RMS profiles between the open-source DNS results \citep{del2003spectra,hoyas2008reynolds} and those generated in this study.}
	\label{fig:Compare_DNSProfile}
\end{figure}

In figure \ref{fig:Compare_DNSProfile}, the DNS dataset generated in this study is compared with the open source DNS database \citep{del2003spectra,hoyas2008reynolds} for the mean streamwise velocity and the RMS velocities. Very small differences are found between the two datasets, which demonstrates that the DNS data generated in this study are reliable.

\section{Discussions on the determination of quasi-reference layer}\label{appC}
The LCS ($\gamma^2$) used to determine the quasi-reference layer $y_{\rm Q}$ is estimated using the WBE in each spatio-temporal scale.
This quantity is obtained by firstly estimating the CSD tensor with the white-noise-assumed forcing through equation (\ref{eq:resolvent_WBE_decompose}), i.e.,
\begin{equation}
	\label{eq:resolvent_CSD_appC}
	S_{\boldsymbol{uu},\boldsymbol{k}}
	=
	\mathcal{R}_{\boldsymbol{k}}
	S_{\boldsymbol{ff},\boldsymbol{k}}
	\mathcal{R}_{\boldsymbol{k}}^{\ast}
	=
	E_{\boldsymbol{k}}
	\cdot
	\mathcal{R}_{\boldsymbol{k}}
	\mathcal{R}_{\boldsymbol{k}}^{\ast}.
\end{equation}
From the definition of $\gamma^2$ in equation (\ref{eq:LCS}), and considering equation (\ref{eq:resolvent_CSD_appC}), $\gamma^2$ is calculated by,
\begin{equation}
	\label{eq:LCS_R_appC}
	\begin{aligned}
		&\gamma^2 (\boldsymbol{k})
		&&= \frac{\left| \left\langle  \hat{u} (y_{\rm Q}^+) \overline{\hat{u} (y_{\rm P}^+)} \right\rangle  \right|^2 }
		{\left\langle \left|  \hat{u} (y_{\rm Q}^+) \right|^2 \right\rangle \left\langle \left|  \hat{u} (y_{\rm P}^+) \right|^2 \right\rangle} \\
		&&&= \frac{S_{\boldsymbol{uu},\boldsymbol{k}}^2 \left( y_{\rm Q}^+ , y_{\rm P}^+ \right)}{S_{\boldsymbol{uu},\boldsymbol{k}} \left( y_{\rm Q}^+ , y_{\rm Q}^+ \right) S_{\boldsymbol{uu},\boldsymbol{k}} \left( y_{\rm P}^+ , y_{\rm P}^+ \right)} \\
		&&&= \frac{\left| \mathscr{C}(u,y_{\rm Q}^+) \mathcal{R}_{\boldsymbol{k}} \mathcal{R}_{\boldsymbol{k}}^{\ast} \mathscr{C}^{\rm T}(u,y_{\rm P}^+)  \right|^2 }
		{\left[ {\mathscr{C}(u,y_{\rm Q}^+) \mathcal{R}_{\boldsymbol{k}} \mathcal{R}_{\boldsymbol{k}}^{\ast} \mathscr{C}^{\rm T}(u,y_{\rm Q}^+)}\right]  \cdot \left[ {\mathscr{C}(u,y_{\rm P}^+) \mathcal{R}_{\boldsymbol{k}} \mathcal{R}_{\boldsymbol{k}}^{\ast} \mathscr{C}^{\rm T}(u,y_{\rm P}^+)}\right] },
	\end{aligned}
\end{equation}
where the expressions of $\mathcal{R}_{\boldsymbol{k}}$ and $\mathscr{C}(u,y^+)$ can be found in equations (\ref{eq:resolventoperator}-\ref{eq:resolventeqn2}) and Appendix \ref{appA}, respectively.

\begin{figure}\centering% Requires \usepackage{graphicx}
	{\includegraphics[width=5.0in, angle=0]{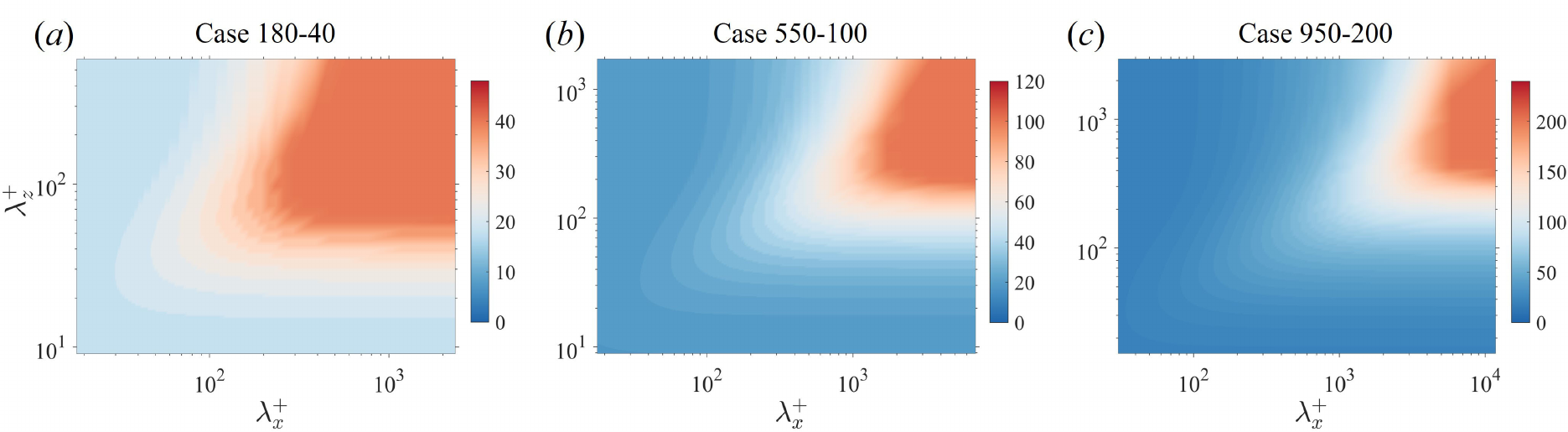}}
	\caption{Distributions of the quasi-reference layer $y_{\rm Q}^+$ at different wavelengths with $\omega = 10 u_{\tau} / k_x$ in Cases 180-40, 550-100, and 950-200.}
	\label{fig:YQ}
\end{figure}

Since the LCS calculated in equation (\ref{eq:LCS_R_appC}) is determined separately in each scale $\boldsymbol{k}$, the height of the quasi-reference layer calculated from the LCS is also determined scale-by-scale. When the threshold LCS is 0.3 as adopted in this study, the value of $y_{\rm Q}^+$ at the wavelength $(\lambda_x , \lambda_z) = (2 \pi/k_x , 2 \pi/k_z)$ with $\omega = 10 u_{\tau} / k_x$ in Cases 180-40, 550-100, and 950-200 are depicted in figure \ref{fig:YQ}. It can be observed that the height of the quasi-reference layer tends to increase when the flow scale is larger.

\begin{figure}\centering% Requires \usepackage{graphicx}
	{\includegraphics[width=2.5in, angle=0]{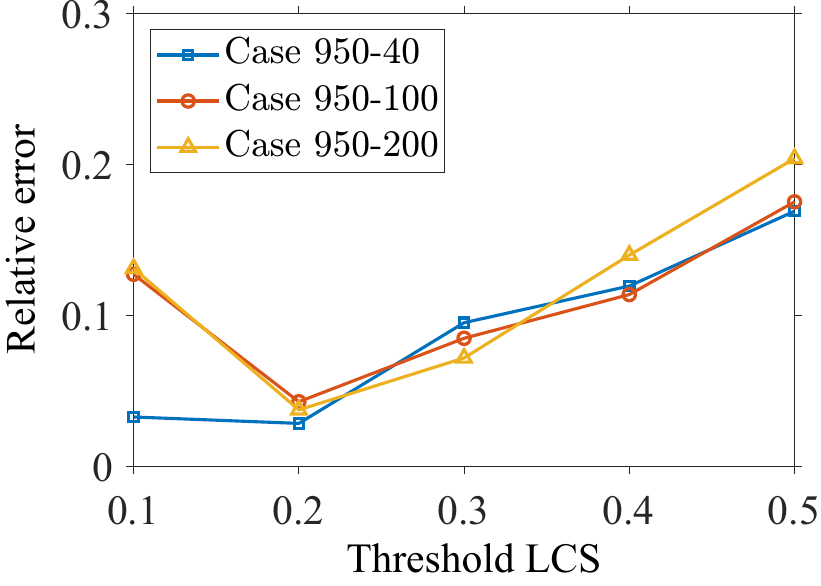}}
	\caption{Impact of the threshold LCS on the estimation error of the streamwise RMS profile.}
	\label{fig:Compare_YQ}
\end{figure}

In this study, we use the threshold LCS value (denoted as ${\rm LCS}_{\rm t}$) to determine the height of the quasi-reference layer, which has an important impact on the estimation result of the RWE. To efficiently modify the forcing profile, an ideal choice of the quasi-reference layer should be as high as possible under the premise of keeping the RBE valid for estimating the near-wall statistics.
To test the impact of ${\rm LCS}_{\rm t}$ on the estimation accuracy, the RWEs with five values of ${\rm LCS}_{\rm t}$ equal to 0.1, 0.2, 0.3, 0.4, and 0.5 are set to estimate the streamwise RMS profiles in Cases 950-40, 950-100, and 950-200. To quantify the relative estimation error in the tested cases, the integrated relative error of the streamwise RMS profile is defined as
\begin{equation}
	\label{eq:relatie_error}
	\epsilon = \left[ \frac{\int_{0}^{y_{\rm R}} \left( u_{\rm RMS, RWE} - u_{\rm RMS, DNS} \right)^2 {\rm d}y}{\int_{0}^{y_{\rm R}} \left( u_{\rm RMS, DNS} \right)^2 {\rm d}y} \right]^{0.5}.
\end{equation}
The estimation errors are summarized in figure \ref{fig:Compare_YQ}.
A universal law between the relative error and ${\rm LCS_t}$ clearly exists in all the tested cases when ${\rm LCS_t} \ge 0.2$.
Although the tested cases with ${\rm LCS_t} = 0.2$ appear to have the minimum errors, we notice that there are sharp increases of the relative error when ${\rm LCS_t}$ decreases from 0.2 to 0.1 in Cases 950-100 and 950-200. Such increases in the estimation error are due to the underestimation of the fluctuation energy in the near-wall region when ${\rm LCS_t}$ is too low, in which case $y_{Q}$ is too high so that the RBE becomes invalid to predict the near-wall statistics. To ensure the estimation accuracy in higher friction Reynolds numbers beyond 950, it is safer to conservatively set the value of ${\rm LCS_t}$ to be 0.3 to avoid the possible sharp increase of estimation error at ${\rm LCS_t} = 0.2$. Thus, the threshold value of ${\rm LCS_t} = 0.3$ is adopted as a default setting for all the cases with RWE in this study.

\section{Relationship between the relative energy profiles of forcing and response}\label{appD}
In this study, the relative forcing energy is modified between the wall and the quasi-reference layer $y_{\rm Q}$, which is set to be unity upon $y_{\rm Q}$ then.
The range of the relative forcing energy to be modified is the same as the range of the target response estimated from RBE, both of which are below the quasi-reference layer. In the optimization problem of this study, the relationship between the forcing and response energy below the quasi-reference layer is expressed by
\begin{equation}
	\label{eq:relation_f_u}
	\hat{\mathcal{S}}_{\boldsymbol{k}}
	=
	\frac{{ \mathscr{R}_{\boldsymbol{k},1} \cdot \hat{\mathcal{W}}_{\boldsymbol{k}} } + { \mathscr{R}_{\boldsymbol{k},2} \cdot \hat{\mathcal{U}} }}
	{{ \mathscr{R}_{\boldsymbol{k},{N_Q},1} \cdot  \hat{\mathcal{W}}_{\boldsymbol{k}} } + { \mathscr{R}_{\boldsymbol{k},{N_Q},2} \cdot \hat{\mathcal{U}} }},
\end{equation}
where $ \hat{\mathcal{S}}_{\boldsymbol{k}} = \frac{{\rm diag}\left[ S_{ {u_i u_i}, \boldsymbol{k}} \right] |_{y < y_{\rm Q}}}
{{\rm diag}\left[ S_{ {u_i u_i}, \boldsymbol{k}} \right] |_{y_{\rm Q}}}  \in \mathbb{C}^{N_{\rm Q}-1} $ is the relative energy profile below the quasi-reference layer. The relative forcing energy profile $W_{\boldsymbol{k}}(y)$ is split in equation (\ref{eq:relation_f_u}) as $W_{\boldsymbol{k}} = [\hat{\mathcal{W}}_{\boldsymbol{k}}^{\rm T} , \hat{\mathcal{U}}^{\rm T}]^{\rm T}$, where $ \hat{\mathcal{W}}_{\boldsymbol{k}}  \in \mathbb{C}^{N_{\rm Q}-1}$ is the relative forcing energy below $y_{\rm Q}$ that is to be determined, $\hat{\mathcal{U}} = [1,...,1]_{N-N_{\rm Q}+1}$ is that beyond $y_{\rm Q}$. The linear operators in equation (\ref{eq:relation_f_u}) are expressed by $\mathscr{R}_{\boldsymbol{k},1} = \mathcal{C}_1 \mathscr{R}_{\boldsymbol{k}} \mathcal{C}_1^{\rm T}$, $\mathscr{R}_{\boldsymbol{k},2} = \mathcal{C}_1 \mathscr{R}_{\boldsymbol{k}} \mathcal{C}_2^{\rm T}$, $\mathscr{R}_{\boldsymbol{k},{N_Q},1} = \mathcal{C}_{N_Q} \mathscr{R}_{\boldsymbol{k}} \mathcal{C}_1^{\rm T}$, $\mathscr{R}_{\boldsymbol{k},{N_Q},2} = \mathcal{C}_{N_Q} \mathscr{R}_{\boldsymbol{k}} \mathcal{C}_2^{\rm T}$, and
\begin{equation}
	\label{eq:Rk_expand}
	\begin{aligned}
		&&&\mathcal{C}_1
		=
		\left[ \begin{array}{cc}
			\mathcal{I}_{(N_Q-1),(N_Q-1)}  & \mathcal{O}_{(N_Q-1),(N-N_Q+1)} \\
		\end{array}  \right]  \in \mathbb{C}^{(N_Q-1) \times N}, \\
		&&&\mathcal{C}_2
		=
		\left[ \begin{array}{cc}
			\mathcal{O}_{(N-N_Q+1),(N_Q-1)} & \mathcal{I}_{(N-N_Q+1),(N-N_Q+1)} \\
		\end{array}  \right]  \in \mathbb{C}^{(N-N_Q+1) \times N}, \\
		&&&\mathcal{C}_{N_Q}
		=
		\left[ \begin{array}{ccc}
			\mathcal{O}_{1,(N_Q-1)} & 1 & \mathcal{O}_{1,(N-N_Q)} \\
		\end{array}  \right]  \in \mathbb{C}^{1 \times N}, \\
		&&&\mathscr{R}_{\boldsymbol{k}} = \sum_{j=1}^{3}\mathcal{R}_{\boldsymbol{k},i,j} \circ \overline{\mathcal{R}_{\boldsymbol{k},i,j}} \in \mathbb{C}^{N \times N},
	\end{aligned}
\end{equation}
where  $\mathcal{R}_{\boldsymbol{k},i,j}$ denotes the sub-matrix of the resolvent operator $\mathcal{R}_{\boldsymbol{k}}$ by selecting its rows for $u_i$ and columns for the forcing in $x_j$ direction,
$\circ$ denotes the element-wise matrix multiplication, and $N_{\rm Q}$ is the node number between $y_{\rm Q}$ and the wall.

\bibliographystyle{jfm}
%\bibliography{jfm}
%Use of the above commands will create a bibliography using the .bib file. Shown below is a bibliography built from individual items.

%\bibliographystyle{jfm}
%\bibliography{jfm2esam}

\begin{thebibliography}{59}
\expandafter\ifx\csname natexlab\endcsname\relax\def\natexlab#1{#1}\fi
\def\au#1{#1} \def\ed#1{#1} \def\yr#1{#1}\def\at#1{#1}\def\jt#1{\textit{#1}}
  \def\bt#1{#1}\def\bvol#1{\textbf{#1}} \def\vol#1{#1} \def\pg#1{#1}
  \def\publ#1{#1}\def\arxiv#1{#1}\def\org#1{#1}\def\st#1{\textit{#1}}

\bibitem[Amaral {\em et~al.\/}(2021)Amaral, Cavalieri, Martini, Jordan \&
  Towne]{amaral2021resolvent}
{\sc \au{Amaral, Filipe~R}, \au{Cavalieri, Andr{\'e}~VG}, \au{Martini,
  Eduardo}, \au{Jordan, Peter} \& \au{Towne, Aaron}} \yr{2021}
  \at{Resolvent-based estimation of turbulent channel flow using wall
  measurements}.  \jt{Journal of Fluid Mechanics}  \bvol{927}.

\bibitem[Baars {\em et~al.\/}(2016)Baars, Hutchins \&
  Marusic]{baars2016spectral}
{\sc \au{Baars, W.J}, \au{Hutchins, N.} \& \au{Marusic, I.}} \yr{2016}
  \at{Spectral stochastic estimation of high-reynolds-number wall-bounded
  turbulence for a refined inner-outer interaction model}.  \jt{Physical Review
  Fluids}  \bvol{1}~(5),  \pg{054406}.

\bibitem[Bae {\em et~al.\/}(2018)Bae, Lozano-Durán, Bose \& Moin]{Bae2018}
{\sc \au{Bae, H.J.}, \au{Lozano-Durán, A.}, \au{Bose, S.T.} \& \au{Moin, P.}}
  \yr{2018}  \at{Turbulence intensities in large-eddy simulation of
  wall-bounded flows}.  \jt{Physical Review Fluids}  \bvol{3},  \pg{014610}.

\bibitem[Beneddine {\em et~al.\/}(2016)Beneddine, Sipp, Arnault, Dandois \&
  Lesshafft]{beneddine2016conditions}
{\sc \au{Beneddine, Samir}, \au{Sipp, Denis}, \au{Arnault, Anthony},
  \au{Dandois, Julien} \& \au{Lesshafft, Lutz}} \yr{2016}  \at{Conditions for
  validity of mean flow stability analysis}.  \jt{Journal of Fluid Mechanics}
  \bvol{798},  \pg{485--504}.

\bibitem[Cess(1958)]{cess1958survey}
{\sc \au{Cess, R.D.}} \yr{1958}  \bt{A survey of the literature on heat
  transfer in turbulent tube flow. rep}. {\em Tech. Rep.\/}.  \org{8-0529-R24.
  Westinghouse Research}.

\bibitem[Cheng \& Fu(2022)]{cheng2022consistency}
{\sc \au{Cheng, Cheng} \& \au{Fu, Lin}} \yr{2022}  \at{Consistency between the
  attached-eddy model and the inner--outer interaction model: a study of
  streamwise wall-shear stress fluctuations in a turbulent channel flow}.
  \jt{Journal of Fluid Mechanics}  \bvol{942}.

\bibitem[Cheng \& Fu(2023)]{cheng2023scale}
{\sc \au{Cheng, Cheng} \& \au{Fu, Lin}} \yr{2023}  \at{{A scale-based study of
  the Reynolds number scaling for the near-wall streamwise turbulence intensity
  in wall turbulence}}.  \jt{International Journal of Heat and Fluid Flow}
  \bvol{101},  \pg{109136}.

\bibitem[Cheng {\em et~al.\/}(2019)Cheng, Li, Lozano-Dur{\'a}n \&
  Liu]{cheng2019identity}
{\sc \au{Cheng, Cheng}, \au{Li, Weipeng}, \au{Lozano-Dur{\'a}n, Adri{\'a}n} \&
  \au{Liu, Hong}} \yr{2019}  \at{Identity of attached eddies in turbulent
  channel flows with bidimensional empirical mode decomposition}.  \jt{Journal
  of fluid mechanics}  \bvol{870},  \pg{1037--1071}.

\bibitem[Cheng {\em et~al.\/}(2022)Cheng, Shyy \& Fu]{cheng2022streamwise}
{\sc \au{Cheng, Cheng}, \au{Shyy, Wei} \& \au{Fu, Lin}} \yr{2022}
  \at{Streamwise inclination angle of wall-attached eddies in turbulent channel
  flows}.  \jt{Journal of Fluid Mechanics}  \bvol{946},  \pg{A49}.

\bibitem[Chevalier {\em et~al.\/}(2006)Chevalier, H{\oe}pffner, Bewley \&
  Henningson]{chevalier2006state}
{\sc \au{Chevalier, Mattias}, \au{H{\oe}pffner, J{\'e}r{\^o}me}, \au{Bewley,
  Thomas~R} \& \au{Henningson, Dan~S}} \yr{2006}  \at{State estimation in
  wall-bounded flow systems. part 2. turbulent flows}.  \jt{Journal of Fluid
  Mechanics}  \bvol{552},  \pg{167--187}.

\bibitem[Cho {\em et~al.\/}(2018)Cho, Hwang \& Choi]{cho2018scale}
{\sc \au{Cho, Minjeong}, \au{Hwang, Yongyun} \& \au{Choi, Haecheon}} \yr{2018}
  \at{Scale interactions and spectral energy transfer in turbulent channel
  flow}.  \jt{Journal of Fluid Mechanics}  \bvol{854},  \pg{474--504}.

\bibitem[Choi \& Moin(1990)]{choi1990space}
{\sc \au{Choi, Haecheon} \& \au{Moin, Parviz}} \yr{1990}  \at{On the space-time
  characteristics of wall-pressure fluctuations}.  \jt{Physics of Fluids A:
  Fluid Dynamics}  \bvol{2}~(8),  \pg{1450--1460}.

\bibitem[Del~Alamo \& Jim{\'e}nez(2003)]{del2003spectra}
{\sc \au{Del~Alamo, Juan~C} \& \au{Jim{\'e}nez, Javier}} \yr{2003}  \at{Spectra
  of the very large anisotropic scales in turbulent channels}.  \jt{Physics of
  Fluids}  \bvol{15}~(6),  \pg{L41--L44}.

\bibitem[Fu {\em et~al.\/}(2022)Fu, Bose \& Moin]{fu2022prediction}
{\sc \au{Fu, Lin}, \au{Bose, Sanjeeb} \& \au{Moin, Parviz}} \yr{2022}
  \at{{Prediction of aerothermal characteristics of a generic hypersonic inlet
  flow}}.  \jt{Theoretical and Computational Fluid Dynamics}  \bvol{36}~(2),
  \pg{345--368}.

\bibitem[Fu {\em et~al.\/}(2021)Fu, Karp, Bose, Moin \& Urzay]{fu2021shock}
{\sc \au{Fu, Lin}, \au{Karp, Michael}, \au{Bose, Sanjeeb~T}, \au{Moin, Parviz}
  \& \au{Urzay, Javier}} \yr{2021}  \at{Shock-induced heating and transition to
  turbulence in a hypersonic boundary layer}.  \jt{Journal of Fluid Mechanics}
  \bvol{909},  \pg{A8}.

\bibitem[Fukagata {\em et~al.\/}(2002)Fukagata, Iwamoto \&
  Kasagi]{fukagata2002contribution}
{\sc \au{Fukagata, Koji}, \au{Iwamoto, Kaoru} \& \au{Kasagi, Nobuhide}}
  \yr{2002}  \at{Contribution of reynolds stress distribution to the skin
  friction in wall-bounded flows}.  \jt{Physics of fluids}  \bvol{14}~(11),
  \pg{L73--L76}.

\bibitem[Guastoni {\em et~al.\/}(2021)Guastoni, G{\"u}emes, Ianiro, Discetti,
  Schlatter, Azizpour \& Vinuesa]{guastoni2021convolutional}
{\sc \au{Guastoni, Luca}, \au{G{\"u}emes, Alejandro}, \au{Ianiro, Andrea},
  \au{Discetti, Stefano}, \au{Schlatter, Philipp}, \au{Azizpour, Hossein} \&
  \au{Vinuesa, Ricardo}} \yr{2021}  \at{Convolutional-network models to predict
  wall-bounded turbulence from wall quantities}.  \jt{Journal of Fluid
  Mechanics}  \bvol{928},  \pg{A27}.

\bibitem[G{\"u}emes {\em et~al.\/}(2021)G{\"u}emes, Discetti, Ianiro, Sirmacek,
  Azizpour \& Vinuesa]{guemes2021coarse}
{\sc \au{G{\"u}emes, Alejandro}, \au{Discetti, Stefano}, \au{Ianiro, Andrea},
  \au{Sirmacek, Beril}, \au{Azizpour, Hossein} \& \au{Vinuesa, Ricardo}}
  \yr{2021}  \at{From coarse wall measurements to turbulent velocity fields
  through deep learning}.  \jt{Physics of fluids}  \bvol{33}~(7),  \pg{075121}.

\bibitem[Gupta {\em et~al.\/}(2021)Gupta, Madhusudanan, Wan, Illingworth \&
  Juniper]{gupta2021linear}
{\sc \au{Gupta, Vikrant}, \au{Madhusudanan, Anagha}, \au{Wan, Minping},
  \au{Illingworth, Simon~J} \& \au{Juniper, Matthew~P}} \yr{2021}
  \at{Linear-model-based estimation in wall turbulence: improved stochastic
  forcing and eddy viscosity terms}.  \jt{Journal of Fluid Mechanics}
  \bvol{925}.

\bibitem[He {\em et~al.\/}(2017)He, Jin \& Yang]{he2017space}
{\sc \au{He, Guowei}, \au{Jin, Guodong} \& \au{Yang, Yue}} \yr{2017}
  \at{Space-time correlations and dynamic coupling in turbulent flows}.
  \jt{Annual Review of Fluid Mechanics}  \bvol{49},  \pg{51--70}.

\bibitem[He \& Zhang(2006)]{he2006elliptic}
{\sc \au{He, Guo-Wei} \& \au{Zhang, Jin-Bai}} \yr{2006}  \at{Elliptic model for
  space-time correlations in turbulent shear flows}.  \jt{Physical Review E}
  \bvol{73}~(5),  \pg{055303}.

\bibitem[H{\oe}pffner {\em et~al.\/}(2005)H{\oe}pffner, Chevalier, Bewley \&
  Henningson]{hoepffner2005state}
{\sc \au{H{\oe}pffner, J{\'e}r{\^o}me}, \au{Chevalier, Mattias}, \au{Bewley,
  Thomas~R} \& \au{Henningson, Dan~S}} \yr{2005}  \at{State estimation in
  wall-bounded flow systems. part 1. perturbed laminar flows}.  \jt{Journal of
  Fluid Mechanics}  \bvol{534},  \pg{263--294}.

\bibitem[Holford {\em et~al.\/}(2023)Holford, Lee \&
  Hwang]{holford_lee_hwang_2023}
{\sc \au{Holford, Jacob~J.}, \au{Lee, Myoungkyu} \& \au{Hwang, Yongyun}}
  \yr{2023}  \at{Optimal white-noise stochastic forcing for linear models of
  turbulent channel flow}.  \jt{Journal of Fluid Mechanics}  \bvol{961},
  \pg{A32}.

\bibitem[Hoyas \& Jim{\'e}nez(2008)]{hoyas2008reynolds}
{\sc \au{Hoyas, Sergio} \& \au{Jim{\'e}nez, Javier}} \yr{2008}  \at{Reynolds
  number effects on the reynolds-stress budgets in turbulent channels}.
  \jt{Physics of Fluids}  \bvol{20}~(10),  \pg{101511}.

\bibitem[Hwang \& Cossu(2010)]{hwang2010linear}
{\sc \au{Hwang, Yongyun} \& \au{Cossu, Carlo}} \yr{2010}  \at{Linear non-normal
  energy amplification of harmonic and stochastic forcing in the turbulent
  channel flow}.  \jt{Journal of Fluid Mechanics}  \bvol{664},  \pg{51--73}.

\bibitem[Illingworth {\em et~al.\/}(2018)Illingworth, Monty \&
  Marusic]{illingworth2018estimating}
{\sc \au{Illingworth, Simon~J}, \au{Monty, Jason~P} \& \au{Marusic, Ivan}}
  \yr{2018}  \at{Estimating large-scale structures in wall turbulence using
  linear models}.  \jt{Journal of Fluid Mechanics}  \bvol{842},  \pg{146--162}.

\bibitem[Jim{\'e}nez(2018)]{jimenez2018coherent}
{\sc \au{Jim{\'e}nez, Javier}} \yr{2018}  \at{Coherent structures in
  wall-bounded turbulence}.  \jt{Journal of Fluid Mechanics}  \bvol{842},
  \pg{P1}.

\bibitem[Jovanovi{\'c} \& Bamieh(2005)]{jovanovic2005componentwise}
{\sc \au{Jovanovi{\'c}, Mihailo~R} \& \au{Bamieh, Bassam}} \yr{2005}
  \at{Componentwise energy amplification in channel flows}.  \jt{Journal of
  Fluid Mechanics}  \bvol{534},  \pg{145--183}.

\bibitem[Karban {\em et~al.\/}(2022)Karban, Martini, Cavalieri, Lesshafft \&
  Jordan]{karban2022self}
{\sc \au{Karban, U}, \au{Martini, E}, \au{Cavalieri, AVG}, \au{Lesshafft, L} \&
  \au{Jordan, P}} \yr{2022}  \at{Self-similar mechanisms in wall turbulence
  studied using resolvent analysis}.  \jt{Journal of Fluid Mechanics}
  \bvol{939}.

\bibitem[Kim \& Hussain(1993)]{kim1993propagation}
{\sc \au{Kim, John} \& \au{Hussain, Fazle}} \yr{1993}  \at{Propagation velocity
  of perturbations in turbulent channel flow}.  \jt{Physics of Fluids A: Fluid
  Dynamics}  \bvol{5}~(3),  \pg{695--706}.

\bibitem[Kunkel \& Marusic(2006)]{kunkel2006study}
{\sc \au{Kunkel, Gary~J} \& \au{Marusic, Ivan}} \yr{2006}  \at{Study of the
  near-wall-turbulent region of the high-reynolds-number boundary layer using
  an atmospheric flow}.  \jt{Journal of Fluid Mechanics}  \bvol{548},
  \pg{375--402}.

\bibitem[Larsson {\em et~al.\/}(2016)Larsson, Kawai, Bodart \&
  Bermejo-Moreno]{larsson2016large}
{\sc \au{Larsson, Johan}, \au{Kawai, Soshi}, \au{Bodart, Julien} \&
  \au{Bermejo-Moreno, Ivan}} \yr{2016}  \at{Large eddy simulation with modeled
  wall-stress: recent progress and future directions}.  \jt{Mechanical
  Engineering Reviews}  \bvol{3}~(1),  \pg{15--00418}.

\bibitem[Lee \& Sung(2013)]{lee2013comparison}
{\sc \au{Lee, Jae~Hwa} \& \au{Sung, Hyung~Jin}} \yr{2013}  \at{Comparison of
  very-large-scale motions of turbulent pipe and boundary layer simulations}.
  \jt{Physics of Fluids}  \bvol{25}~(4).

\bibitem[Luhar {\em et~al.\/}(2015)Luhar, Sharma \& McKeon]{luhar2015framework}
{\sc \au{Luhar, Mitul}, \au{Sharma, Ati~S} \& \au{McKeon, BJ}} \yr{2015}  \at{A
  framework for studying the effect of compliant surfaces on wall turbulence}.
  \jt{Journal of Fluid Mechanics}  \bvol{768},  \pg{415--441}.

\bibitem[Madhusudanan {\em et~al.\/}(2019)Madhusudanan, Illingworth \&
  Marusic]{madhusudanan2019coherent}
{\sc \au{Madhusudanan, Anagha}, \au{Illingworth, Simon~J} \& \au{Marusic,
  Ivan}} \yr{2019}  \at{Coherent large-scale structures from the linearized
  navier--stokes equations}.  \jt{Journal of Fluid Mechanics}  \bvol{873},
  \pg{89--109}.

\bibitem[Martini {\em et~al.\/}(2020)Martini, Cavalieri, Jordan, Towne \&
  Lesshafft]{martini2020resolvent}
{\sc \au{Martini, Eduardo}, \au{Cavalieri, Andr{\'e}~VG}, \au{Jordan, Peter},
  \au{Towne, Aaron} \& \au{Lesshafft, Lutz}} \yr{2020}  \at{Resolvent-based
  optimal estimation of transitional and turbulent flows}.  \jt{Journal of
  Fluid Mechanics}  \bvol{900}.

\bibitem[Marusic \& Heuer(2007)]{marusic2007reynolds}
{\sc \au{Marusic, Ivan} \& \au{Heuer, Weston~DC}} \yr{2007}  \at{Reynolds
  number invariance of the structure inclination angle in wall turbulence}.
  \jt{Physical review letters}  \bvol{99}~(11),  \pg{114504}.

\bibitem[Marusic {\em et~al.\/}(2010)Marusic, Mathis \&
  Hutchins]{marusic2010predictive}
{\sc \au{Marusic, I.}, \au{Mathis, R.} \& \au{Hutchins, N}} \yr{2010}
  \at{Predictive model for wall-bounded turbulent flow}.  \jt{Science}
  \bvol{329}~(5988),  \pg{193--196}.

\bibitem[McKeon(2017)]{mckeon2017engine}
{\sc \au{McKeon, BJ}} \yr{2017}  \at{The engine behind (wall) turbulence:
  perspectives on scale interactions}.  \jt{Journal of Fluid Mechanics}
  \bvol{817}.

\bibitem[McKeon \& Sharma(2010)]{mckeon2010critical}
{\sc \au{McKeon, Beverley~J} \& \au{Sharma, Ati~S}} \yr{2010}  \at{A
  critical-layer framework for turbulent pipe flow}.  \jt{Journal of Fluid
  Mechanics}  \bvol{658},  \pg{336--382}.

\bibitem[Meneveau \& Marusic(2013)]{meneveau_marusic_2013}
{\sc \au{Meneveau, Charles} \& \au{Marusic, Ivan}} \yr{2013}  \at{Generalized
  logarithmic law for high-order moments in turbulent boundary layers}.
  \jt{Journal of Fluid Mechanics.}  \bvol{719},  \pg{R1}.

\bibitem[Moarref {\em et~al.\/}(2013)Moarref, Sharma, Tropp \&
  McKeon]{moarref2013model}
{\sc \au{Moarref, Rashad}, \au{Sharma, Ati~S}, \au{Tropp, Joel~A} \&
  \au{McKeon, Beverley~J}} \yr{2013}  \at{Model-based scaling of the streamwise
  energy density in high-reynolds-number turbulent channels}.  \jt{Journal of
  Fluid Mechanics}  \bvol{734},  \pg{275--316}.

\bibitem[Momoh {\em et~al.\/}(1999)Momoh, El-Hawary \& Adapa]{momoh1999review}
{\sc \au{Momoh, James~A}, \au{El-Hawary, ME} \& \au{Adapa, Ramababu}} \yr{1999}
   \at{A review of selected optimal power flow literature to 1993. ii. newton,
  linear programming and interior point methods}.  \jt{IEEE transactions on
  power systems}  \bvol{14}~(1),  \pg{105--111}.

\bibitem[Monty {\em et~al.\/}(2009)Monty, Hutchins, Ng, Marusic \&
  Chong]{monty2009comparison}
{\sc \au{Monty, JP}, \au{Hutchins, N}, \au{Ng, HCH}, \au{Marusic, I} \&
  \au{Chong, MS}} \yr{2009}  \at{A comparison of turbulent pipe, channel and
  boundary layer flows}.  \jt{Journal of fluid mechanics}  \bvol{632},
  \pg{431--442}.

\bibitem[Morra {\em et~al.\/}(2021)Morra, Nogueira, Cavalieri \&
  Henningson]{morra2021colour}
{\sc \au{Morra, Pierluigi}, \au{Nogueira, Petr{\^o}nio~AS}, \au{Cavalieri,
  Andr{\'e}~VG} \& \au{Henningson, Dan~S}} \yr{2021}  \at{The colour of forcing
  statistics in resolvent analyses of turbulent channel flows}.  \jt{Journal of
  Fluid Mechanics}  \bvol{907}.

\bibitem[Morra {\em et~al.\/}(2019)Morra, Semeraro, Henningson \&
  Cossu]{morra2019relevance}
{\sc \au{Morra, Pierluigi}, \au{Semeraro, Onofrio}, \au{Henningson, Dan~S} \&
  \au{Cossu, Carlo}} \yr{2019}  \at{On the relevance of reynolds stresses in
  resolvent analyses of turbulent wall-bounded flows}.  \jt{Journal of Fluid
  Mechanics}  \bvol{867},  \pg{969--984}.

\bibitem[Nakashima {\em et~al.\/}(2017)Nakashima, Fukagata \&
  Luhar]{nakashima2017assessment}
{\sc \au{Nakashima, Satoshi}, \au{Fukagata, Koji} \& \au{Luhar, Mitul}}
  \yr{2017}  \at{Assessment of suboptimal control for turbulent skin friction
  reduction via resolvent analysis}.  \jt{Journal of Fluid Mechanics}
  \bvol{828},  \pg{496--526}.

\bibitem[Pickering {\em et~al.\/}(2021)Pickering, Rigas, Schmidt, Sipp \&
  Colonius]{pickering2021optimal}
{\sc \au{Pickering, Ethan}, \au{Rigas, Georgios}, \au{Schmidt, Oliver~T},
  \au{Sipp, Denis} \& \au{Colonius, Tim}} \yr{2021}  \at{Optimal eddy viscosity
  for resolvent-based models of coherent structures in turbulent jets}.
  \jt{Journal of Fluid Mechanics}  \bvol{917},  \pg{A29}.

\bibitem[Reynolds \& Hussain(1972)]{reynolds1972mechanics}
{\sc \au{Reynolds, WC} \& \au{Hussain, AKMF}} \yr{1972}  \at{The mechanics of
  an organized wave in turbulent shear flow. part 3. theoretical models and
  comparisons with experiments}.  \jt{Journal of Fluid Mechanics}
  \bvol{54}~(2),  \pg{263--288}.

\bibitem[Schmidt {\em et~al.\/}(2018)Schmidt, Towne, Rigas, Colonius \&
  Br{\`e}s]{schmidt2018spectral}
{\sc \au{Schmidt, Oliver~T}, \au{Towne, Aaron}, \au{Rigas, Georgios},
  \au{Colonius, Tim} \& \au{Br{\`e}s, Guillaume~A}} \yr{2018}  \at{Spectral
  analysis of jet turbulence}.  \jt{Journal of Fluid Mechanics}  \bvol{855},
  \pg{953--982}.

\bibitem[Sharma \& McKeon(2013)]{sharma2013coherent}
{\sc \au{Sharma, AS} \& \au{McKeon, BJ}} \yr{2013}  \at{On coherent structure
  in wall turbulence}.  \jt{Journal of Fluid Mechanics}  \bvol{728},
  \pg{196--238}.

\bibitem[de~Silva {\em et~al.\/}(2015)de~Silva, Marusic, Woodcock \&
  Meneveau]{de2015scaling}
{\sc \au{de~Silva, CM}, \au{Marusic, I}, \au{Woodcock, JD} \& \au{Meneveau, C}}
  \yr{2015}  \at{Scaling of second-and higher-order structure functions in
  turbulent boundary layers}.  \jt{Journal of Fluid Mechanics.}  \bvol{769},
  \pg{654--686}.

\bibitem[Smits {\em et~al.\/}(2011)Smits, McKeon \& Marusic]{smits2011high}
{\sc \au{Smits, Alexander~J}, \au{McKeon, Beverley~J} \& \au{Marusic, Ivan}}
  \yr{2011}  \at{High--reynolds number wall turbulence}.  \jt{Annual Review of
  Fluid Mechanics}  \bvol{43},  \pg{353--375}.

\bibitem[Taylor(1938)]{taylor1938spectrum}
{\sc \au{Taylor, Geoffrey~Ingram}} \yr{1938}  \at{The spectrum of turbulence}.
  \jt{Proceedings of the Royal Society of London. Series A-Mathematical and
  Physical Sciences}  \bvol{164}~(919),  \pg{476--490}.

\bibitem[Towne {\em et~al.\/}(2020)Towne, Lozano-Dur{\'a}n \&
  Yang]{towne2020resolvent}
{\sc \au{Towne, Aaron}, \au{Lozano-Dur{\'a}n, Adri{\'a}n} \& \au{Yang, Xiang}}
  \yr{2020}  \at{Resolvent-based estimation of space--time flow statistics}.
  \jt{Journal of Fluid Mechanics}  \bvol{883}.

\bibitem[Towne {\em et~al.\/}(2018)Towne, Schmidt \&
  Colonius]{towne2018spectral}
{\sc \au{Towne, Aaron}, \au{Schmidt, Oliver~T} \& \au{Colonius, Tim}} \yr{2018}
   \at{Spectral proper orthogonal decomposition and its relationship to dynamic
  mode decomposition and resolvent analysis}.  \jt{Journal of Fluid Mechanics}
  \bvol{847},  \pg{821--867}.

\bibitem[Townsend(1976)]{Townsend1976}
{\sc \au{Townsend, A.A.}} \yr{1976} {\em The structure of turbulent shear
  flow\/}, 2nd edn.  \publ{Cambridge University Press}.

\bibitem[Yang {\em et~al.\/}(2020)Yang, Jin, Wu, Yang \& He]{yang2020numerical}
{\sc \au{Yang, Bowen}, \au{Jin, Guodong}, \au{Wu, Ting}, \au{Yang, Zixuan} \&
  \au{He, Guowei}} \yr{2020}  \at{Numerical implementation and evaluation of
  resolvent-based estimation for space--time energy spectra in turbulent
  channel flows}.  \jt{Acta Mechanica Sinica}  \bvol{36},  \pg{775--788}.

\bibitem[Yang {\em et~al.\/}(2016)Yang, Marusic \&
  Meneveau]{yang2016hierarchical}
{\sc \au{Yang, XIA}, \au{Marusic, I} \& \au{Meneveau, C}} \yr{2016}
  \at{Hierarchical random additive process and logarithmic scaling of
  generalized high order, two-point correlations in turbulent boundary layer
  flow}.  \jt{Physical Review Fluids}  \bvol{1}~(2),  \pg{024402}.

\end{thebibliography}

%% End of file `jfm2esam.bib'.

\end{document}